\documentclass[useAMS,fleqn,usenatbib]{mnras}
\usepackage{newtxtext,newtxmath}
\usepackage{ae,aecompl}
\usepackage[T1]{fontenc}

\usepackage{amssymb,amsmath}
\usepackage{wasysym}
\usepackage{color}
\usepackage{epsfig}
\usepackage{longtable}
\usepackage{pdflscape}
\usepackage{mathtools}
\usepackage{float}
\usepackage{footnote}
\usepackage{epstopdf}
\usepackage{units}
\usepackage{enumerate}
\DeclareRobustCommand{\ion}[2]{\textup{#1\,\textsc{\lowercase{#2}}}}

\title[OCCASO II. Physical parameters and Fe abundances of Red Clump stars in 18 Open Clusters]
{OCCASO II. Physical parameters and Fe abundances of Red Clump stars in 18 Open Clusters\footnote{Based on observations made
with the Nordic Optical Telescope, operated by the Nordic Optical Telescope
Scientific Association, and the Mercator Telescope, operated by the Flemish Community, both at the Observatorio del Roque de los Muchachos,
La Palma, Spain, of the Instituto de Astrof\'isica de Canarias.}
\footnote{Based on observations collected at the Centro Astron\'omico Hispano
Alem\'an (CAHA) at Calar Alto, operated jointly by the Max-Planck Institut f\"ur
Astronomie and the Instituto de Astrof\'isica de Andaluc\'ia (CSIC)}}
\author[L. Casamiquela et al.]{L. Casamiquela$^{1}$\thanks{E-mail:
laiacf@fqa.ub.edu} , R. Carrera$^{2,3}$, S. Blanco-Cuaresma$^{4}$, C. Jordi$^{1}$,\newauthor
L. Balaguer-N\'u\~nez$^{1}$, E. Pancino$^{5,6}$, F. Anders$^{7}$, C. Chiappini$^{7}$, L. D\'iaz-P\'erez$^{2,3}$,\newauthor
D. S. Aguado$^{2,3}$, A. Aparicio$^{2,3}$, R. Garcia-Dias$^{2,3}$, U. Heiter$^{8}$,\newauthor
C. E. Mart\'inez-V\'azquez$^{2,3}$, S. Murabito$^{2,3}$, A. del Pino$^{9}$\\
$^{1}$Departament de F\'isica Qu\`antica i Astrof\'isica, Universitat de Barcelona, ICC/IEEC, 08007 Barcelona, Spain\\
$^{2}$Instituto de Astrof\'isica de Canarias, La Laguna, 38205 Tenerife, Spain\\
$^{3}$Departamento de Astrof\'isica, Universidad de La Laguna, 38207 Tenerife, Spain\\
$^{4}$Observatoire de Gen\`eve, Universit\'e de Gen\`eve, 1290, Versoix, Switzerland\\
$^{5}$INAF - Osservatorio Astrofisico di Arcetri, Largo Enrico Fermi 5, 50125 Firenze, Italy\\
$^{6}$ASI Science Data Center, Via del Politecnico SNC, 00133 Roma, Italy\\
$^{7}$Leibniz-Institut f\"ur Astrophysik Potsdam (AIP), An der Sternwarte 16, 14482 Potsdam, Germany\\
$^{8}$Observational Astrophysics, Department of Physics and Astronomy, Uppsala University, Box 516, 75120 Uppsala, Sweden\\
$^{9}$Nicolaus Copernicus Astronomical Centre of the Polish Academy of Sciences. ul. Bartycka 18 00-716, Warsaw\\}

\date{Accepted --. --; }
\pubyear{2016}

\begin{document}
\label{firstpage}
\pagerange{\pageref{firstpage}--\pageref{lastpage}} \pubyear{2016}
\maketitle

\begin{abstract}
\textit{Open Clusters have long been used to study the chemo-dynamical evolution of the
Galactic disk. This requires an homogeneously analysed sample covering a wide range of ages and distances. In this paper we present the OCCASO second data release. This comprises a sample of high-resolution ($R>65,000$) and high signal-to-noise spectra of 115 Red Clump stars in 18 Open Clusters. We derive atmospheric parameters ($T_{\mathrm{eff}}$, $\log g$, $\xi$), and [Fe/H] abundances using two analysis techniques: equivalent widths and spectral synthesis. A detailed comparison and a critical review of the results of the two methods are made. Both methods are carefully tested between them, with the \emph{Gaia} FGK Benchmark stars, and with an extensive sample of literature values. We perform a membership study using radial velocities and the resulting abundances. Finally, we compare our results with a chemo-dynamical model of the Milky Way thin disk concluding that the oldest Open Clusters are consistent with the models only when dynamical effects are taken into account.}
\end{abstract}



\section{Introduction}
The Open Clusters Chemical Abundances from Spanish Observatories (OCCASO) survey \citep[Paper I hereafter]{Casamiquela+2016} is a high-resolution spectroscopic survey of Open Clusters (OCs). It was designed to obtain accurate radial velocities and homogeneous chemical abundances for around 30 different species in Northern OCs. A list of 25 candidate OCs were selected taking into account ages, metallicities, and positions in the Galactic disk. In Paper I there is a full description of the motivation, design and strategy of the survey. Also radial velocities for 77 stars in 12 OCs were analysed to obtain an accurate membership selection. 
We included a very detailed description of the used instruments and the observational strategy. In brief, OCCASO observations are performed with high-resolution echelle spectrographs available at Spanish observatories: CAFE at the 2.2 m telescope in the Centro Astron\'omico Hispano-Alem\'an (CAHA), FIES at the 2.5 m NOT telescope in the Observatorio del Roque de los Muchachos (ORM) and HERMES at the 1.2 m Mercator telescope also in the ORM. These instruments have similar resolution $R\geq 65,000$ and wavelength range coverages $4000\leq \lambda \leq 9000$ \AA. The typical obtained signal-to-noise ratios (SNR) were around 70.

In this paper we present the analysis of atmospheric parameters and iron abundances for the whole sample of stars in 18 OCs: 12 OCs from Paper I (observations completed by January 2015), plus 6 new OCs (38 stars) finished until August 2016. The analysis is done using two different methods widely used in the literature: equivalent widths (EW), and spectral synthesis (SS). A detailed analysis of the differences found using both methods is performed as well as a wide comparison with the literature.
 
The analysed OCs cover Galactocentric distances between 6.8 and 10.7 kpc, and ages between 300 Myr and 10.2 Gyr. This coverage allows a first investigation of the iron abundance gradient in the Milky Way disk and its change with time. Our sample has the advantage that is done from high-resolution spectra, it is large and has been analysed homogeneously. Our data allows the study of up to 35 chemical species, which will be analysed in a further paper in preparation.
 
This paper is organized as follows: we present an overview of the used data in Section 2, the analysis strategy is detailed in Section 3, which includes the used line list in Section 3.1 and model atmosphere in Section 3.2, and the description of the analysis methods in Section 3.3. The calculation of the atmospheric parameters is detailed in Section 4, where we include the comparison between the two methods (Section 4.1), the results for the  Benchmark Stars (Section 4.2), and an external check with photometric parameters (Section 4.3). Results on iron abundances are presented in Section 5, where we include an analysis of the performance of the methods (Section 5.1). An analysis cluster-by-cluster is done in Section 6, and an extensive comparison with the literature in Section 7. Finally, a preliminary discussion related to the Galactic disk gradients is presented in Section 8, and the summary is provided in Section 9.

\section{OCCASO second data release}
\begin{figure*}
\begin{minipage}{140mm}
\begin{centering}
\includegraphics[width=\textwidth]{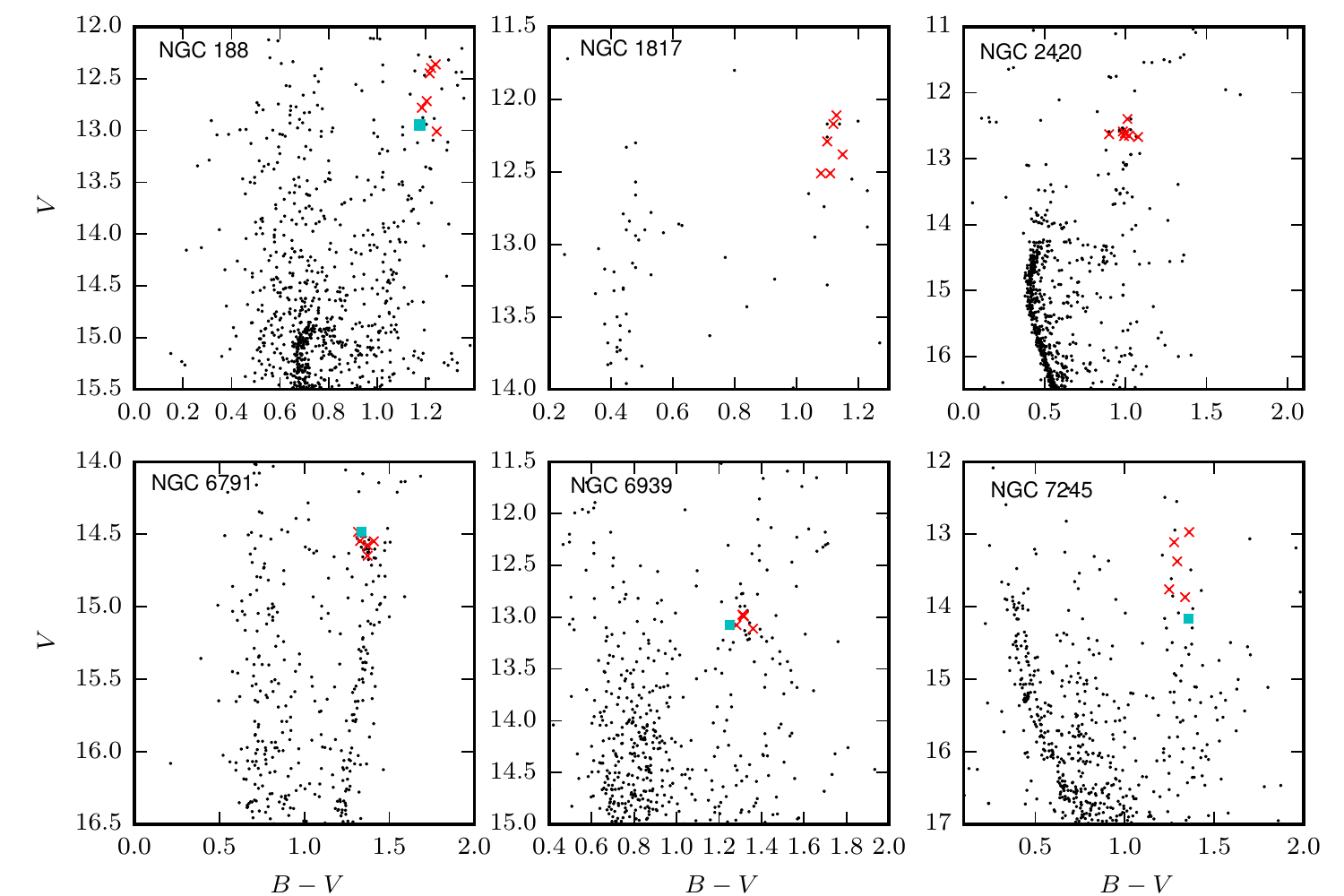}
\caption{(B-V), V colour-magnitude diagrams of the newly completed clusters (references are listed in Table~\ref{clusters}). The red crosses indicate target stars, and cyan squares indicate stars that we have found to be probably non-members or spectroscopic binaries from the radial velocity study.}\label{fig:cmd}
\end{centering}
\end{minipage}
\end{figure*}

The second data release of OCCASO includes the analysis of high resolution spectra of 115 stars belonging to 18 OCs. The details of the observational material can be found in Sec.~\ref{sec:observ}.The general properties of the 18 OCs are summarized in Table~\ref{clusters}, where the 6 added clusters with respect to Paper I are marked in bold. Colour-magnitude diagrams (CMDs) from the available photometries for these 6 OCs are plotted in Fig.~\ref{fig:cmd}. CMDs for the previous 12 OCs were presented in Paper I.

Radial velocity measurements for the 38 stars in the 6 added OCs will be detailed in a future paper (Casamiquela et al. 2017, in prep.). We have made a membership analysis of these OCs using the same criteria as in Paper I. That is, rejecting those stars that have a $v_{\mathrm{r}}$ not compatible at the 3$\sigma$ level of the radial velocity of the cluster. We have found 3 probable non-member stars or spectroscopic binaries: NGC~6791 W3899, NGC~6939 W130 and NGC~7245 W045.

\begin{table}
  \caption{\label{clusters}Clusters of OCCASO completed by the end of August 2016. Newly added clusters to those of Paper I are marked in bold. distance from the Sun $D$, $R_{\text{GC}}$, $z$ are from \citet{Dias+2002}. We list the $V$ magnitude of the Red Clump and the number of stars observed. The photometry used to select the target stars is indicated as a footnote.}
\setlength\tabcolsep{3.5pt}
\begin{centering}
\begin{tabular}{ccccccc}
\hline 
Cluster & $D$ & $R_{\text{GC}}$ & $z$ & Age & $V_{\text{RC}}$ &Num. Stars \\
 & (kpc) & (kpc) & (pc) & (Gyr) & &  \\
\hline 
IC 4756$^1$ & 0.48 & 8.14 & +41 & 0.8$^\text{a}$ & 9 & 8 \\
\textbf{NGC 188}$^2$ & 1.71 & 9.45 & +651 & 6.3$^\text{a}$ & 12.5 & 6 \\
NGC 752$^3$  & 0.46 & 8.80 & -160 & 1.2$^\text{a}$ & 9 & 7 \\
\textbf{NGC 1817}$^4$ & 1.97 & 10.41 & -446 & 1.1$^\text{a}$ & 12.5 & 5\\
NGC 1907$^5$  & 1.80 & 10.24 & +9 & 0.4$^\text{b}$ & 9 & 6 \\
NGC 2099$^6$ & 1.38 & 9.87 & +74 & 0.4$^\text{c}$ & 12 & 7 \\
\textbf{NGC 2420}$^7$ & 2.48 & 10.74 & +833 & 2.2$^\text{a}$ & 12.5 & 7\\
NGC 2539$^8$ & 1.36 & 9.37 & +250 & 0.7$^\text{d}$ & 11 & 6 \\
NGC 2682$^9$ & 0.81 & 9.16 & +426 & 4.3$^\text{a}$ & 10.5 & 8 \\
NGC 6633$^{10}$ & 0.38 & 8.20 & +54 & 0.6$^\text{e}$ & 8.5 & 4$^{\star}$ \\
NGC 6705$^{11}$ & 1.88 & 6.83 & -90 & 0.3$^\text{f}$ & 11.5 & 8 \\
\textbf{NGC 6791}$^{12}$& 5.04 & 8.24 & +953 & 10.2$^\text{a}$ & 14.5 & 7\\
NGC 6819$^{13}$ & 2.51 & 8.17 & +370 & 2.9$^\text{a}$ & 13 & 6 \\
\textbf{NGC 6939}$^{14}$ & 1.80 & 8.86 & +384 & 1.3$^\text{g}$ & 13 & 6 \\
NGC 6991$^{15}$ & 0.70 & 8.47 & +19 & 1.3$^\text{h}$ & 10 & 6 \\
\textbf{NGC 7245}$^{16}$ & 3.47 & 9.79 & -112 & 0.4$^\text{i}$ & 13 & 6 \\
NGC 7762$^{14}$ & 0.78 & 8.86 & +79 & 2.5$^\text{j}$ & 12.5 & 6 \\
NGC 7789$^{17}$ & 1.80 & 9.41 & -168 & 1.8$^\text{a}$ & 13 & 7 \\
\hline
\end{tabular}
\end{centering}
$^1$\citet{Alcaino1965}; $^2$\citet{Platais+2003}; $^3$\citet{Johnson1953}; $^4$\citet{Harris+1977}; $^5$\citet{Pandey+2007}; 
$^6$\citet{Kiss+2001}; $^7$\citet{AnthonyTwarog+1990}; $^8$\citet{Choo+2003}; $^9$\citet{Montgomery+1993};
$^{10}$\citet{Harmer+2001}; $^{11}$\citet{Sung+1999}; $^{12}$\citet{Stetson+2003}; $^{13}$\citet{Rosvick+1998}; 
$^{14}$\citet{Maciejewski+2007}; $^{15}$\citet{Kharachenko+2005}; $^{16}$\citet{Subramaniam+2007}; $^{17}$\citet{Mochejska+1999,McNamara+1981}.\\
$^\text{a}$\citet{Salaris+2004}; $^\text{b}$\citet{Subramaniam+1999}; $^\text{c}$\citet{Nilakshi+2002}; $^\text{d}$\citet{Vogel+2003}; $^\text{e}$\citet{Jeffries+2002}; $^\text{f}$\citet{cantatgaudin+2014b}; $^\text{g}$\citet{Andreuzzi+2004}; $^\text{h}$\citet{Karchenko+2005}; $^\text{i}$\citet{Subramaniam+2007}; $^\text{j}$\citet{Carraro+2016}\\
$^{\star}$It has only 4 stars in the RC but was included for observation in a night
with non optimal weather conditions.
\end{table}

\subsection{Observational material}\label{sec:observ}

The current work uses observations of the runs described in Paper I (53 nights of observations between January 2013 - January 2015), which include data for 12 OCs. And also we incorporate five additional runs: 28 nights between April 2015 and August 2016. This makes a total of 81 nights of observations. With the whole set of data we are capable to analyse 115 stars in 18 OCs. Additionally, Arcturus ($\alpha$-Bootes) and $\mu$-Leo, two extensively studied stars, part of the \textit{Gaia} FGK Benchmark Stars (GBS) \citep{Heiter+2015} and of the APOGEE \citep{Frinchaboy+2013} reference stars, were observed with the three telescopes for the sake of comparison. Details of the runs (April 2015 - August 2016), dates, instruments and radial velocity accuracies will be described in Casamiquela et al. 2017 (in prep.).

We have modified the data reduction strategy with respect to the one explained in Paper I to improve the quality of the final spectra. We have built our own pipeline (see Appendix) to perform skyline subtraction, telluric correction, normalization and order merging. These improvements do not change the radial velocities from Paper I, but they are important for the atmospheric parameters and the abundances determination.

\subsubsection{Benchmark stars}\label{sec:BS}
Aside of our own observational material, we also analyse a sample of GBS. The GBS are a set of calibration stars, covering different regions of the HR diagram and spanning a wide range in metallicity. For these stars there exists enough data to determine effective temperature and surface gravity independently from spectroscopy by using their angular diameter measurements and bolometric fluxes. These determinations and related uncertainties are fully described in \citet{Heiter+2015}. Reference metallicities also exist for these stars, and are determined from a careful spectroscopic study by \citet{Jofre+2014}.

We retrieved the data from the library of high-resolution optical spectra of the GBS \citep{Blanco+2014}. This library includes 100 high SNR spectra of 34 stars from the spectrographs HARPS, NARVAL, UVES, and ESPaDOnS, which cover the visual spectral range ($4800\leq \lambda \leq 6800$ \AA). Taking into account our target stars, we have selected the GBS that covered the appropriate range of the parameter space: $4000\leq T_{\mathrm{eff}}\leq6650$ (K), $1.1\leq \log g \leq 4.5$, [Fe/H]$\geq-1.5$, with 23 GBS fulfilling these criteria. We have degraded the resolution of the spectra to a common resolution of $62,000$ to analyse them homogeneously with our OCCASO spectra.

\section{Analysis strategy}
The high-resolution and large wavelength coverage of the spectra allows for the determination of a large number of astrophysical quantities: effective temperature ($T_{\text{eff}}$), surface gravity ($\log g$), microturbulence ($\xi$), overall stellar metallicity [M/H], and individual abundances for more than 30 chemical species.

In this section we summarize the analysis strategy: line list used, adopted model atmospheres, and analysis methods.

\subsection{Line list}
We used the Gaia-ESO Survey line list which is a compilation of experimental and theoretical atomic and molecular data that is being updated and improved regularly. It is convenient for our study because it covers the wavelength range of our instruments, it has been extensively used in the literature, and its atomic parameters are recent. Details of this compilation are provided in \citet{Heiter+2015b}.

In the present work, we have used version 5, which covers a wavelength range between $4200 \le \lambda \le 9200$ \AA. Collisional broadening by hydrogen is treated considering the theory by Anstee, Barklem and O'Mara \citep{AnsteeOmara1991,BarklemOmara1998}. It contains atomic information for 35 different chemical species: Li, C, N, O, Na, Mg, Al, Si, S, K, Ca, Sc, Ti, V, Cr, Mn, Fe, Co, Ni, Cu, Zn, Rb, Sr, Y, Zr, Mo, Ru, Ba, La, Ce, Pr, Nd, Sm, Eu, Dy.

We have used two different analysis methods (see Sec.~\ref{sec:analysis}), so, even though the master line list is the same, each method chooses independently the most suitable lines. The line selection by each method is explained in Sec.~\ref{sec:analysis}.

\subsection{Model atmospheres}
We adopted the MARCS grid\footnote{\url{http://marcs.astro.uu.se/}} model atmospheres of \citet{Gustafsson2008}. It is an extensive grid of $10^4$ spherically-symmetric models (supplemented with plane-parallel for the highest surface gravities)  for stars with 2500 $\le T_{\text{eff}} \le$ 8000 K, 0 $\le \log g \le$ 5 (cgs) with various masses and radii, and -5 $\le \text{[M/H]} \le$ +1. Underlying assumptions in addition to 1D stratification (spherical or plane-parallel) include hydrostatic equilibrium, mixing-length convection and local thermodynamic equilibrium. The standard MARCS models assume Solar abundances of \citet{Grevesse+2007} and $\alpha-$enhancement at low metallicities.

\subsection{Analysis methods}\label{sec:analysis}
There are two state-of-the-art methodologies currently employed in the literature: EW and SS. We used these two approaches to determine atmospheric parameters and abundances. The strategy of applying multiple pipelines to determine atmospheric parameters and abundances is applied in other surveys such as the Gaia-ESO Survey \citep[GES][]{Gilmore+2012}, as explained in \citet{Smiljanic+2014}. This strategy has the advantage that allows the investigation of method-dependent effects, different sources of uncertainty, and provides an estimation of the accuracy of the derived parameters and abundances.

Both methods ran independently on the same spectra, with a common master line list and model atmospheres to guarantee some internal consistency.

\subsubsection*{EW: DAOSPEC+GALA}
DAOSPEC+GALA is our EW method. It consists in two steps performed by two different codes.

First, EWs were measured using DOOp \citep{cantatgaudin+2014} which is an automatic wrapper for DAOSPEC \citep{Stetson+2008}. DAOSPEC is a Fortran code that finds absorption lines in a stellar spectrum, fits the continuum, measures EWs, identifies lines from a provided line list, and gives a radial velocity estimate. DOOp optimizes the most critical DAOSPEC parameters in order to obtain the best measurements of EWs. In brief, it fine tunes the FWHM and the continuum placement among other parameters, through a fully automatic and iterative procedure.

The determination of the atmospheric parameters was done with the GALA code \citep{Mucciarelli+2013}. It is based on the set of Kurucz abundance calculation codes \citep[WIDTH9, ][]{Sbordone+2004,Kurucz2005}. GALA optimizes atmospheric parameters ($T_{\text{eff}}$, $\log g$, $\xi$, [M/H]) using the classical spectroscopic method based on iron lines. The $T_{\text{eff}}$ is optimized by minimizing the slope of the iron abundance versus excitation potential. The difference of abundances between neutral iron \ion{Fe}{I} and ionised iron \ion{Fe}{II} lines is used to constrain the surface gravity. The angular coefficient in the iron abundance-EW is used to optimize the microturbulence. And the average Fe abundance to constrain the global metallicty of the model. GALA measures the line abundances and performs a rejection of lines of the same chemical species using a threshold on too weak or too strong lines (we use $-5.9 \lesssim \log (\frac{\text{EW}}{\lambda}) \lesssim -4.7$), a limit in the EW error measured by DAOSPEC (we choose $\sim 15 \%$ depending on the SNR of the star), and finally performing a $\sigma$ clipping rejection in abundance (we choose $2.5\sigma$).

To select the lines for this method we use pre-selection of the Gaia-ESO v5 master line list. This compilation is done by one of the GES nodes (P. Donati, priv. comm.), and it is suitable for an EW analysis since lines are checked for blends with synthesis. We further perform a cleaning process to select lines that provide consistent abundances, and to get rid of blends or lines with bad atomic parameters. This process is divided in two steps. Firstly, \ion{Fe}{I} and \ion{Fe}{II} lines detected by DAOSPEC in less than three stars were rejected. This provides a better determination of the FWHM and the continuum placement. Afterwards, \ion{Fe}{I} and \ion{Fe}{II} lines that were rejected by GALA in all the stars, or that gave systematically discrepant abundances with respect to the mean Fe abundance were discarded. The cleaned line list fed to DAOSPEC is detailed in Table~\ref{tab:linelist}.

\begin{table}
\begin{centering}
  \small
  \centering
  \caption{\label{tab:linelist}\ion{Fe}{I} and \ion{Fe}{II} lines within our line list, used by the EW analysis method. Excitation potential $\chi$, and oscillator strengths $\log gf$ are listed. References for the $\log gf$ are listed in the last column. When two references separated by comma are listed, it means that the mean value of the $\log gf$ is taken. When two references separated by "|" are listed, it means that the $\log gf$ from the first source was brought onto the same scale as the second. The complete version of the table is available as online data. Here only few lines are shown.}
  \setlength\tabcolsep{3.1pt}
\begin{tabular}{ccccc}
\hline 
$\lambda$ (\AA) & Element& $\chi\,\left( \mathrm{eV} \right)$ & $\log gf$ & Ref \\
\hline
5012.695 & \ion{Fe}{I} & 4.283 & -1.690 & MRW\\
5044.211 & \ion{Fe}{I} & 2.851 & -2.038 & BK, BWL\\
5058.496 & \ion{Fe}{I} & 3.642 & -2.830 & RW70 | FMW\\
5088.153 & \ion{Fe}{I} & 4.154 & -1.680 & MRW\\
\hline
\end{tabular}
\flushleft References. MRW: \citet{MRW},  R14: \citet{R14},  K07: \citet{K07},  BWL: \citet{BWL},  BK: \citet{BK},  GESHRL14: \citet{GESHRL14},  RW70: \citet{RW70},  FMW: \citet{FMW},  GESB82c: \citet{GESB82c},  GESB79c: \citet{GESB79b},  BKK: \citet{BKK},  WBW: \citet{WBW},  WBW70: \citet{WBW70},  BIPS: \citet{BIPS},  GESHRL14: \citet{GESHRL14},  GESB82d: \citet{GESB82d}, GESB86: \citet{GESB86},  FW06: \cite{FW06},  KKS84: \citet{KKS84},  RU: \citet{RU},  MB09: \citet{MB09}
\end{centering}
\end{table}

\subsubsection*{SS: iSpec}

iSpec \citep{BlancoCuaresma+2014} is a tool that can be used to perform spectroscopic manipulations such as determine/correct radial velocities, normalize and degrade the spectral resolution. And more importantly, it also offers the possibility to derive atmospheric parameters and chemical abundances by using the EW method and the SS fitting technique with many different atomic line lists, model atmosphere and radiative transfer codes.

In this work, iSpec was used to prepare the custom library of GBS (as described in Section~\ref{sec:BS}) and a customized pipeline was developed to analyse OCCASO targets using the SS technique. iSpec compares regions of the observed spectrum with synthetic ones generated on-the-fly using SPECTRUM \citep{Gray+1994}. A least-square algorithm minimizes the differences between the synthetic and observed spectra until it converges into a final set of atmospheric parameters.

In the analysis by iSpec, the line selection was done based on the automatic detection  of absorption lines in the NARVAL solar spectrum included in the GBS library. Each line was cross-matched with the atomic line list and we derived solar line-by-line chemical abundances using the reference atmospheric parameters for the Sun. Good lines lead to abundances similar to the solar ones \citep[i.e.][]{Grevesse+2007}, thus we selected all lines with an abundance that falls in the range $\pm$0.05 dex. Additionally, in our analysis we used the wings of H$\alpha$/$\beta$ and Mg triplet, which helps us to break degeneracies.

\section{Atmospheric parameters}\label{sec:AP}
Our final goal is to calculate detailed abundances from the spectra. To do so, one has to first determine atmospheric parameters $T_{\text{eff}}$, $\log g$, $\xi$ and [M/H] to then derive individual abundances from a fixed model atmosphere for each line/species.

\subsection{Results from GALA and iSpec}
Both methods analysed the same dataset of 115 stars corresponding to 18 OCs, as well as the reference stars Arcturus and $\mu$-Leo observed with every instrument. For 17 out of these 117 stars we repeated observations with more than one instrument, for comparison purposes. In total, we analysed 154 spectra, 62 corresponding to FIES, 81 to HERMES, and 11 to CAFE.

\begin{figure}
 \centering
\includegraphics[width=0.4\textwidth]{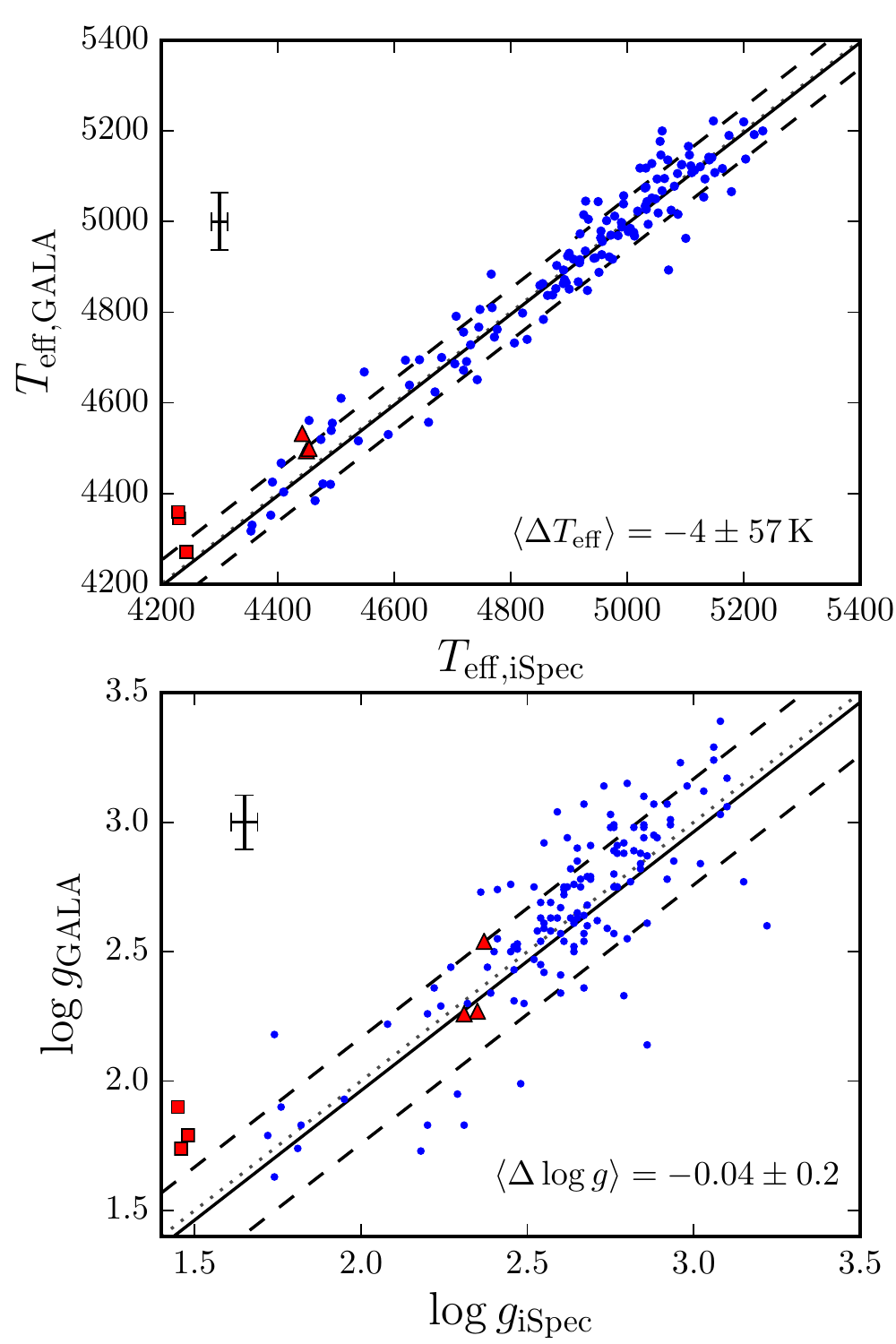}
 \caption{Comparison of the effective temperature and surface gravity from GALA and iSpec analysis. Red symbols indicate the values of Arcturus (squares) and $\mu$-Leo (triangles). The solid line stands for the mean difference, and the dashed lines indicate the $1\sigma$ level. The dotted line is the 1:1 relation. In the top left corner of each panel we plot the mean errors in X and Y axis.}\label{fig:APcompare}
\end{figure}

The two pipelines have run letting all the atmospheric parameters free for the 154 spectra. Fig.~\ref{fig:APcompare} shows the comparison of the resulting $T_{\text{eff}}$ and $\log g$ with GALA and iSpec. The dispersion in effective temperature (57 K) is compatible with the errors estimated by the GALA, 68 K in average, but not with iSpec ones, 14 K (mean errors are drawn in the plot). The dispersion in surface gravity (0.2 dex) is large considering the mean errors (0.11 and 0.04, respectively, drawn in the plot). It is well known that surface gravity is the most difficult quantity to derive from spectroscopy. Comparing the results of GALA and iSpec we obtain differences similar with other studies in the literature, like GES iDR1 and iDR2 node-to-node dispersions \citep{Smiljanic+2014}.

In Table~\ref{tab:finpar} we list the $T_{\text{eff}}$, $\log g$ and $\xi$ and their errors, derived by the two methods. If we compare between methods we see that the $T_{\text{eff}}$ dispersion is consistent with the uncertainties. For $\log g$, at least one of the error estimations is too optimistic.

\newpage
\onecolumn
\begin{landscape}
\centering
\tiny
\LTcapwidth=\textwidth
\setlength\tabcolsep{2pt}
\begin{longtable}{cccccccccccccccccccccccccccccccc}
  \caption{\label{tab:finpar}Atmospheric parameters and iron abundances obtained for the stars analysed in OCCASO. Basic data of each star, SNR and the instrument used, is listed in the first 7 columns. $T_{\mathrm{eff}}$, $\log g$ and $\xi$ derived with each method are in columns 8-13. Average effective temperature $T_{\mathrm{eff}}$ and surface gravity $\log g$ in columns 14 and 17. Two errors are given: the mean of the errors quoted by both methods $\delta_1$, and the standard deviation between the two values $\delta_2$. [Fe/H] derived with each method with the errors as described in the text (Section~\ref{sec:FeH}), is listed in columns 20-21. $\sigma$[Fe/H] stands for the standard deviation of the two [Fe/H] determinations.}\\
Cluster & Star & RA & DEC & V & SNR & Instr & $T_{\mathrm{eff}}$ & $\log g$ & $\xi$ & $T_{\mathrm{eff}}$ & $\log g$ & $\xi$ & $T_{\mathrm{eff}}$ & $\delta_1 T$ & $\delta_2 T$ & $\log g$ & $\delta_1 \log g$ & $\delta_2 \log g$ & [Fe/H]$_{\text{EW}}$ & [Fe/H]$_{\text{SS}}$ & $\sigma$[Fe/H] \\
  & & & & & & & \multicolumn{3}{|c|}{EW} & \multicolumn{3}{|c|}{SS} & & & & & & & & & &  \\ 
\hline 
- & Arcturus & 14:15:39.672 & +19:10:56.67 & -0.05 &  715 & CAFE & $4359\pm56$ & $1.90\pm0.18$ & $1.38\pm0.05$ & $4228\pm6$ & $1.45\pm0.03$ & $1.67\pm0.01$ & 4293 & 31 & 92 & 1.68 & 0.10 & 0.32 & $-0.56\pm0.04$ & $-0.51\pm0.06$ & 0.03 \\ 
  &   &   &   &   &  399 & FIES & $4345\pm53$ & $1.79\pm0.14$ & $1.49\pm0.06$ & $4230\pm4$ & $1.48\pm0.02$ & $1.68\pm0.01$ & 4287 & 28 & 80 & 1.64 & 0.08 & 0.22 & $-0.55\pm0.05$ & $-0.55\pm0.05$ & 0.00 \\ 
  &   &   &   &   &  414 & HERMES & $4271\pm55$ & $1.74\pm0.10$ & $1.35\pm0.06$ & $4243\pm3$ & $1.46\pm0.01$ & $1.66\pm0.01$ & 4257 & 29 & 19 & 1.60 & 0.06 & 0.20 & $-0.51\pm0.05$ & $-0.58\pm0.05$ & 0.03 \\ 
- & MuLeo & 09:52:45.817 & +26:00:25.03 & 3.88 &  149 & CAFE & $4499\pm93$ & $2.54\pm0.24$ & $1.46\pm0.18$ & $4453\pm8$ & $2.37\pm0.03$ & $1.48\pm0.02$ & 4476 & 50 & 31 & 2.46 & 0.14 & 0.12 & $0.21\pm0.05$ & $0.07\pm0.07$ & 0.07 \\ 
  &   &   &   &   &  396 & FIES & $4532\pm112$ & $2.27\pm0.21$ & $1.15\pm0.16$ & $4442\pm3$ & $2.35\pm0.01$ & $1.47\pm0.01$ & 4487 & 57 & 63 & 2.31 & 0.11 & 0.06 & $0.31\pm0.05$ & $0.17\pm0.06$ & 0.07 \\ 
  &   &   &   &   &  161 & HERMES & $4494\pm88$ & $2.26\pm0.15$ & $1.14\pm0.09$ & $4449\pm4$ & $2.31\pm0.02$ & $1.46\pm0.01$ & 4471 & 46 & 31 & 2.28 & 0.08 & 0.04 & $0.30\pm0.05$ & $0.14\pm0.05$ & 0.08 \\ 
IC4756 & W0038 & 18:37:05.22 & +05:17:31.6 & 9.77 &  81 & FIES & $5136\pm52$ & $3.10\pm0.07$ & $1.43\pm0.13$ & $5069\pm12$ & $2.85\pm0.04$ & $1.45\pm0.02$ & 5102 & 32 & 46 & 2.98 & 0.06 & 0.18 & $-0.04\pm0.05$ & $-0.04\pm0.05$ & 0.00 \\ 
  & W0042 & 18:37:20.77 & +05:53:43.1 & 9.46 &  106 & HERMES & $5200\pm33$ & $3.06\pm0.06$ & $1.06\pm0.04$ & $5232\pm14$ & $3.10\pm0.03$ & $1.31\pm0.02$ & 5216 & 23 & 23 & 3.08 & 0.04 & 0.03 & $0.03\pm0.04$ & $-0.01\pm0.05$ & 0.02 \\ 
  & W0044 & 18:37:29.72 & +05:12:15.5 & 9.79 &  68 & HERMES & $5222\pm60$ & $3.29\pm0.07$ & $1.16\pm0.04$ & $5147\pm17$ & $3.06\pm0.03$ & $1.41\pm0.02$ & 5184 & 38 & 52 & 3.18 & 0.05 & 0.16 & $0.04\pm0.04$ & $-0.03\pm0.05$ & 0.04 \\ 
  & W0049 & 18:37:34.22 & +05:28:33.5 & 9.43 &  68 & HERMES & $5126\pm45$ & $2.89\pm0.06$ & $1.28\pm0.05$ & $5093\pm13$ & $2.76\pm0.04$ & $1.46\pm0.02$ & 5109 & 29 & 22 & 2.82 & 0.05 & 0.09 & $-0.02\pm0.04$ & $-0.07\pm0.05$ & 0.02 \\ 
  & W0081 & 18:38:20.76 & +05:26:02.3 & 9.38 &  72 & HERMES & $5220\pm44$ & $3.12\pm0.07$ & $1.11\pm0.05$ & $5200\pm18$ & $3.03\pm0.03$ & $1.27\pm0.02$ & 5210 & 31 & 14 & 3.08 & 0.05 & 0.06 & $0.02\pm0.04$ & $-0.05\pm0.05$ & 0.04 \\ 
  & W0101 & 18:38:43.79 & +05:14:20.0 & 9.38 &  78 & HERMES & $5136\pm36$ & $3.07\pm0.07$ & $1.26\pm0.05$ & $5141\pm11$ & $2.88\pm0.04$ & $1.45\pm0.02$ & 5138 & 23 & 3 & 2.98 & 0.06 & 0.13 & $0.02\pm0.04$ & $-0.04\pm0.05$ & 0.03 \\ 
  & W0109 & 18:38:52.93 & +05:20:16.5 & 9.02 &  87 & CAFE & $4973\pm43$ & $2.55\pm0.12$ & $1.34\pm0.06$ & $4919\pm10$ & $2.41\pm0.04$ & $1.59\pm0.02$ & 4946 & 26 & 38 & 2.48 & 0.08 & 0.10 & $-0.07\pm0.04$ & $-0.07\pm0.05$ & 0.00 \\ 
  &   &   &   &   &  114 & FIES & $4917\pm49$ & $2.52\pm0.08$ & $1.23\pm0.11$ & $4975\pm12$ & $2.64\pm0.03$ & $1.57\pm0.02$ & 4946 & 30 & 41 & 2.58 & 0.06 & 0.08 & $0.02\pm0.05$ & $-0.05\pm0.05$ & 0.04 \\ 
  &   &   &   &   &  87 & HERMES & $4969\pm45$ & $2.64\pm0.05$ & $1.33\pm0.05$ & $4984\pm12$ & $2.67\pm0.03$ & $1.50\pm0.02$ & 4976 & 28 & 10 & 2.66 & 0.04 & 0.02 & $-0.02\pm0.04$ & $-0.07\pm0.05$ & 0.02 \\ 
  & W0125 & 18:39:17.88 & +05:13:48.8 & 9.29 &  75 & CAFE & $5123\pm56$ & $2.80\pm0.09$ & $1.30\pm0.07$ & $5109\pm13$ & $2.76\pm0.04$ & $1.54\pm0.02$ & 5116 & 34 & 9 & 2.78 & 0.06 & 0.03 & $-0.02\pm0.04$ & $-0.06\pm0.05$ & 0.02 \\ 
  &   &   &   &   &  82 & FIES & $5108\pm46$ & $2.88\pm0.05$ & $1.22\pm0.07$ & $5110\pm13$ & $2.77\pm0.04$ & $1.51\pm0.02$ & 5109 & 29 & 2 & 2.82 & 0.04 & 0.08 & $0.02\pm0.05$ & $-0.03\pm0.05$ & 0.02 \\ 
  &   &   &   &   &  75 & HERMES & $5121\pm41$ & $2.87\pm0.05$ & $1.32\pm0.05$ & $5125\pm11$ & $2.86\pm0.04$ & $1.47\pm0.02$ & 5123 & 26 & 3 & 2.86 & 0.04 & 0.01 & $-0.04\pm0.04$ & $-0.09\pm0.05$ & 0.02 \\ 
NGC188 & W1105 & 0:46:59.62 & 85:13:15.80 & 12.36 &  64 & HERMES & $4530\pm114$ & $2.29\pm0.14$ & $1.25\pm0.07$ & $4589\pm10$ & $2.24\pm0.04$ & $1.36\pm0.02$ & 4559 & 62 & 42 & 2.26 & 0.09 & 0.04 & $-0.01\pm0.04$ & $-0.10\pm0.06$ & 0.04 \\ 
  & W2051 & 0:42:25.55 & 85:16:22.03 & 12.95 &  50 & HERMES & $4668\pm63$ & $2.92\pm0.16$ & $0.87\pm0.10$ & $4548\pm15$ & $2.55\pm0.04$ & $1.14\pm0.02$ & 4608 & 39 & 84 & 2.74 & 0.10 & 0.26 & $0.22\pm0.05$ & $-0.01\pm0.06$ & 0.12 \\ 
  & W2088 & 0:47:18.42 & 85:19:45.78 & 13.01 &  49 & HERMES & $4516\pm60$ & $2.44\pm0.15$ & $1.14\pm0.07$ & $4538\pm10$ & $2.38\pm0.04$ & $1.22\pm0.02$ & 4527 & 35 & 15 & 2.41 & 0.10 & 0.04 & $0.04\pm0.04$ & $-0.08\pm0.06$ & 0.06 \\ 
  & W5217 & 0:54:11.48 & 85:15:23.19 & 12.40 &  56 & HERMES & $4639\pm65$ & $2.30\pm0.13$ & $1.30\pm0.07$ & $4626\pm10$ & $2.32\pm0.05$ & $1.47\pm0.02$ & 4632 & 37 & 9 & 2.31 & 0.09 & 0.01 & $0.03\pm0.04$ & $-0.08\pm0.06$ & 0.06 \\ 
  & W5224 & 0:54:36.60 & 85:1:15.31 & 12.45 &  60 & HERMES & $4695\pm48$ & $2.31\pm0.42$ & $1.33\pm0.07$ & $4643\pm11$ & $2.46\pm0.04$ & $1.40\pm0.02$ & 4669 & 29 & 36 & 2.38 & 0.23 & 0.11 & $0.04\pm0.04$ & $0.01\pm0.05$ & 0.02 \\ 
  & W7323 & 0:49:5.60 & 85:26:7.78 & 12.72 &  71 & FIES & $4519\pm103$ & $2.74\pm0.21$ & $1.53\pm0.11$ & $4474\pm8$ & $2.41\pm0.03$ & $1.34\pm0.02$ & 4496 & 55 & 31 & 2.58 & 0.12 & 0.23 & $0.01\pm0.05$ & $0.02\pm0.06$ & 0.00 \\ 
NGC752 & W0001 & 01:55:12.60 & +37:50:14.60 & 9.48 &  71 & FIES & $5044\pm49$ & $3.24\pm0.07$ & $1.22\pm0.10$ & $5033\pm17$ & $3.06\pm0.03$ & $1.37\pm0.02$ & 5038 & 33 & 7 & 3.15 & 0.05 & 0.13 & $-0.01\pm0.05$ & $-0.03\pm0.05$ & 0.01 \\ 
  & W0024 & 01:55:39.35 & +37:52:52.69 & 8.91 &  89 & FIES & $5044\pm67$ & $3.03\pm0.06$ & $1.35\pm0.11$ & $4950\pm11$ & $2.75\pm0.03$ & $1.47\pm0.02$ & 4997 & 39 & 66 & 2.89 & 0.04 & 0.20 & $0.03\pm0.05$ & $0.01\pm0.05$ & 0.01 \\ 
  &   &   &   &   &  72 & HERMES & $4964\pm31$ & $2.79\pm0.07$ & $1.19\pm0.05$ & $4954\pm13$ & $2.69\pm0.04$ & $1.43\pm0.02$ & 4959 & 22 & 6 & 2.74 & 0.06 & 0.07 & $0.02\pm0.04$ & $-0.05\pm0.05$ & 0.04 \\ 
  & W0027 & 01:55:42.39 & +37:37:54.66 & 9.17 &  71 & FIES & $4920\pm49$ & $2.77\pm0.05$ & $1.32\pm0.07$ & $4945\pm11$ & $2.81\pm0.04$ & $1.45\pm0.02$ & 4932 & 30 & 17 & 2.79 & 0.04 & 0.03 & $0.00\pm0.05$ & $-0.04\pm0.05$ & 0.02 \\ 
  &   &   &   &   &  67 & HERMES & $4956\pm32$ & $2.98\pm0.05$ & $1.15\pm0.04$ & $4957\pm13$ & $2.76\pm0.04$ & $1.36\pm0.02$ & 4956 & 22 & 0 & 2.87 & 0.04 & 0.16 & $0.04\pm0.04$ & $-0.04\pm0.05$ & 0.04 \\ 
  & W0077 & 01:56:21.63 & +37:36:08.53 & 9.38 &  69 & HERMES & $4837\pm40$ & $2.92\pm0.05$ & $1.04\pm0.05$ & $4863\pm11$ & $2.79\pm0.04$ & $1.30\pm0.02$ & 4850 & 25 & 18 & 2.86 & 0.04 & 0.09 & $0.04\pm0.04$ & $-0.06\pm0.05$ & 0.05 \\ 
  & W0137 & 01:57:03.12 & +38:08:02.73 & 8.90 &  75 & FIES & $4909\pm63$ & $2.79\pm0.10$ & $1.36\pm0.12$ & $4918\pm13$ & $2.68\pm0.04$ & $1.47\pm0.02$ & 4913 & 38 & 6 & 2.74 & 0.07 & 0.08 & $0.01\pm0.05$ & $-0.03\pm0.05$ & 0.02 \\ 
  &   &   &   &   &  72 & HERMES & $4848\pm63$ & $2.57\pm0.06$ & $1.17\pm0.04$ & $4931\pm13$ & $2.67\pm0.04$ & $1.41\pm0.02$ & 4889 & 38 & 59 & 2.62 & 0.05 & 0.07 & $-0.03\pm0.04$ & $-0.09\pm0.05$ & 0.03 \\ 
  & W0295 & 01:58:29.81 & +37:51:37.68 & 9.30 &  69 & FIES & $5074\pm65$ & $2.94\pm0.11$ & $1.15\pm0.09$ & $5030\pm12$ & $2.89\pm0.04$ & $1.41\pm0.02$ & 5052 & 38 & 30 & 2.92 & 0.08 & 0.04 & $0.06\pm0.05$ & $-0.01\pm0.05$ & 0.03 \\ 
  & W0311 & 01:58:52.90 & +37:48:57.30 & 9.06 &  74 & HERMES & $4851\pm36$ & $2.78\pm0.05$ & $1.18\pm0.06$ & $4900\pm12$ & $2.69\pm0.04$ & $1.38\pm0.02$ & 4875 & 24 & 34 & 2.74 & 0.04 & 0.06 & $0.01\pm0.04$ & $-0.04\pm0.05$ & 0.02 \\ 
NGC1817 & W0008 & 5:12:19.39 & 16:40:48.64 & 12.12 &  92 & FIES & $5016\pm54$ & $2.60\pm0.05$ & $1.28\pm0.05$ & $5087\pm15$ & $2.68\pm0.04$ & $1.57\pm0.02$ & 5051 & 34 & 50 & 2.64 & 0.04 & 0.06 & $-0.12\pm0.05$ & $-0.16\pm0.05$ & 0.02 \\ 
  & W0022 & 5:12:38.44 & 16:42:23.12 & 12.34 &  66 & FIES & $5094\pm45$ & $2.59\pm0.09$ & $1.31\pm0.10$ & $5133\pm16$ & $2.74\pm0.04$ & $1.55\pm0.02$ & 5113 & 30 & 27 & 2.66 & 0.06 & 0.11 & $-0.07\pm0.05$ & $-0.13\pm0.05$ & 0.03 \\ 
  & W0073 & 5:12:24.65 & 16:35:48.84 & 12.04 &  66 & FIES & $4863\pm53$ & $2.74\pm0.05$ & $1.22\pm0.08$ & $4854\pm15$ & $2.61\pm0.04$ & $1.46\pm0.02$ & 4858 & 34 & 5 & 2.68 & 0.04 & 0.09 & $-0.07\pm0.05$ & $-0.08\pm0.05$ & 0.00 \\ 
  & W0079 & 5:12:10.68 & 16:38:31.15 & 12.49 &  57 & FIES & $5117\pm43$ & $2.94\pm0.09$ & $1.27\pm0.12$ & $5163\pm15$ & $2.85\pm0.05$ & $1.50\pm0.03$ & 5140 & 29 & 32 & 2.90 & 0.07 & 0.06 & $-0.06\pm0.05$ & $-0.08\pm0.05$ & 0.01 \\ 
  & W0127 & 5:12:50.10 & 16:40:49.73 & 12.25 &  52 & FIES & $5200\pm75$ & $3.07\pm0.06$ & $1.48\pm0.10$ & $5060\pm21$ & $2.67\pm0.05$ & $1.49\pm0.03$ & 5130 & 48 & 98 & 2.87 & 0.06 & 0.28 & $-0.09\pm0.05$ & $-0.08\pm0.05$ & 0.00 \\ 
NGC1907 & W0062 & 05:27:49.053 & +35:20:10.13 & 12.41 &  54 & HERMES & $5066\pm66$ & $2.33\pm0.16$ & $1.39\pm0.08$ & $5179\pm19$ & $2.79\pm0.06$ & $1.48\pm0.03$ & 5122 & 42 & 79 & 2.56 & 0.11 & 0.33 & $-0.04\pm0.05$ & $-0.11\pm0.05$ & 0.03 \\ 
  & W0113 & 05:28:04.207 & +35:19:16.32 & 11.81 &  88 & HERMES & $4919\pm37$ & $2.50\pm0.04$ & $1.27\pm0.05$ & $4942\pm9$ & $2.40\pm0.04$ & $1.56\pm0.02$ & 4930 & 23 & 16 & 2.45 & 0.04 & 0.07 & $-0.03\pm0.04$ & $-0.17\pm0.05$ & 0.07 \\ 
  & W0131 & 05:28:05.276 & +35:19:49.64 & 12.30 &  63 & HERMES & $5108\pm30$ & $2.36\pm0.09$ & $1.37\pm0.06$ & $5150\pm19$ & $2.67\pm0.05$ & $1.60\pm0.02$ & 5129 & 24 & 29 & 2.52 & 0.07 & 0.22 & $-0.10\pm0.04$ & $-0.18\pm0.05$ & 0.04 \\ 
  & W0133 & 05:28:05.863 & +35:19:38.87 & 12.74 &  91 & HERMES & $5141\pm48$ & $2.84\pm0.12$ & $0.69\pm0.06$ & $5145\pm16$ & $2.84\pm0.04$ & $1.03\pm0.03$ & 5143 & 32 & 3 & 2.84 & 0.08 & 0.00 & $-0.06\pm0.04$ & $-0.20\pm0.05$ & 0.07 \\ 
  & W0256 & 05:28:01.783 & +35:21:14.89 & 11.23 &  92 & HERMES & $4539\pm58$ & $2.18\pm0.08$ & $1.42\pm0.10$ & $4491\pm8$ & $1.74\pm0.04$ & $1.69\pm0.01$ & 4515 & 33 & 33 & 1.96 & 0.06 & 0.31 & $-0.08\pm0.05$ & $-0.18\pm0.05$ & 0.05 \\ 
  & W2087 & 05:27:38.899 & +35:17:18.04 & 13.09 &  52 & HERMES & $4694\pm61$ & $2.51\pm0.13$ & $0.95\pm0.06$ & $4619\pm16$ & $2.47\pm0.05$ & $1.13\pm0.03$ & 4656 & 38 & 52 & 2.49 & 0.09 & 0.03 & $-0.53\pm0.04$ & $-0.62\pm0.06$ & 0.04 \\ 
NGC2099 & W007 & 05:52:20.31 & +32:33:49.3 & 11.42 &  59 & HERMES & $5025\pm50$ & $2.57\pm0.09$ & $1.36\pm0.06$ & $5075\pm16$ & $2.76\pm0.05$ & $1.54\pm0.03$ & 5050 & 33 & 35 & 2.66 & 0.07 & 0.13 & $0.04\pm0.04$ & $-0.02\pm0.05$ & 0.03 \\ 
  & W016 & 05:52:17.26 & +32:32:56.5 & 11.26 &  60 & HERMES & $5019\pm72$ & $2.54\pm0.07$ & $1.43\pm0.08$ & $5053\pm17$ & $2.67\pm0.05$ & $1.62\pm0.02$ & 5036 & 44 & 24 & 2.60 & 0.06 & 0.09 & $0.09\pm0.05$ & $0.03\pm0.05$ & 0.03 \\ 
  & W031 & 05:52:16.68 & +32:31:39.3 & 11.52 &  62 & HERMES & $5125\pm47$ & $2.88\pm0.07$ & $1.30\pm0.06$ & $5093\pm15$ & $2.84\pm0.05$ & $1.54\pm0.03$ & 5109 & 31 & 22 & 2.86 & 0.06 & 0.03 & $0.14\pm0.04$ & $0.03\pm0.05$ & 0.06 \\ 
  & W148 & 05:52:08.10 & +32:30:33.1 & 11.09 &  64 & HERMES & $4970\pm48$ & $2.54\pm0.08$ & $1.53\pm0.08$ & $4971\pm17$ & $2.54\pm0.04$ & $1.69\pm0.02$ & 4970 & 32 & 1 & 2.54 & 0.06 & 0.00 & $0.08\pm0.05$ & $-0.01\pm0.05$ & 0.04 \\ 
  & W172 & 05:52:04.89 & +32:33:18.3 & 11.45 &  61 & HERMES & $5078\pm51$ & $2.62\pm0.09$ & $1.36\pm0.06$ & $5080\pm17$ & $2.71\pm0.05$ & $1.59\pm0.03$ & 5079 & 34 & 2 & 2.66 & 0.07 & 0.06 & $0.06\pm0.04$ & $-0.04\pm0.05$ & 0.05 \\ 
  & W401 & 05:51:55.14 & +32:30:03.0 & 11.36 &  65 & HERMES & $4994\pm42$ & $2.68\pm0.04$ & $1.46\pm0.05$ & $5035\pm15$ & $2.68\pm0.04$ & $1.57\pm0.02$ & 5014 & 28 & 29 & 2.68 & 0.04 & 0.00 & $0.08\pm0.04$ & $0.02\pm0.05$ & 0.03 \\ 
  & W488 & 05:52:46.97 & +32:33:19.4 & 11.17 &  62 & HERMES & $4998\pm44$ & $2.72\pm0.09$ & $1.40\pm0.06$ & $4990\pm17$ & $2.61\pm0.04$ & $1.57\pm0.02$ & 4994 & 30 & 5 & 2.66 & 0.06 & 0.08 & $0.07\pm0.04$ & $0.01\pm0.05$ & 0.03 \\ 
NGC2420 & W041 & 7:38:6.27 & 21:36:54.60 & 12.67 &  58 & FIES & $4732\pm65$ & $2.41\pm0.11$ & $1.34\pm0.07$ & $4806\pm16$ & $2.60\pm0.04$ & $1.50\pm0.02$ & 4769 & 40 & 52 & 2.50 & 0.08 & 0.13 & $-0.18\pm0.05$ & $-0.21\pm0.05$ & 0.02 \\ 
  & W076 & 7:38:15.50 & 21:38:1.80 & 12.66 &  78 & FIES & $5002\pm63$ & $3.04\pm0.06$ & $1.32\pm0.10$ & $4964\pm16$ & $2.59\pm0.04$ & $1.51\pm0.02$ & 4983 & 39 & 26 & 2.82 & 0.05 & 0.32 & $-0.07\pm0.05$ & $-0.11\pm0.05$ & 0.02 \\ 
  & W091 & 7:38:18.17 & 21:32:6.80 & 12.61 &  74 & FIES & $4922\pm56$ & $2.50\pm0.12$ & $1.37\pm0.09$ & $4969\pm15$ & $2.64\pm0.04$ & $1.55\pm0.02$ & 4945 & 35 & 33 & 2.57 & 0.08 & 0.10 & $-0.08\pm0.05$ & $-0.12\pm0.06$ & 0.02 \\ 
  & W111 & 7:38:21.43 & 21:35:5.60 & 12.60 &  72 & FIES & $4888\pm63$ & $2.78\pm0.08$ & $1.08\pm0.11$ & $4951\pm12$ & $2.92\pm0.04$ & $1.41\pm0.02$ & 4919 & 37 & 44 & 2.85 & 0.06 & 0.10 & $-0.05\pm0.05$ & $-0.12\pm0.05$ & 0.03 \\ 
  & W118 & 7:38:21.90 & 21:35:50.90 & 12.57 &  60 & FIES & $4863\pm55$ & $2.47\pm0.10$ & $1.33\pm0.06$ & $4890\pm17$ & $2.52\pm0.03$ & $1.49\pm0.02$ & 4876 & 36 & 19 & 2.50 & 0.06 & 0.04 & $-0.13\pm0.05$ & $-0.17\pm0.05$ & 0.02 \\ 
  & W174 & 7:38:26.93 & 21:38:24.80 & 12.40 &  65 & FIES & $4872\pm50$ & $2.63\pm0.06$ & $1.24\pm0.07$ & $4892\pm15$ & $2.57\pm0.04$ & $1.59\pm0.02$ & 4882 & 32 & 14 & 2.60 & 0.05 & 0.04 & $-0.05\pm0.05$ & $-0.15\pm0.05$ & 0.05 \\ 
  & W236 & 7:38:37.59 & 21:34:12.40 & 12.58 &  71 & FIES & $4978\pm49$ & $2.75\pm0.11$ & $1.41\pm0.09$ & $5001\pm16$ & $2.66\pm0.04$ & $1.57\pm0.02$ & 4989 & 32 & 16 & 2.70 & 0.08 & 0.06 & $-0.10\pm0.05$ & $-0.13\pm0.05$ & 0.02 \\ 
NGC2539 & W229 & 08:10:33.80 & -12:51:48.9 & 11.20 &  73 & HERMES & $5050\pm68$ & $2.98\pm0.12$ & $1.26\pm0.05$ & $5048\pm12$ & $2.75\pm0.04$ & $1.41\pm0.02$ & 5049 & 40 & 0 & 2.86 & 0.08 & 0.16 & $0.06\pm0.04$ & $0.01\pm0.05$ & 0.02 \\ 
  & W251 & 08:10:38.99 & -12:44:44.7 & 11.23 &  70 & HERMES & $5106\pm49$ & $2.61\pm0.08$ & $1.31\pm0.08$ & $5086\pm13$ & $2.86\pm0.04$ & $1.43\pm0.03$ & 5096 & 31 & 13 & 2.74 & 0.06 & 0.18 & $0.05\pm0.05$ & $-0.06\pm0.05$ & 0.06 \\ 
  & W346 & 08:10:23.02 & -12:50:43.3 & 10.92 &  101 & HERMES & $5094\pm39$ & $2.91\pm0.07$ & $1.23\pm0.05$ & $5051\pm10$ & $2.77\pm0.03$ & $1.48\pm0.02$ & 5072 & 24 & 30 & 2.84 & 0.05 & 0.10 & $0.07\pm0.04$ & $-0.02\pm0.05$ & 0.04 \\ 
  & W463 & 08:10:42.87 & -12:40:11.8 & 10.69 &  99 & HERMES & $4979\pm38$ & $2.58\pm0.06$ & $1.32\pm0.06$ & $4954\pm13$ & $2.57\pm0.03$ & $1.56\pm0.02$ & 4966 & 25 & 17 & 2.58 & 0.04 & 0.01 & $0.07\pm0.04$ & $-0.01\pm0.05$ & 0.04 \\ 
  & W502 & 08:11:27.67 & -12:41:06.8 & 11.03 &  76 & HERMES & $5147\pm50$ & $3.14\pm0.10$ & $1.36\pm0.07$ & $5057\pm13$ & $2.73\pm0.04$ & $1.46\pm0.02$ & 5102 & 31 & 63 & 2.94 & 0.07 & 0.29 & $0.08\pm0.04$ & $0.02\pm0.05$ & 0.03 \\ 
NGC2682 & W084 & 08:51:12.73 & +11:52:42.7 & 10.52 &  64 & HERMES & $4728\pm45$ & $2.52\pm0.12$ & $1.11\pm0.06$ & $4731\pm12$ & $2.46\pm0.04$ & $1.44\pm0.02$ & 4729 & 28 & 2 & 2.49 & 0.08 & 0.04 & $0.08\pm0.04$ & $-0.08\pm0.06$ & 0.08 \\ 
  & W141 & 08:51:22.83 & +11:48:02.0 & 10.48 &  70 & HERMES & $4691\pm34$ & $2.58\pm0.13$ & $1.26\pm0.09$ & $4724\pm13$ & $2.53\pm0.03$ & $1.45\pm0.02$ & 4707 & 23 & 23 & 2.56 & 0.08 & 0.04 & $0.05\pm0.05$ & $-0.05\pm0.05$ & 0.05 \\ 
  & W151 & 08:51:26.22 & +11:53:52.2 & 10.48 &  65 & HERMES & $4745\pm58$ & $2.59\pm0.09$ & $1.19\pm0.07$ & $4771\pm13$ & $2.55\pm0.03$ & $1.43\pm0.02$ & 4758 & 35 & 19 & 2.57 & 0.06 & 0.03 & $0.04\pm0.04$ & $-0.04\pm0.05$ & 0.04 \\ 
  & W164 & 08:51:29.03 & +11:50:33.4 & 10.52 &  63 & HERMES & $4686\pm48$ & $2.50\pm0.08$ & $1.20\pm0.06$ & $4704\pm11$ & $2.45\pm0.04$ & $1.43\pm0.02$ & 4695 & 29 & 12 & 2.48 & 0.06 & 0.04 & $0.01\pm0.04$ & $-0.07\pm0.05$ & 0.04 \\ 
  & W223 & 08:51:43.91 & +11:56:42.9 & 10.58 &  55 & HERMES & $4651\pm52$ & $2.43\pm0.16$ & $1.16\pm0.08$ & $4742\pm13$ & $2.46\pm0.04$ & $1.45\pm0.02$ & 4696 & 32 & 64 & 2.44 & 0.10 & 0.02 & $0.00\pm0.05$ & $-0.08\pm0.05$ & 0.04 \\ 
  & W224 & 08:51:43.55 & +11:44:26.8 & 10.76 &  61 & HERMES & $4557\pm89$ & $2.42\pm0.09$ & $0.99\pm0.07$ & $4658\pm13$ & $2.55\pm0.03$ & $1.29\pm0.02$ & 4607 & 51 & 72 & 2.48 & 0.06 & 0.09 & $0.09\pm0.04$ & $-0.05\pm0.05$ & 0.07 \\ 
  & W266 & 08:51:59.56 & +11:55:05.2 & 10.55 &  67 & HERMES & $4762\pm37$ & $2.63\pm0.07$ & $1.20\pm0.05$ & $4776\pm13$ & $2.54\pm0.03$ & $1.45\pm0.02$ & 4769 & 25 & 10 & 2.58 & 0.05 & 0.06 & $0.03\pm0.04$ & $-0.06\pm0.05$ & 0.04 \\ 
  & W286 & 08:52:18.61 & +11:44:26.5 & 10.47 &  94 & HERMES & $4672\pm116$ & $2.34\pm0.10$ & $1.08\pm0.07$ & $4719\pm7$ & $2.39\pm0.03$ & $1.45\pm0.01$ & 4695 & 61 & 33 & 2.37 & 0.06 & 0.04 & $0.04\pm0.04$ & $-0.08\pm0.05$ & 0.06 \\ 
NGC6633 & W100 & 18:27:54.73 & +06:36:00.3 & 8.30 &  86 & CAFE & $4976\pm61$ & $2.57\pm0.07$ & $1.39\pm0.07$ & $5011\pm15$ & $2.60\pm0.04$ & $1.64\pm0.02$ & 4993 & 38 & 24 & 2.58 & 0.06 & 0.02 & $-0.05\pm0.04$ & $-0.03\pm0.05$ & 0.01 \\ 
  &   &   &   &   &  72 & FIES & $4968\pm78$ & $2.61\pm0.06$ & $1.49\pm0.12$ & $5012\pm15$ & $2.64\pm0.04$ & $1.69\pm0.02$ & 4990 & 46 & 31 & 2.62 & 0.05 & 0.02 & $0.07\pm0.05$ & $-0.03\pm0.05$ & 0.05 \\ 
  &   &   &   &   &  74 & HERMES & $5034\pm38$ & $2.85\pm0.07$ & $1.33\pm0.05$ & $5030\pm13$ & $2.65\pm0.03$ & $1.56\pm0.02$ & 5032 & 25 & 2 & 2.75 & 0.05 & 0.14 & $0.04\pm0.04$ & $-0.04\pm0.05$ & 0.04 \\ 
  & W106 & 18:28:00.18 & +06:54:51.5 & 8.67 &  102 & FIES & $5113\pm41$ & $2.82\pm0.05$ & $1.22\pm0.07$ & $5115\pm11$ & $2.84\pm0.04$ & $1.52\pm0.02$ & 5114 & 26 & 1 & 2.83 & 0.04 & 0.01 & $0.07\pm0.05$ & $0.02\pm0.05$ & 0.02 \\ 
  &   &   &   &   &  65 & HERMES & $5147\pm46$ & $2.98\pm0.07$ & $1.15\pm0.06$ & $5106\pm12$ & $2.85\pm0.04$ & $1.43\pm0.02$ & 5126 & 29 & 28 & 2.92 & 0.06 & 0.09 & $0.08\pm0.04$ & $-0.02\pm0.05$ & 0.05 \\ 
  & W119 & 18:28:17.64 & +06:46:00.1 & 8.95 &  67 & FIES & $5138\pm42$ & $2.84\pm0.06$ & $1.28\pm0.06$ & $5203\pm21$ & $3.02\pm0.04$ & $1.47\pm0.03$ & 5170 & 31 & 46 & 2.93 & 0.05 & 0.13 & $0.03\pm0.05$ & $-0.02\pm0.05$ & 0.02 \\ 
  &   &   &   &   &  70 & HERMES & $5192\pm56$ & $3.03\pm0.07$ & $1.09\pm0.05$ & $5218\pm18$ & $3.08\pm0.04$ & $1.44\pm0.02$ & 5205 & 37 & 18 & 3.06 & 0.06 & 0.04 & $0.03\pm0.04$ & $-0.07\pm0.05$ & 0.05 \\ 
  & W126 & 18:28:22.97 & +06:42:29.3 & 8.77 &  95 & FIES & $5054\pm50$ & $2.55\pm0.06$ & $1.28\pm0.09$ & $5131\pm12$ & $2.80\pm0.04$ & $1.53\pm0.02$ & 5092 & 31 & 54 & 2.68 & 0.05 & 0.18 & $-0.01\pm0.05$ & $-0.04\pm0.05$ & 0.02 \\ 
  &   &   &   &   &  78 & HERMES & $5190\pm37$ & $3.07\pm0.09$ & $1.23\pm0.05$ & $5174\pm12$ & $2.92\pm0.03$ & $1.44\pm0.02$ & 5182 & 24 & 10 & 3.00 & 0.06 & 0.11 & $0.04\pm0.04$ & $-0.02\pm0.05$ & 0.03 \\ 
NGC6705 & W0660 & 18:51:15.691 & -06:18:14.47 & 11.81 &  56 & HERMES & $4756\pm79$ & $2.36\pm0.10$ & $1.60\pm0.09$ & $4719\pm13$ & $2.22\pm0.05$ & $1.80\pm0.02$ & 4737 & 46 & 26 & 2.29 & 0.08 & 0.10 & $0.20\pm0.05$ & $0.05\pm0.06$ & 0.08 \\ 
  & W0669 & 18:51:15.318 & -06:18:35.51 & 11.97 &  54 & HERMES & $4791\pm79$ & $2.26\pm0.15$ & $1.72\pm0.13$ & $4706\pm16$ & $2.20\pm0.06$ & $1.77\pm0.03$ & 4748 & 47 & 59 & 2.23 & 0.10 & 0.04 & $0.21\pm0.05$ & $0.08\pm0.06$ & 0.06 \\ 
  & W0686 & 18:51:14.507 & -06:16:54.74 & 11.92 &  59 & HERMES & $4884\pm69$ & $2.44\pm0.14$ & $1.85\pm0.12$ & $4766\pm16$ & $2.27\pm0.06$ & $1.73\pm0.03$ & 4825 & 42 & 83 & 2.36 & 0.10 & 0.12 & $0.14\pm0.05$ & $0.09\pm0.06$ & 0.03 \\ 
  & W0779 & 18:51:11.141 & -06:14:33.76 & 11.47 &  92 & FIES & $4330\pm162$ & $1.83\pm0.23$ & $1.47\pm0.15$ & $4355\pm6$ & $1.82\pm0.04$ & $1.86\pm0.01$ & 4342 & 84 & 18 & 1.82 & 0.14 & 0.01 & $0.18\pm0.05$ & $0.05\pm0.06$ & 0.06 \\ 
  &   &   &   &   &  64 & HERMES & $4317\pm77$ & $1.63\pm0.20$ & $1.45\pm0.17$ & $4354\pm9$ & $1.74\pm0.05$ & $1.78\pm0.02$ & 4335 & 43 & 26 & 1.68 & 0.12 & 0.08 & $0.19\pm0.05$ & $-0.03\pm0.06$ & 0.11 \\ 
  & W0916 & 18:51:07.847 & -06:17:11.89 & 11.62 &  73 & HERMES & $4810\pm73$ & $1.95\pm0.20$ & $1.76\pm0.13$ & $4768\pm13$ & $2.29\pm0.05$ & $1.81\pm0.02$ & 4789 & 43 & 29 & 2.12 & 0.12 & 0.24 & $0.17\pm0.05$ & $0.04\pm0.06$ & 0.06 \\ 
  & W1184 & 18:51:01.989 & -06:17:26.50 & 11.43 &  74 & FIES & $4352\pm125$ & $1.74\pm0.15$ & $1.66\pm0.10$ & $4388\pm7$ & $1.81\pm0.04$ & $1.83\pm0.02$ & 4370 & 66 & 25 & 1.78 & 0.10 & 0.05 & $0.03\pm0.05$ & $-0.02\pm0.06$ & 0.02 \\ 
  &   &   &   &   &  69 & HERMES & $4425\pm85$ & $1.79\pm0.15$ & $1.34\pm0.08$ & $4390\pm8$ & $1.72\pm0.04$ & $1.71\pm0.02$ & 4407 & 46 & 24 & 1.76 & 0.10 & 0.05 & $0.13\pm0.05$ & $-0.09\pm0.06$ & 0.11 \\ 
  & W1256 & 18:51:00.194 & -06:16:59.06 & 11.59 &  84 & HERMES & $4467\pm92$ & $1.90\pm0.15$ & $1.59\pm0.12$ & $4405\pm7$ & $1.76\pm0.04$ & $1.76\pm0.01$ & 4436 & 49 & 43 & 1.83 & 0.10 & 0.10 & $0.07\pm0.05$ & $-0.05\pm0.06$ & 0.06 \\ 
  & W1423 & 18:50:55.789 & -06:18:14.26 & 11.41 &  78 & FIES & $4555\pm202$ & $2.22\pm0.15$ & $1.47\pm0.11$ & $4493\pm12$ & $2.08\pm0.03$ & $1.88\pm0.02$ & 4524 & 107 & 43 & 2.15 & 0.09 & 0.10 & $0.22\pm0.05$ & $0.12\pm0.06$ & 0.05 \\ 
  &   &   &   &   &  65 & HERMES & $4384\pm63$ & $1.93\pm0.19$ & $1.54\pm0.10$ & $4464\pm9$ & $1.95\pm0.04$ & $1.81\pm0.02$ & 4424 & 36 & 56 & 1.94 & 0.12 & 0.01 & $0.16\pm0.05$ & $0.00\pm0.06$ & 0.08 \\ 
NGC6791 & W1794 & 19:21:6.31 & 37:44:59.90 & 14.48 &  56 & FIES & $4421\pm68$ & $1.73\pm0.14$ & $1.48\pm0.16$ & $4477\pm14$ & $2.18\pm0.05$ & $1.40\pm0.02$ & 4449 & 41 & 39 & 1.96 & 0.10 & 0.32 & $0.04\pm0.06$ & $0.01\pm0.07$ & 0.02 \\ 
  & W2562 & 19:21:0.87 & 37:46:39.90 & 14.58 &  62 & FIES & $4610\pm167$ & $2.30\pm0.18$ & $1.85\pm0.24$ & $4508\pm17$ & $2.49\pm0.03$ & $1.46\pm0.02$ & 4559 & 92 & 71 & 2.40 & 0.10 & 0.13 & $0.24\pm0.06$ & $0.28\pm0.08$ & 0.02 \\ 
  & W2579 & 19:21:0.87 & 37:45:34.10 & 14.55 &  64 & FIES & $4403\pm158$ & $1.83\pm0.21$ & $1.55\pm0.25$ & $4410\pm12$ & $2.20\pm0.04$ & $1.44\pm0.02$ & 4406 & 85 & 5 & 2.02 & 0.12 & 0.26 & $0.17\pm0.06$ & $0.02\pm0.07$ & 0.08 \\ 
  & W3363 & 19:20:56.31 & 37:44:33.70 & 14.65 &  53 & FIES & $4561\pm142$ & $2.73\pm0.25$ & $1.67\pm0.17$ & $4453\pm11$ & $2.36\pm0.04$ & $1.40\pm0.02$ & 4507 & 76 & 75 & 2.54 & 0.14 & 0.26 & $0.27\pm0.06$ & $0.22\pm0.07$ & 0.03 \\ 
  & W3899 & 19:20:52.47 & 37:50:15.80 & 14.48 &  50 & FIES & $4624\pm167$ & $1.83\pm0.26$ & $1.30\pm0.10$ & $4670\pm20$ & $2.31\pm0.07$ & $0.79\pm0.04$ & 4647 & 93 & 32 & 2.07 & 0.16 & 0.34 & $0.24\pm0.08$ & $0.30\pm0.08$ & 0.03 \\ 
  & W3926 & 19:20:52.89 & 37:45:33.40 & 14.55 &  53 & FIES & $4420\pm99$ & $1.99\pm0.25$ & $1.49\pm0.14$ & $4490\pm15$ & $2.48\pm0.04$ & $1.50\pm0.02$ & 4455 & 57 & 49 & 2.24 & 0.14 & 0.35 & $0.22\pm0.06$ & $0.13\pm0.07$ & 0.04 \\ 
NGC6819 & W333 & 19:41:13.55 & +40:12:20.5 & 13.069 &  66 & HERMES & $4740\pm92$ & $2.63\pm0.08$ & $1.22\pm0.06$ & $4828\pm13$ & $2.63\pm0.04$ & $1.39\pm0.02$ & 4784 & 52 & 62 & 2.63 & 0.06 & 0.00 & $0.05\pm0.04$ & $-0.06\pm0.05$ & 0.06 \\ 
  & W386 & 19:41:22.45 & +40:12:05.3 & 13.016 &  57 & HERMES & $4927\pm52$ & $2.99\pm0.07$ & $0.88\pm0.08$ & $4956\pm14$ & $2.93\pm0.04$ & $1.34\pm0.02$ & 4941 & 33 & 20 & 2.96 & 0.06 & 0.04 & $0.09\pm0.04$ & $-0.07\pm0.05$ & 0.08 \\ 
  & W398 & 19:41:13.45 & +40:11:57.9 & 13.119 &  51 & HERMES & $4767\pm52$ & $2.61\pm0.08$ & $1.24\pm0.07$ & $4745\pm16$ & $2.55\pm0.04$ & $1.41\pm0.02$ & 4756 & 34 & 15 & 2.58 & 0.06 & 0.04 & $0.07\pm0.04$ & $-0.06\pm0.05$ & 0.06 \\ 
  & W978 & 19:41:14.76 & +40:11:00.8 & 12.869 &  62 & HERMES & $4852\pm51$ & $2.65\pm0.09$ & $1.32\pm0.07$ & $4877\pm13$ & $2.65\pm0.04$ & $1.43\pm0.02$ & 4864 & 32 & 18 & 2.65 & 0.06 & 0.00 & $0.06\pm0.04$ & $-0.01\pm0.05$ & 0.03 \\ 
  & W979 & 19:41:15.93 & +40:11:11.5 & 12.956 &  61 & HERMES & $5027\pm59$ & $2.95\pm0.06$ & $1.18\pm0.08$ & $5032\pm12$ & $2.88\pm0.04$ & $1.30\pm0.02$ & 5029 & 35 & 4 & 2.92 & 0.05 & 0.05 & $0.14\pm0.05$ & $0.04\pm0.05$ & 0.05 \\ 
  & W983 & 19:41:09.91 & +40:15:49.5 & 12.928 &  57 & HERMES & $4806\pm48$ & $2.75\pm0.09$ & $1.24\pm0.06$ & $4747\pm15$ & $2.52\pm0.03$ & $1.44\pm0.02$ & 4776 & 31 & 41 & 2.64 & 0.06 & 0.16 & $0.13\pm0.04$ & $-0.03\pm0.05$ & 0.08 \\ 
NGC6939 & W130 & 20:31:25.43 & 60:41:16.67 & 13.07 &  42 & HERMES & $5142\pm125$ & $2.77\pm0.19$ & $1.28\pm0.13$ & $5140\pm25$ & $3.15\pm0.06$ & $1.21\pm0.03$ & 5141 & 75 & 1 & 2.96 & 0.12 & 0.27 & $0.22\pm0.05$ & $0.11\pm0.07$ & 0.06 \\ 
  & W145 & 20:31:28.55 & 60:40:7.82 & 12.97 &  56 & HERMES & $4865\pm53$ & $2.67\pm0.10$ & $1.03\pm0.06$ & $4895\pm16$ & $2.60\pm0.04$ & $1.33\pm0.02$ & 4880 & 34 & 21 & 2.64 & 0.07 & 0.05 & $0.11\pm0.04$ & $-0.06\pm0.06$ & 0.08 \\ 
  & W170 & 20:31:32.04 & 60:39:27.37 & 12.99 &  60 & HERMES & $4924\pm48$ & $2.76\pm0.10$ & $1.13\pm0.06$ & $4897\pm14$ & $2.64\pm0.04$ & $1.26\pm0.02$ & 4910 & 31 & 18 & 2.70 & 0.07 & 0.08 & $0.10\pm0.04$ & $0.01\pm0.05$ & 0.04 \\ 
  & W214 & 20:31:40.18 & 60:41:31.69 & 13.08 &  57 & HERMES & $5039\pm71$ & $2.99\pm0.07$ & $1.14\pm0.08$ & $4993\pm12$ & $2.85\pm0.04$ & $1.36\pm0.02$ & 5016 & 41 & 32 & 2.92 & 0.06 & 0.10 & $0.17\pm0.05$ & $0.01\pm0.05$ & 0.08 \\ 
  & W230 & 20:31:43.42 & 60:40:38.82 & 12.99 &  59 & HERMES & $4893\pm53$ & $2.82\pm0.07$ & $1.12\pm0.06$ & $4890\pm15$ & $2.63\pm0.04$ & $1.31\pm0.02$ & 4891 & 34 & 1 & 2.72 & 0.06 & 0.13 & $0.09\pm0.04$ & $-0.03\pm0.05$ & 0.06 \\ 
  & W292 & 20:31:59.11 & 60:42:4.76 & 13.11 &  59 & HERMES & $4916\pm43$ & $2.75\pm0.11$ & $1.18\pm0.06$ & $4918\pm15$ & $2.61\pm0.04$ & $1.43\pm0.02$ & 4917 & 29 & 1 & 2.68 & 0.08 & 0.10 & $0.04\pm0.04$ & $-0.08\pm0.05$ & 0.06 \\ 
NGC6991 & W034 & 20:53:37.68 & +47:12:23.66 & 10.30 &  80 & CAFE & $5076\pm47$ & $3.14\pm0.09$ & $1.15\pm0.06$ & $5032\pm15$ & $2.98\pm0.03$ & $1.44\pm0.02$ & 5054 & 31 & 30 & 3.06 & 0.06 & 0.11 & $-0.07\pm0.04$ & $-0.09\pm0.05$ & 0.01 \\ 
  &   &   &   &   &  73 & FIES & $5128\pm63$ & $3.17\pm0.09$ & $1.09\pm0.08$ & $5042\pm16$ & $3.10\pm0.04$ & $1.33\pm0.02$ & 5085 & 39 & 60 & 3.14 & 0.06 & 0.05 & $0.05\pm0.05$ & $-0.02\pm0.05$ & 0.04 \\ 
  & W043 & 20:53:50.82 & +47:05:06.75 & 10.08 &  71 & CAFE & $5068\pm45$ & $2.85\pm0.09$ & $1.35\pm0.07$ & $5059\pm15$ & $2.94\pm0.04$ & $1.54\pm0.02$ & 5063 & 30 & 5 & 2.90 & 0.06 & 0.06 & $-0.02\pm0.04$ & $-0.05\pm0.05$ & 0.02 \\ 
  & W049 & 20:54:01.74 & +47:25:49.16 & 10.17 &  95 & CAFE & $5118\pm58$ & $3.23\pm0.05$ & $1.04\pm0.07$ & $5021\pm14$ & $2.96\pm0.03$ & $1.54\pm0.02$ & 5069 & 36 & 67 & 3.10 & 0.04 & 0.19 & $-0.04\pm0.04$ & $-0.08\pm0.05$ & 0.02 \\ 
  &   &   &   &   &  90 & FIES & $5177\pm49$ & $3.39\pm0.04$ & $1.15\pm0.08$ & $5056\pm15$ & $3.08\pm0.03$ & $1.34\pm0.02$ & 5116 & 32 & 85 & 3.24 & 0.04 & 0.22 & $0.04\pm0.05$ & $0.01\pm0.05$ & 0.02 \\ 
  & W067 & 20:54:29.81 & +47:28:03.15 & 9.43 &  70 & CAFE & $4917\pm47$ & $2.54\pm0.13$ & $1.31\pm0.07$ & $4907\pm15$ & $2.61\pm0.04$ & $1.63\pm0.02$ & 4912 & 31 & 6 & 2.58 & 0.08 & 0.05 & $-0.06\pm0.04$ & $-0.11\pm0.05$ & 0.02 \\ 
  &   &   &   &   &  70 & FIES & $4930\pm64$ & $2.78\pm0.12$ & $1.27\pm0.14$ & $4900\pm13$ & $2.66\pm0.04$ & $1.51\pm0.02$ & 4915 & 38 & 20 & 2.72 & 0.08 & 0.08 & $0.00\pm0.05$ & $-0.04\pm0.05$ & 0.02 \\ 
  & W100 & 20:55.03.98 & +47:19:20.03 & 9.87 &  77 & CAFE & $5095\pm58$ & $3.01\pm0.10$ & $1.20\pm0.09$ & $5064\pm14$ & $2.93\pm0.04$ & $1.48\pm0.02$ & 5079 & 36 & 21 & 2.97 & 0.07 & 0.06 & $0.02\pm0.04$ & $-0.03\pm0.05$ & 0.02 \\ 
  & W131 & 20:55:42.69 & +47:22:32.60 & 9.66 &  71 & CAFE & $5118\pm47$ & $3.15\pm0.05$ & $1.21\pm0.08$ & $5032\pm12$ & $2.80\pm0.04$ & $1.49\pm0.02$ & 5075 & 29 & 60 & 2.98 & 0.04 & 0.25 & $0.01\pm0.04$ & $0.00\pm0.05$ & 0.00 \\ 
  &   &   &   &   &  81 & FIES & $5057\pm48$ & $2.99\pm0.05$ & $1.21\pm0.06$ & $4993\pm13$ & $2.76\pm0.04$ & $1.45\pm0.02$ & 5025 & 30 & 44 & 2.88 & 0.04 & 0.16 & $0.03\pm0.05$ & $-0.03\pm0.05$ & 0.03 \\ 
NGC7245 & W0205 & 22:15:14.90 & 54:20:4.10 & 13.87 &  64 & FIES & $4893\pm101$ & $2.14\pm0.16$ & $1.44\pm0.17$ & $5071\pm14$ & $2.86\pm0.04$ & $1.61\pm0.02$ & 4982 & 57 & 125 & 2.50 & 0.10 & 0.51 & $0.12\pm0.06$ & $0.01\pm0.06$ & 0.06 \\ 
  & W045 & 22:15:7.80 & 54:18:26.90 & 14.16 &  66 & FIES & $4963\pm110$ & $2.60\pm0.12$ & $1.33\pm0.14$ & $5100\pm16$ & $3.22\pm0.04$ & $1.43\pm0.02$ & 5031 & 63 & 97 & 2.91 & 0.08 & 0.44 & $0.23\pm0.05$ & $0.27\pm0.07$ & 0.02 \\ 
  & W055 & 22:15:17.50 & 54:18:12.60 & 13.11 &  75 & FIES & $5005\pm71$ & $2.45\pm0.09$ & $1.21\pm0.07$ & $4933\pm16$ & $2.54\pm0.04$ & $1.47\pm0.02$ & 4969 & 43 & 50 & 2.50 & 0.06 & 0.06 & $0.09\pm0.05$ & $0.01\pm0.05$ & 0.04 \\ 
  & W095 & 22:15:12.00 & 54:21:11.40 & 13.37 &  74 & FIES & $5023\pm58$ & $2.69\pm0.08$ & $1.62\pm0.13$ & $5017\pm16$ & $2.57\pm0.04$ & $1.61\pm0.02$ & 5020 & 37 & 3 & 2.63 & 0.06 & 0.08 & $-0.02\pm0.05$ & $0.01\pm0.05$ & 0.02 \\ 
  & W178 & 22:15:5.40 & 54:22:43.60 & 13.76 &  72 & FIES & $5166\pm75$ & $2.75\pm0.12$ & $1.26\pm0.08$ & $5105\pm16$ & $2.77\pm0.05$ & $1.26\pm0.03$ & 5135 & 45 & 42 & 2.76 & 0.08 & 0.01 & $0.04\pm0.05$ & $0.06\pm0.06$ & 0.01 \\ 
  & W179 & 22:15:5.40 & 54:22:49.40 & 12.97 &  78 & FIES & $5045\pm51$ & $2.76\pm0.08$ & $1.68\pm0.13$ & $4928\pm13$ & $2.45\pm0.04$ & $1.66\pm0.02$ & 4986 & 32 & 82 & 2.60 & 0.06 & 0.22 & $0.08\pm0.05$ & $0.10\pm0.05$ & 0.01 \\ 
NGC7762 & W0002 & 23:49:48.40 & +68:01:35.14 & 12.56 &  66 & HERMES & $4798\pm67$ & $2.69\pm0.08$ & $1.23\pm0.07$ & $4820\pm14$ & $2.54\pm0.03$ & $1.40\pm0.02$ & 4809 & 40 & 15 & 2.62 & 0.06 & 0.11 & $0.04\pm0.04$ & $-0.02\pm0.05$ & 0.03 \\ 
  & W0003 & 23:49:49.26 & +68:01:07.35 & 12.88 &  58 & HERMES & $4700\pm48$ & $2.53\pm0.14$ & $1.19\pm0.08$ & $4681\pm13$ & $2.47\pm0.04$ & $1.33\pm0.02$ & 4690 & 30 & 13 & 2.50 & 0.09 & 0.04 & $-0.01\pm0.05$ & $-0.07\pm0.05$ & 0.03 \\ 
  & W0084 & 23:50:13.52 & +68:03:02.57 & 12.24 &  67 & HERMES & $5052\pm56$ & $2.88\pm0.07$ & $1.27\pm0.06$ & $5042\pm12$ & $2.79\pm0.04$ & $1.42\pm0.02$ & 5047 & 34 & 6 & 2.84 & 0.06 & 0.06 & $0.06\pm0.04$ & $0.02\pm0.05$ & 0.02 \\ 
  & W0110 & 23:49:06.13 & +67:59:08.58 & 12.56 &  63 & HERMES & $4859\pm46$ & $2.94\pm0.07$ & $1.15\pm0.06$ & $4850\pm13$ & $2.62\pm0.04$ & $1.38\pm0.02$ & 4854 & 29 & 6 & 2.78 & 0.06 & 0.23 & $0.09\pm0.04$ & $0.01\pm0.05$ & 0.04 \\ 
  & W0125 & 23:49:15.74 & +68:05:32.14 & 12.57 &  68 & HERMES & $4838\pm34$ & $2.63\pm0.10$ & $1.16\pm0.06$ & $4871\pm13$ & $2.59\pm0.03$ & $1.32\pm0.02$ & 4854 & 23 & 24 & 2.61 & 0.06 & 0.03 & $0.01\pm0.04$ & $-0.04\pm0.05$ & 0.02 \\ 
  & W0139 & 23:50:59.35 & +68:00:36.61 & 12.80 &  56 & HERMES & $4784\pm35$ & $2.34\pm0.09$ & $1.17\pm0.04$ & $4856\pm16$ & $2.60\pm0.04$ & $1.39\pm0.02$ & 4820 & 25 & 50 & 2.47 & 0.06 & 0.18 & $0.01\pm0.04$ & $-0.11\pm0.05$ & 0.06 \\ 
NGC7789 & W05862 & 23:56:57.38 & +56:36:54.69 & 12.98 &  46 & HERMES & $4988\pm49$ & $2.75\pm0.08$ & $1.27\pm0.05$ & $4990\pm17$ & $2.76\pm0.05$ & $1.37\pm0.03$ & 4989 & 33 & 1 & 2.76 & 0.06 & 0.01 & $-0.01\pm0.04$ & $-0.11\pm0.05$ & 0.05 \\ 
  & W07176 & 23:57:12.50 & +56:50:00.41 & 12.84 &  50 & HERMES & $4935\pm60$ & $2.90\pm0.09$ & $1.15\pm0.06$ & $4928\pm16$ & $2.65\pm0.05$ & $1.42\pm0.02$ & 4931 & 38 & 4 & 2.78 & 0.07 & 0.18 & $0.08\pm0.04$ & $-0.03\pm0.05$ & 0.06 \\ 
  & W07714 & 23:57:18.57 & +56:50:26.72 & 13.01 &  46 & HERMES & $4903\pm63$ & $2.75\pm0.08$ & $1.05\pm0.07$ & $4879\pm17$ & $2.62\pm0.05$ & $1.39\pm0.03$ & 4891 & 40 & 16 & 2.68 & 0.06 & 0.09 & $0.05\pm0.04$ & $-0.07\pm0.05$ & 0.06 \\ 
  & W08260 & 23:57:24.05 & +56:45:33.53 & 12.84 &  48 & HERMES & $4867\pm71$ & $2.63\pm0.10$ & $0.99\pm0.09$ & $4915\pm18$ & $2.65\pm0.05$ & $1.15\pm0.03$ & 4891 & 44 & 34 & 2.64 & 0.08 & 0.01 & $0.03\pm0.05$ & $-0.09\pm0.05$ & 0.06 \\ 
  & W08556 & 23:57:27.60 & +56:45:39.20 & 12.97 &  88 & FIES & $5012\pm68$ & $2.98\pm0.08$ & $1.33\pm0.09$ & $4978\pm11$ & $2.82\pm0.03$ & $1.46\pm0.02$ & 4995 & 39 & 23 & 2.90 & 0.06 & 0.11 & $0.02\pm0.05$ & $0.00\pm0.05$ & 0.01 \\ 
  & W08734 & 23:57:29.65 & +56:42:23.23 & 12.69 &  90 & FIES & $5015\pm56$ & $2.91\pm0.09$ & $1.05\pm0.10$ & $4925\pm12$ & $2.69\pm0.03$ & $1.39\pm0.02$ & 4970 & 34 & 63 & 2.80 & 0.06 & 0.16 & $0.15\pm0.05$ & $0.04\pm0.05$ & 0.06 \\ 
  & W10915 & 23:57:54.51 & +56:47:43.46 & 12.82 &  75 & FIES & $4985\pm74$ & $2.89\pm0.10$ & $1.07\pm0.12$ & $5005\pm14$ & $2.82\pm0.04$ & $1.38\pm0.02$ & 4995 & 44 & 14 & 2.86 & 0.07 & 0.05 & $0.08\pm0.05$ & $-0.01\pm0.05$ & 0.04 \\ 
  &   &   &   &   &  49 & HERMES & $4975\pm38$ & $2.75\pm0.08$ & $1.04\pm0.05$ & $5010\pm17$ & $2.77\pm0.05$ & $1.13\pm0.03$ & 4992 & 27 & 24 & 2.76 & 0.06 & 0.01 & $0.07\pm0.04$ & $-0.02\pm0.05$ & 0.04 \\ 
\hline
\end{longtable}
\end{landscape}
\twocolumn

 \subsubsection*{Arcturus and $\mu$-Leo}
Among the OCCASO data we have observations of two GBS (Arcturus and $\mu$-Leo) representative of the parameter space covered by the targeted OCs. Both stars were observed with the three instruments as well. As explained in Section~\ref{sec:BS} the GBS have determinations of atmospheric parameters independently from spectroscopy, and reference metallicities. We compare the results obtained from the two methods with the reference values in Table~\ref{tab:Arcmuleo}. We computed the mean value and standard deviation for each parameter from all the observed spectra. We also list in parentheses the mean error reported by each method. These two determinations of the internal error of the method are roughly of the same order. For both stars, GALA is reporting larger errors and also finds larger dispersions than iSpec in $T_{\mathrm{eff}}$ and $\log g$, but not in metallicity.

From the comparison with the reference values from \citet{Heiter+2015} we obtain an excellent agreement in effective temperature. Differences in gravity are of the same order in both methods: for $\mu$-Leo both methods underestimate by approximately the same amount; for Arcturus, iSpec underestimates it but GALA overestimates it. However, Arcturus has a large uncertainty in $\log g$ as a GBS, and as quoted by the authors \citep{Heiter+2015} it can be used for validation purposes only if the large error is taken into account. The differences found in atmospheric parameters are compatible with the quoted errors.

The differences in iron abundances are compatible within 3$\sigma$ with the dispersions found between the three instruments but not compatible with the mean errors quoted by the methods. In the case of Arcturus both methods slightly underestimate the abundance. For $\mu$-Leo GALA slightly overestimates the abundance and iSpec underestimates it by 0.12 dex. It is a metal-rich star with many blended lines, thus, EW methods which are not able to reproduce blends as well as SS methods, tend to provide higher abundances. Still, the EW method matches the reference value while the SS method gives a lower abundance than the reference. It is worth noting that the GBS reference metallicities were obtained based on a spectroscopic analysis where several methods were averaged, which can bias the reference result to one analysis methodology.

\begin{table*}
\centering
  \small
  \caption{\label{tab:Arcmuleo}Effective temperature, surface gravity and metallicity for Arcturus and $\mu$-Leo obtained from OCCASO data using GALA and iSpec. The errors indicate the dispersion found between the three instruments, and in parenthesis the mean of the errors reported by the methods. Reference values are from \citet{Heiter+2015,Jofre+2014}. The differences (ours-reference) are
  in the last three columns.}
  \setlength\tabcolsep{3.1pt}
\begin{tabular}{ccccccccccc}
\hline 
Star & $T_{\mathrm{eff,ref}}$ (K) & $\log g_{\mathrm{ref}}$ & [Fe/H]$_{\mathrm{ref}}$ & Method &$T_{\mathrm{eff}}$ (K) & $\log g$ & [Fe/H] & $\Delta T{\mathrm{eff}}$ (K)& $\Delta \log g$ & $\Delta$[Fe/H]\\
\hline 
Arcturus  & 4286$\pm$35 & 1.64$\pm$0.20 & $-0.52\pm0.08$ & iSpec & $4234\pm8$  (5)  & $1.46\pm0.02$ (0.02) & $-0.55\pm0.04$ (0.05)  & -52 & -0.18 & -0.03\\
          &      &      &                & GALA  & $4325\pm47$ (54) & $1.81\pm0.08$ (0.14) & $-0.54\pm0.03$ (0.05) &  39 &  0.17 & -0.02\\
\\
$\mu$-Leo & 4474$\pm$60 & 2.51$\pm$0.11 & $ 0.25\pm0.15$ & iSpec & $4448\pm6$  (5)  & $2.34\pm0.03$ (0.02) & $ 0.13\pm0.05$ (0.06)  & -26 & -0.17 & -0.12\\
          &      &      &                & GALA  & $4508\pm20$ (98) & $2.36\pm0.16$ (0.20) & $ 0.27\pm0.06$ (0.05)  &  34 & -0.15 &  0.02\\
\hline
\end{tabular}
\end{table*}

\subsection{Benchmark stars}\label{sec:BSres}
As a sanity check to ensure the validity of our analysis we analysed 67 spectra from 23 GBS using the same line list, atmosphere models and strategy as in the case of OCCASO stars.

\begin{figure}
 \centering
\includegraphics[width=0.5\textwidth]{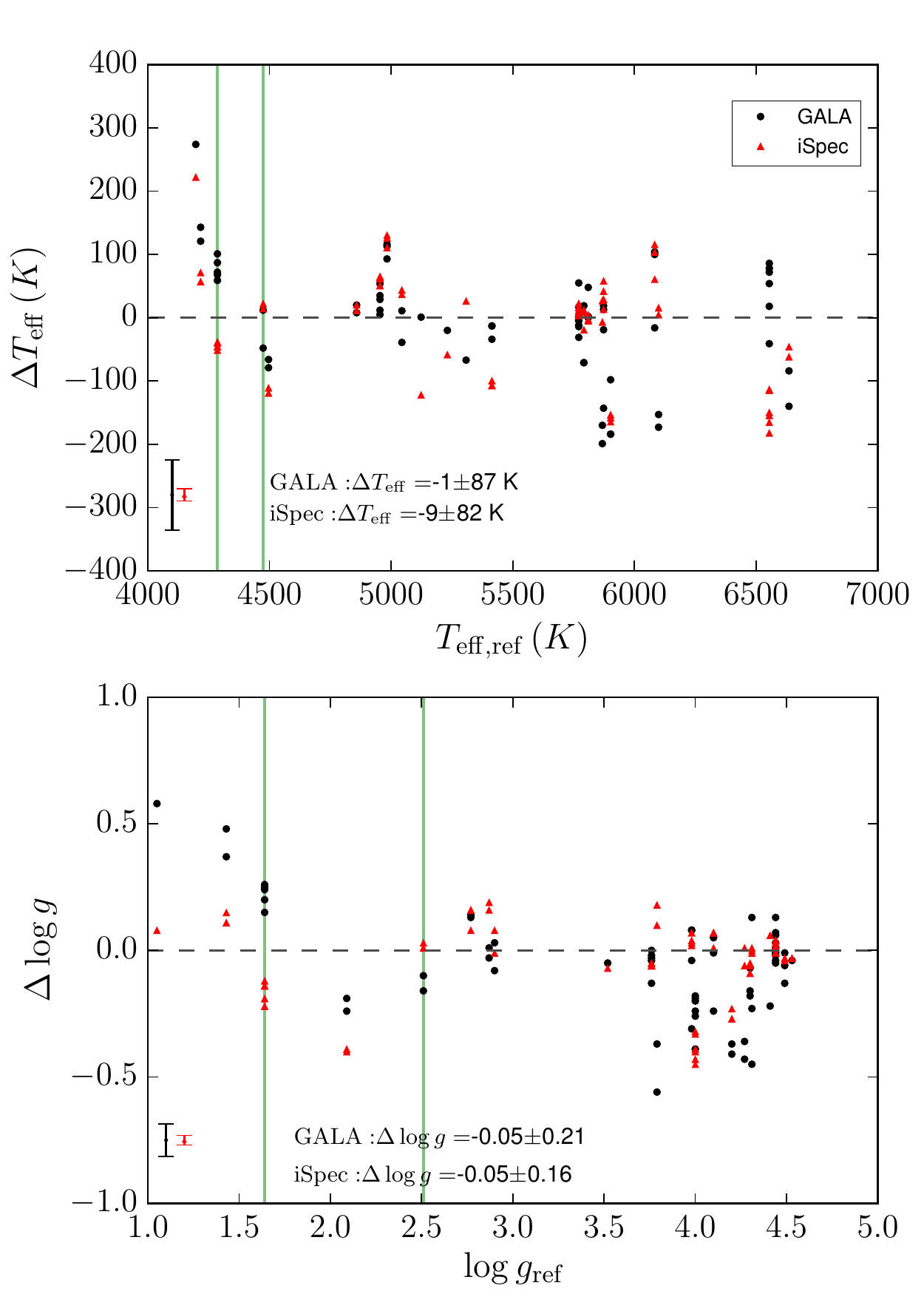}
 \caption{Differences in effective temperature (top panel) and surface gravity (bottom panel), between GALA and reference value (black dots), and iSpec and reference value (red triangles), for GBS library spectra described in Sec~\ref{sec:BS}. The two vertical green lines indicate Arcturus and $\mu$-Leo. Mean errorbars are plotted on the bottom-left of each pannel. Differences are calculated in the sense: this study - reference. Reference values are taken from \citet{Heiter+2015}.}\label{fig:BS}
\end{figure}

We compare the results of our analysis with the reference ones described in \citet{Heiter+2015} in Fig.~\ref{fig:BS}. We remark with vertical green lines the Arcturus and $\mu$-Leo spectra, the two GBS also observed in OCCASO. We obtain overall offsets which are compatible at 1$\sigma$ level with the dispersions in both $T_{\mathrm{eff}}$ and $\log g$. The results are available in Table~\ref{tab:resultsBS}. The highest differences are found for $\beta$-Ara and $\eta$-Boo. For $\beta$-Ara its reference parameters are uncertain and should not be used as a reference for calibration or validation purposes \citep[see Table 10 of ]{Heiter+2015}. $\eta$-Boo has the highest rotational velocity of all GBS (12.7 km s$^{-1}$), see Table 1 of \citet{Jofre+2014}, which makes the spectroscopic analysis more uncertain.

\begin{figure}
 \centering
\includegraphics[width=0.45\textwidth]{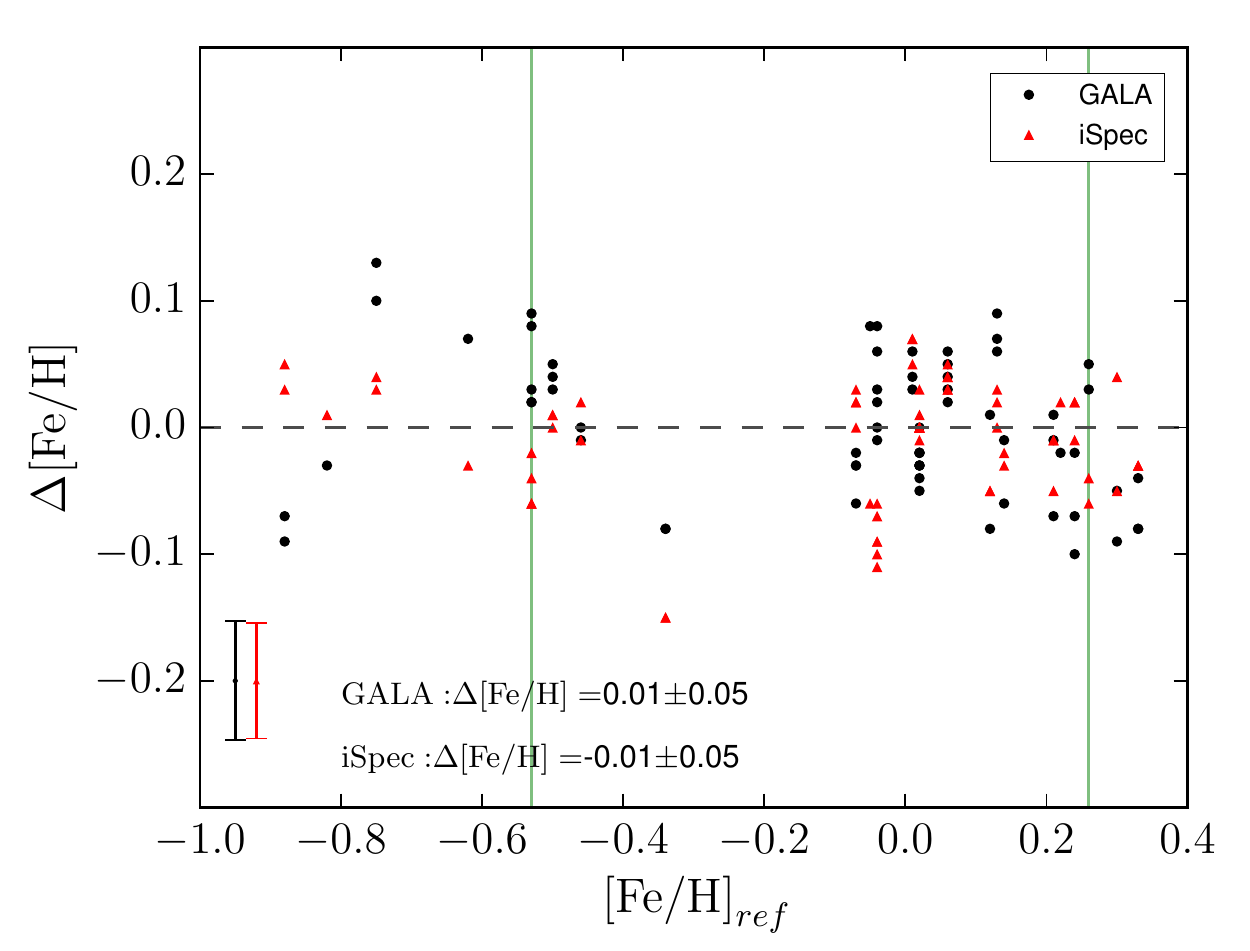}
 \caption{Differences in iron abundances between
GALA and reference value (black dots), and iSpec and reference value (red triangles), for GBS library spectra described in Sec~\ref{sec:BS}. The two vertical green lines correspond to Arcturus and $\mu$-Leo. Mean errorbars are plotted on the left of the pannel. Differences are calculated in the sense: this study - reference. Reference values are taken from \citet{Jofre+2014}.}\label{fig:BS_Fe}
\end{figure}

We also tested the iron abundances derived by the two methods with the GBS sample. Each pipeline analysed the spectra of the selected GBS using its own atmospheric parameters. In Fig.~\ref{fig:BS_Fe} we compare the Fe abundance results from GALA and iSpec, with the reference values in \citet{Jofre+2014}. We assign the internal dispersion given by all the lines divided by the square root of the number of lines, plus a fixed quantity that comes from the dispersion between both methods. Both methods show good agreement.

We calculated the dispersion in each parameter of the different observations of the stars that have more than one spectra. The mean value of these dispersions are $T_{\mathrm{eff}}$: 9 K, 24 K; $\log g$: 0.02 dex, 0.06 dex; and [Fe/H]: 0.01 dex, 0.01 dex (iSpec and GALA, respectively). All are smaller than the dispersions of the comparison with reference values.

\subsection{Photometric parameters}
We did an additional independent check of the spectroscopic results by performing a comparison with photometric $T_{\text{eff}}$ and $\log g$. We used precise BVI Johnson photometry (P. Stetson, priv. comm.; \citet{Stetson2000}) for two clusters in the sample, NGC~2420 and NGC~6791. These are one of the most metal-rich and one of the most metal-poor clusters in the sample.

Photometric $T_{\text{eff}}$ was obtained using \cite{Alonso+1999} colour-temperature empirical relations as a function of the dereddened colour $\left( B-V \right)_{0}$ and the metallicity (Eq. 4 from their Table 2). Photometric surface gravity is derived from $T_{\text{eff}}$ using fundamental relations: 

\begin{align}
\log \left( \frac{g}{g_{\odot}} \right)=0.4(M_{\text{bol}}-M_{\text{bol,}\odot})+\log \left( \frac{m}{m_{\odot}} \right)+4\log \left( \frac{T_{\text{eff}}}{T_{\text{eff},\odot}} \right)
\end{align}

\noindent where $\log g_{\odot}$, $M_{\text{bol,}\odot}$, $m_{\odot}$ and $T_{\text{eff},\odot}$ are the surface gravity, bolometric magnitude, mass and effective temperature of the Sun respectively\footnote{We assume $\log g_{\odot}=4.438$, $M_{\text{bol,}\odot}=4.74$ and $T_{\text{eff},\odot}=5772$ K following the IAU recommendations \citep{Prvsa+2016}.}, and $m$ is the mass of the star derived from the isochrone fitting\footnote{We have used PARSEC isochrones \citep{Bressan+2012}.}. The bolometric magnitude of the star was calculated from the bolometric correction for giants using \cite{Alonso+1999} prescriptions $M_{\text{bol}}=V_0+BC_{\text{V}}$.

We also derived parameters from $\left( V-I \right)$ colour. To do so we calculated extinction in $V-I$ assuming $\frac{A_{\mathrm{I}}}{A_{\mathrm{V}}}=0.479$ \citep{Cardelli+1989}. A similar relation as for $\left( B-V \right)_{0}$ is provided for $\left(V-I \right)_{0}$ by \cite{Alonso+1999} to derive $T_{\text{eff}}$. Surface gravity was derived in the same way using these temperatures.

\begin{figure}
 \centering
 \includegraphics[width=0.4\textwidth]{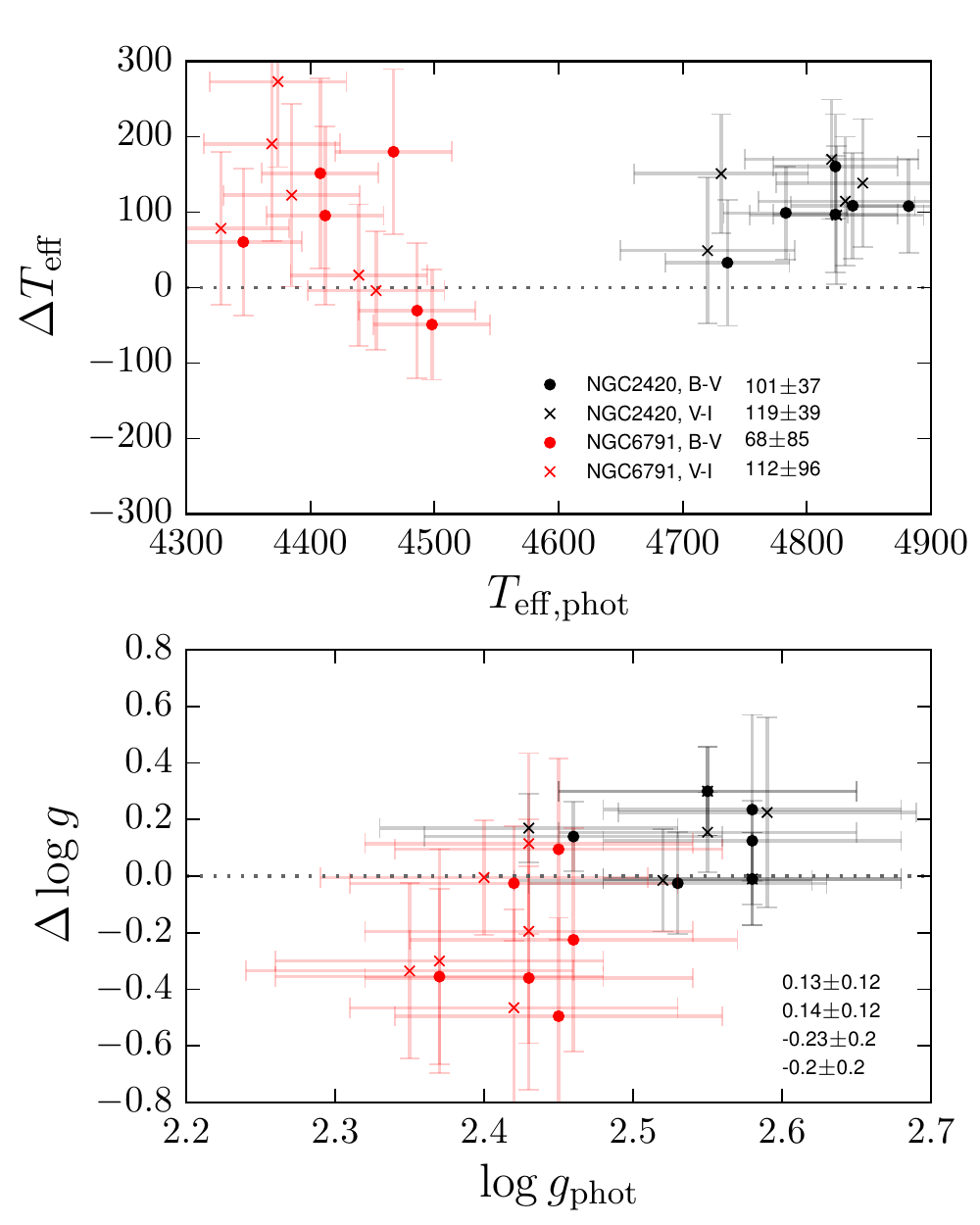}
 \caption{Differences in effective temperature and surface gravity from spectroscopy (mean of GALA and iSpec results) and from photometry for the individual stars in NGC~2420 and NGC~6791. Mean differences and dispersions for the two OCs and the two photometries are in the bottom right.}\label{fig:APcompare_phot}
\end{figure}

We compare the photometric results with the spectroscopic ones in Fig.~\ref{fig:APcompare_phot}. The adopted input parameters for the two clusters: reddening $E \left( B-V \right)$, distance modulus $\left( V_0-M_{\text{V}} \right)$, age and metallicity, are indicated in Table~\ref{tab:photpar}. For the two clusters we compute the mean $T_{\text{eff}}$ and $\log g$, from the spectroscopic and photometric analysis in Table~\ref{tab:photparres}. The dispersion of the spectroscopic parameters within each cluster is around 1.7 and 5.7 times higher (in  $T_{\text{eff}}$ and $\log g$, respectively) than the photometric one. This is compatible within 1$\sigma$ and 2-3$\sigma$, respectively, with the mean uncertainties of the methods.

Both determinations are compatible within $1-2\sigma$, though we find systematic differences which are not the same for the two analysed clusters. Photometric results are very sensitive to the assumed cluster parameters. Any variation in reddening, distance or age within the given errors change the overall offset with respect to spectroscopic parameters. However, the internal dispersion among the stars of the same cluster remains constant. We have assigned as error the dispersion in photometric parameters when changing $E \left( B-V \right)$, $\left( V_0-M_{\text{V}} \right)$ and [Fe/H] by $\pm \sigma_i$.

\begin{table}
\begin{centering}
  \footnotesize
  \caption{\label{tab:photpar}Adopted input cluster parameters to calculate photometric temperatures and surface gravities.}
  \setlength\tabcolsep{3.5pt}
\begin{tabular}{ccccccc}
\hline 
Cluster & $E \left( B-V \right)$ & $\left( V_0-M_{\text{V}} \right)$ & $\log$Age (Gyr) & [Fe/H] \\
\hline 
NGC~2420$^1$ & 0.04$\pm$0.03 & 11.88$\pm$0.27 & 9.47$\pm$0.17 & -0.20$\pm$0.06 \\
NGC~6791$^2$ & 0.12$\pm$0.03 & 13.25$\pm$0.35 & 9.91$\pm$0.20 & +0.29$\pm$0.08 \\
\hline
\end{tabular}
\end{centering}

$^1$Reddening, distance modulus and age from \citet{Pancino+2010}, calculated as average measurements of different authors, metallicity from \citet{Jacobson+2011b}, calculated as average of 9 stars.\\
$^2$Reddening as a mean of all previous determinations (\citet{Sandage+2003}, \citet{Stetson+2003}, \citet{Anthonytwarog+2007}, \citet{Brogaard+2012}, \citet{Geisler+2012}), distance modulus from \citet{Sandage+2003}, age and metallicity from \citet{Brogaard+2012}.\\
\end{table}

\begin{table}
\centering
  \footnotesize
  \caption{\label{tab:photparres}Means and standard deviations of effective temperatures and gravities for the two clusters analysed with photometry. Results from spectroscopy of GALA and iSpec, and from B-V and V-I photometry.}
  \setlength\tabcolsep{2.8pt}
\begin{tabular}{ccccccc}
\hline 
Cluster & & $T_{\mathrm{eff,spectr}}$ & $\log g_{\mathrm{spectr}}$ & & $T_{\mathrm{eff,phot}}$ & $\log g_{\mathrm{phot}}$ \\
 & & (K) & (dex) & & (K) & (dex)  \\
\hline 
NGC~2420 &GALA: & 4899$\pm$87 & 2.69$\pm$0.20 & \textit{B-V:} & 4814$\pm$45 & 2.55$\pm$0.04\\
         &iSpec:& 4931$\pm$64 & 2.66$\pm$0.12 & \textit{V-I:} & 4795$\pm$50 & 2.54$\pm$0.05\\
NGC~6791 &GALA: & 4507$\pm$94 & 2.07$\pm$0.34 & \textit{B-V:} & 4436$\pm$53 & 2.43$\pm$0.03\\
         &iSpec:& 4502$\pm$81 & 2.34$\pm$0.12 & \textit{V-I:} & 4391$\pm$43 & 2.40$\pm$0.03\\
\hline
\end{tabular}
\end{table}

\subsection{Adopted $T_{\text{eff}}$ and $\log g$}
We checked the consistency of the stars repeated with the three instruments and we do not find any significant systematic offset.

Since the results are compatible and the differences are at the level of the expected uncertainties of the analysis we decided to fix $T_{\text{eff}}$ and $\log g$ to the average results from both methods to do the chemical analysis. This approach is a statistically consistent way to combine two results of the same physical quantity that do not show any systematic offset. Moreover, this helps to disentangle the discrepancies in the determination of chemical abundances from the discrepancies due to the propagation of errors from different $T_{\text{eff}}$ or $\log g$. Additionally, this strategy allows us to provide an estimation of the external uncertainty (method-dependent) for each star, aside of the error quoted by each pipeline in the derivation of the parameters.

In Table~\ref{tab:finpar} we list the average results of the two parameters. We indicate two sources of errors: the mean of the errors quoted by the methods $\delta_1$, and the standard deviation between the two values $\delta_2$. In general the dispersion between the methods is similar to the mean of the errors, with mean values of $\delta_1$ and $\delta_2$ of: 38 and 31 K ($T_{\text{eff}}$), and 0.07 and 0.11 ($\log g$).

\section{Iron abundances}\label{sec:FeH}
We used the average values of $T_{\mathrm{eff}}$ and $\log g$ shown in Table~\ref{tab:finpar} to calculate the chemical abundances of the whole sample of 154 spectra in a second step. 

We have followed a global differential approach relative to the Sun with the two methods. That is, subtracting the mean abundance of all lines measured in the Sun from the mean abundance of all lines observed in the star spectrum (not line-to-line). As Solar abundance we derived $A\left(\ion{Fe}{I}\right)_{\odot,\mathrm{GALA}}=7.46\pm0.01$, $A\left(\ion{Fe}{I}\right)_{\odot,\mathrm{iSpec}}=7.39\pm0.02$ using the solar spectra provided in the GBS library \citep{Blanco+2014}. In this way we are sure that the two methods have the same internal scale.

The iron abundances derived from each method are listed in Table~\ref{tab:finpar}. The error assigned by the methods is the standard deviation of the abundances from each line divided by the square root of the number of used lines. We also include in Table~\ref{tab:finpar} the standard deviation of the abundance derived by the two methods $\sigma$[Fe/H]. This last value provides a less model-dependent estimation of the error, and its mean is 0.04 dex. We consider that a good approximation of the error in [Fe/H] derived by each method is the squared sum of the spread of line-by-line abundance divided by the square root of the number of lines, and this value of $0.04$ dex. Therefore the mean errors are $0.047$ in EW, and $0.052$ in SS.

We calculated the errors in the [Fe/H] due to the choice of the parameters: $T_{\mathrm{eff}}$, $\log g$ and $\xi$. To do so, we varied the three parameters by $\pm\sigma_i$ and recomputed the abundance for 5 representative stars. We  used as errors of $T_{\mathrm{eff}}$ and $\log g$ the quadratic sum of $\delta_1$ and $\delta_2$ in Table~\ref{tab:finpar}. We did this process with the two methods. The results are summarized in Table~\ref{tab:errparams}. These uncertainties range from $-0.04$ to $0.03$ dex in \texttt{GALA}, and $-0.02$ to $0.02$ dex in \texttt{iSpec} well within the mean uncertainties of the methods.

\begin{figure}
 \centering
 \includegraphics[width=0.4\textwidth]{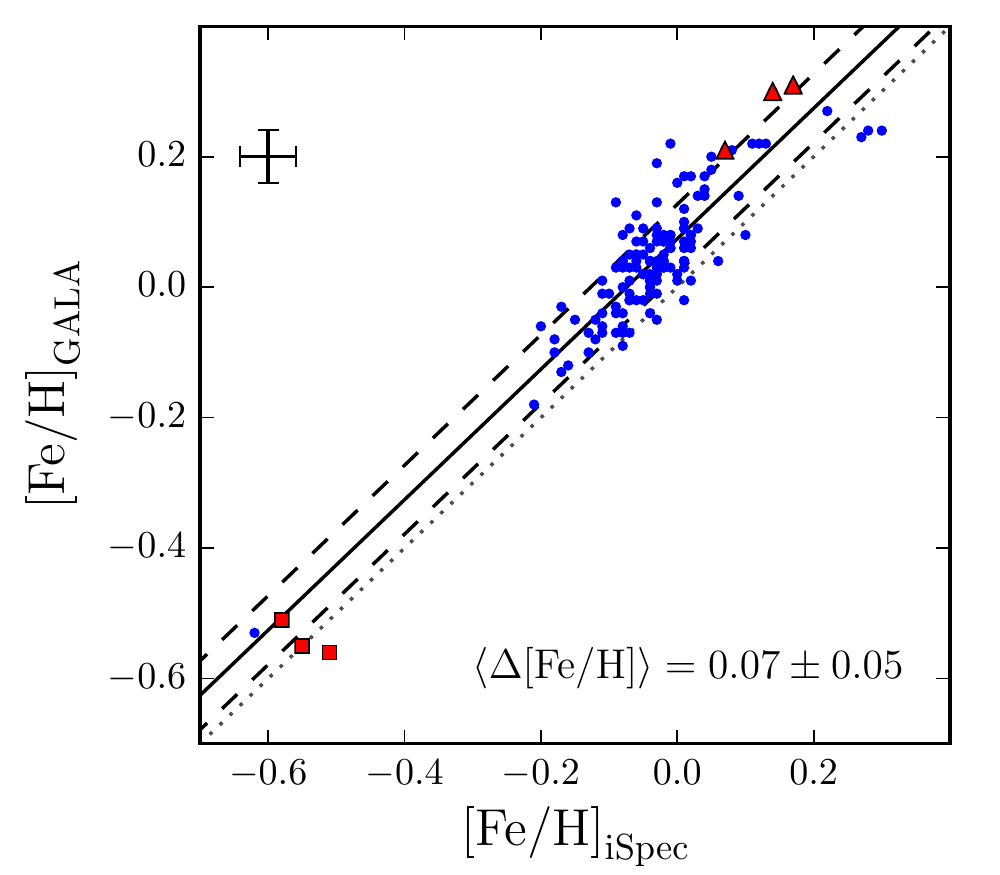}
 \caption{\label{FeHcomp}Results of iron abundance from GALA and iSpec analysis. Red squares and red triangles indicate the values of Arcturus and $\mu$-Leo (3 spectra each), respectively. The solid line indicates the mean differences, and the dashed lines indicate the $1\sigma$ level. The dotted line is the 1:1 relation. In the top left corner we plot the mean errors in X and Y axis.}
\end{figure}

The comparison of the iron abundances obtained by the two methods is plotted in Fig.~\ref{FeHcomp}. The plotted errorbars are the mean of the errors quoted by the methods, plus the mean of the $\sigma$[Fe/H]$=0.04$ from Table~\ref{tab:finpar}. There exists an offset between the two determinations of $0.07\pm0.05$.

\begin{table}
\centering
  \caption{Variation in the [Fe/H] calculated by both methods when altering atmospheric parameters by $\pm \sigma$.  \label{tab:errparams}}
  \renewcommand{\arraystretch}{2.2}
\begin{tabular}{ccccccc}
\hline 
 Parameter & GALA & iSpec \\
\hline 
\(\displaystyle\frac{\mathrm{d}\text{[Fe/H]}}{\mathrm{d}T_{\text{eff}}} \) & $^{+0.027} _{-0.023}$ & $^{+0.022} _{-0.018}$\\
\(\displaystyle\frac{\mathrm{d}\text{[Fe/H]}}{\mathrm{d}\log g}\)          & $^{+0.019} _{-0.024}$ & $^{+0.012} _{-0.010}$\\
\(\displaystyle\frac{\mathrm{d}\text{[Fe/H]}}{\mathrm{d}\xi}\)             & $^{-0.036} _{+0.034}$ & $^{+0.004} _{-0.010}$\\
\hline
\end{tabular}
\end{table}
\newpage

\subsection{Performance of the methods}\label{performance}
The scale of the difference found between our spectroscopic methods is compatible with previous works that already studied this in detail \citep{Hinkel+2016,Blanco-Cuaresma+2016,Blanco-Cuaresma2016b,Jofre+2016b}.

\begin{figure}
 \centering
 \includegraphics[width=0.5\textwidth]{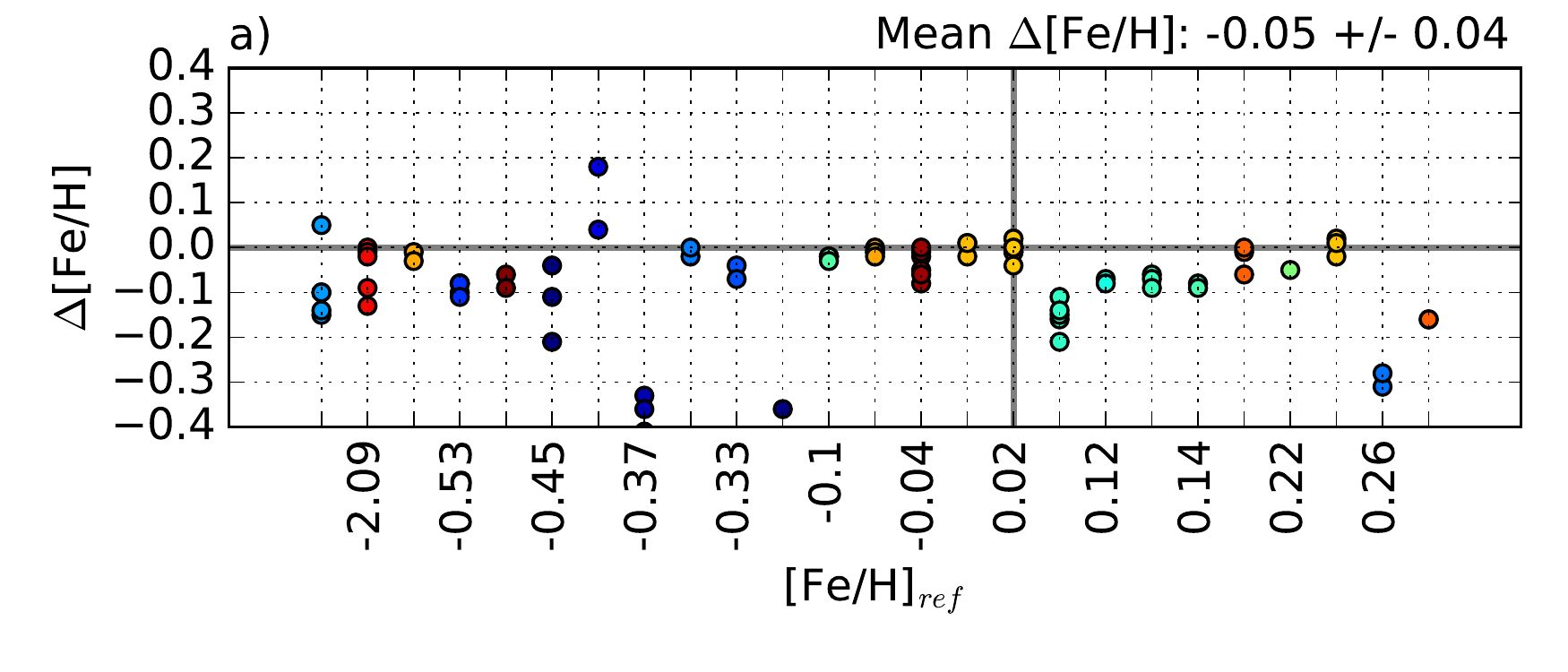}
 \includegraphics[width=0.5\textwidth]{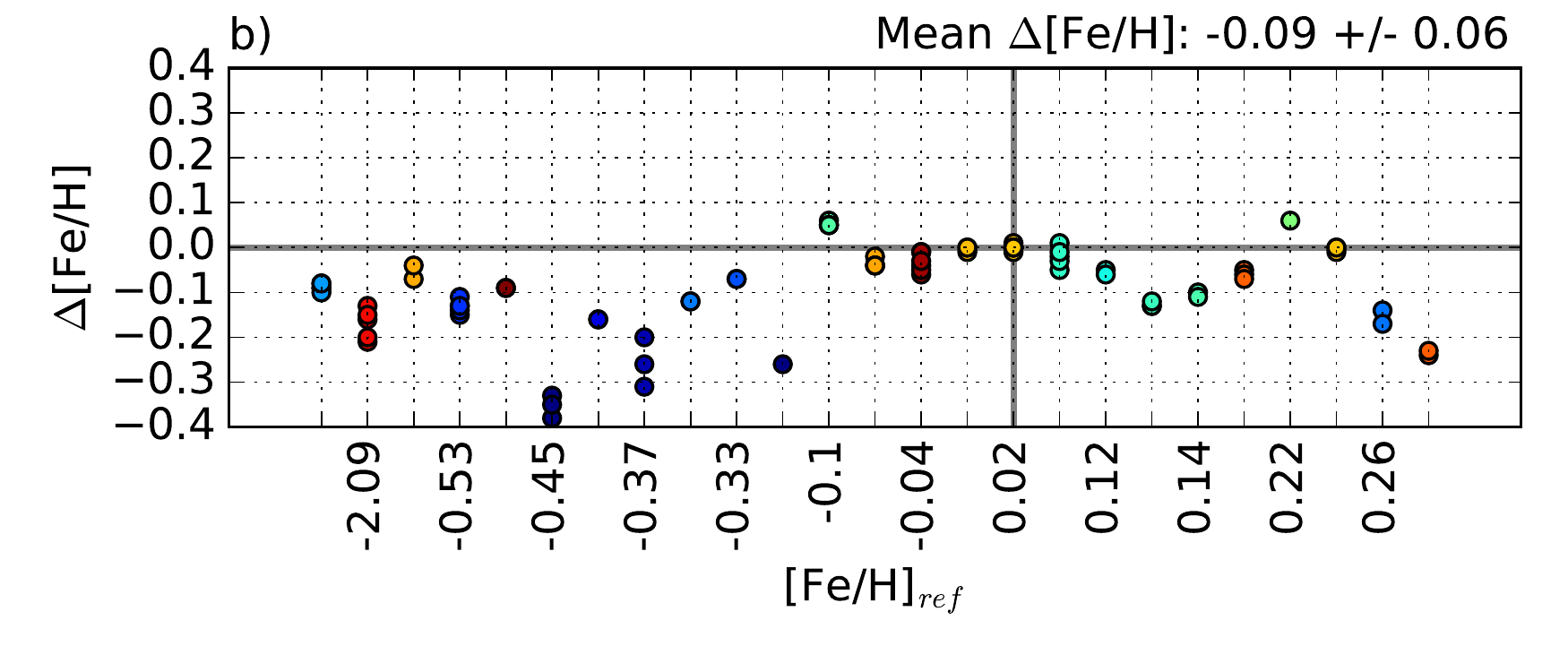}
 \includegraphics[width=0.5\textwidth]{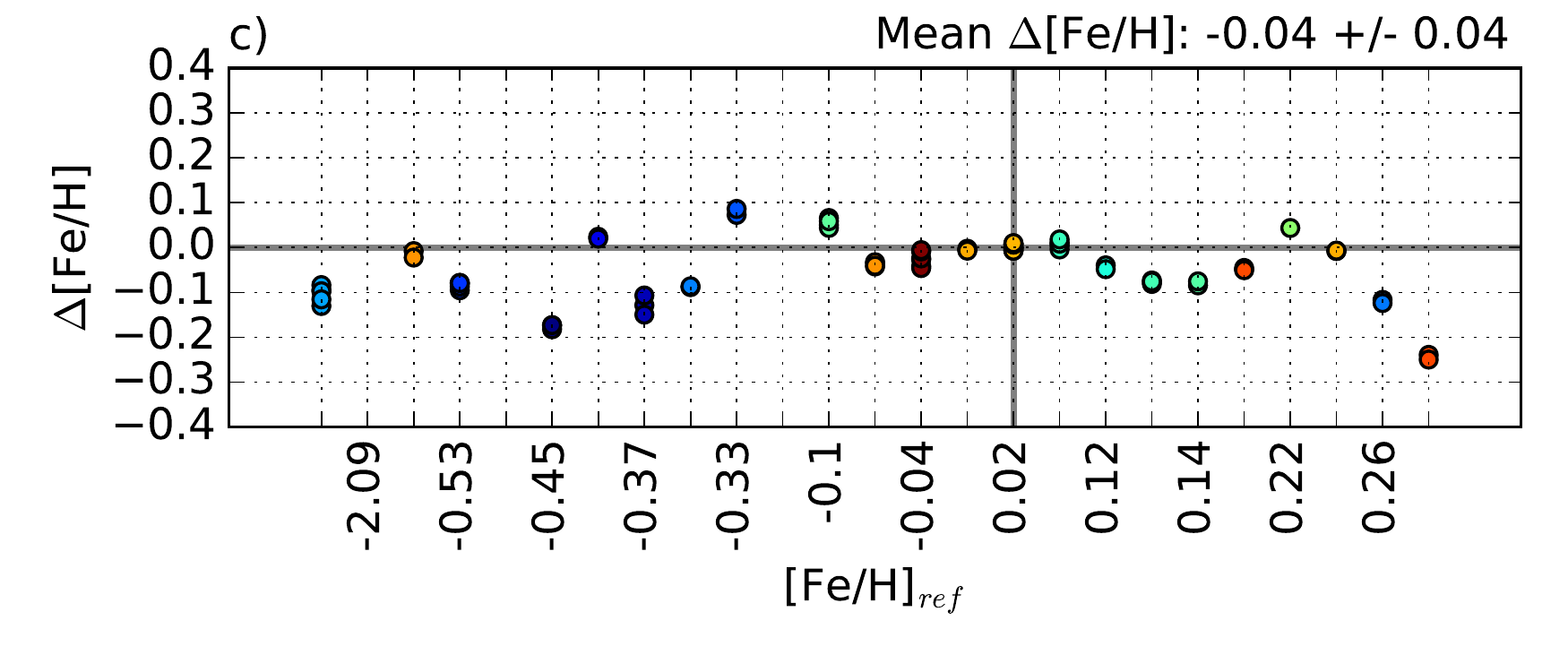}
 \includegraphics[width=0.5\textwidth]{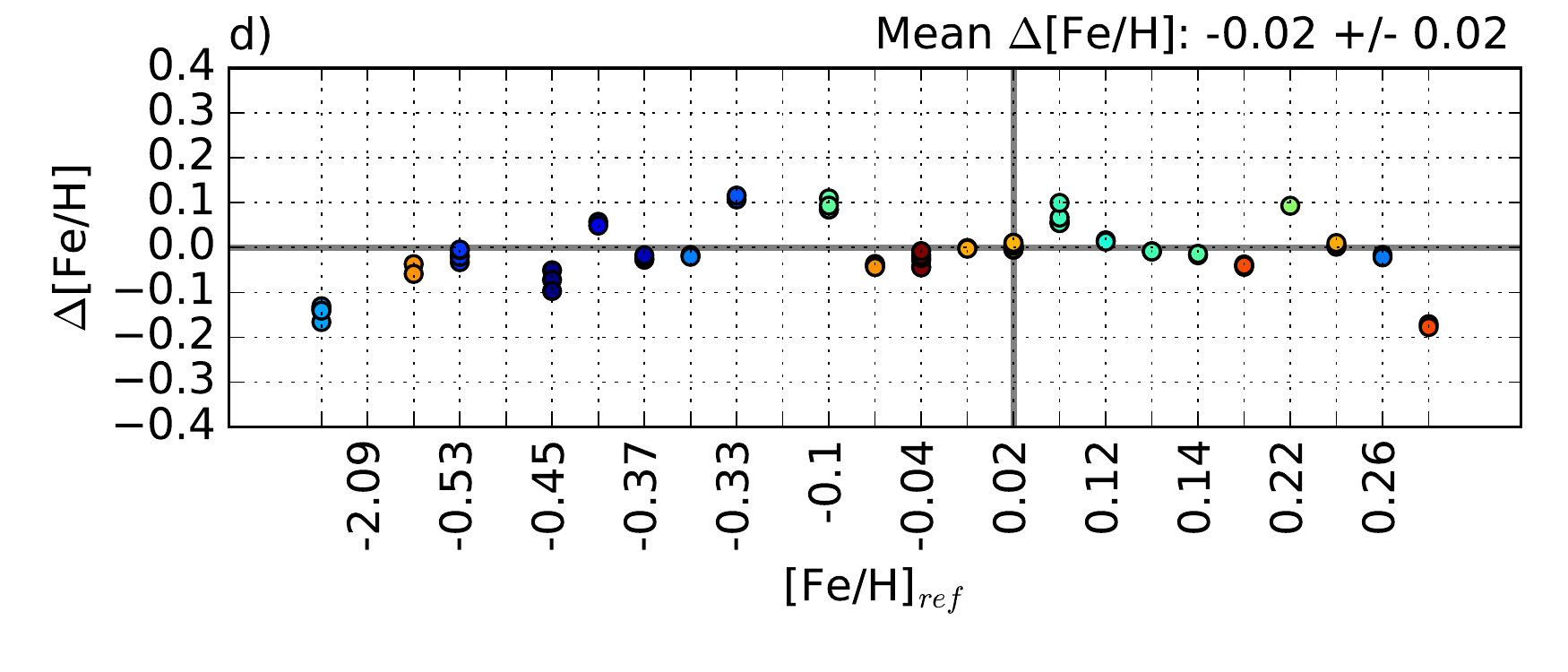}
 \caption{\label{BSperformance}Differences in iron abundance obtained for the GBS analysis between SPECTRUM - WIDTH9. The four panels stand for the four cases mentioned in the text. The colours represent the reference temperature where blue is cold and red is hot. The Sun is indicated with a vertical grey line.}
\end{figure}

To better illustrate the differences between the methods for our particular case, we used iSpec capabilities to perform synthesis and equivalent width analysis. We configured iSpec to use SPECTRUM \citep{Gray+1994} for SS (the same radiative transfer code used in this study) and WIDTH9 \citep{Kurucz1993,Sbordone+2004} for the EW method (which is the one used by GALA). Then we derived the [Fe/H] for the GBS considering four different scenarios as shown in Fig.~\ref{BSperformance}:

\begin{enumerate}[(a)]
\item We fix $T_{\mathrm{eff}}$ and $\log g$ to their reference value \citep{Heiter+2015} and we derive the rest of parameters with each method independently.
\item Like in the previous case but we also fix the microturbulence.
\item Like in the previous case but we use only lines in common between both methods.
\item Like in the previous case and we force the synthesis method not to synthesize blends.
\end{enumerate}

The first case coincides with the strategy followed in our study and its average difference is comparable to our results. If we fix the microturbulence parameter, the dispersion per star does not improve and the overall mean difference worsen. The microturbulence is a parameter used to compensate errors and assumptions in the models and this compensation depends on the method, thus fixing it does not improve the agreement between both methods.

When we use only lines in common, there are three stars that get excluded from the analysis because no overlapping lines were found. Lines that are good enough for methods based on SS might not be convenient for EW (e.g., blended lines), thus the line selection is different for each method and it can be challenging to find lines in common (specially for metal-pool stars). Nevertheless, the agreement between both methods improves when using the same absorption lines.

In the fourth case, we forced the synthetic method to only synthesize the lines being analysed (ignoring the atomic lines around it) to make it more similar to the equivalent width method. This is the case with a higher level of agreement.

This analysis shows that the differences between methods are intrinsic to how each technique works. A further quantitative study of this can be found in \citet{Jofre+2016b}. In this paper it is shown that EW and SS methods can be affected differently by the different assumptions (e.g. blends, wavelength shifts, continuum).\\
Based on our analysis we argue that the derivation of abundances must be properly documented, where input parameters and method assumptions have to be provided to the community for better reproducibility of results, understanding of uncertainties and correct use of the data. Among the scientific community there is no consensus on which methods are better or worse to derive spectroscopic abundances. Thus, we include the results derived by the two methods, and we are fully transparent in how our calculations are done. The reader can choose whichever he or she trusts more.

\section{Cluster-by-cluster analysis}

For each OC we took into account the membership selection from the radial velocities done in Paper I and Casamiquela et al. (2017, in prep). The membership was reanalysed taking into account the metallicities derived in this work (Table~\ref{tab:finpar}).

\begin{figure*}
\centering
\includegraphics[width=0.9\textwidth]{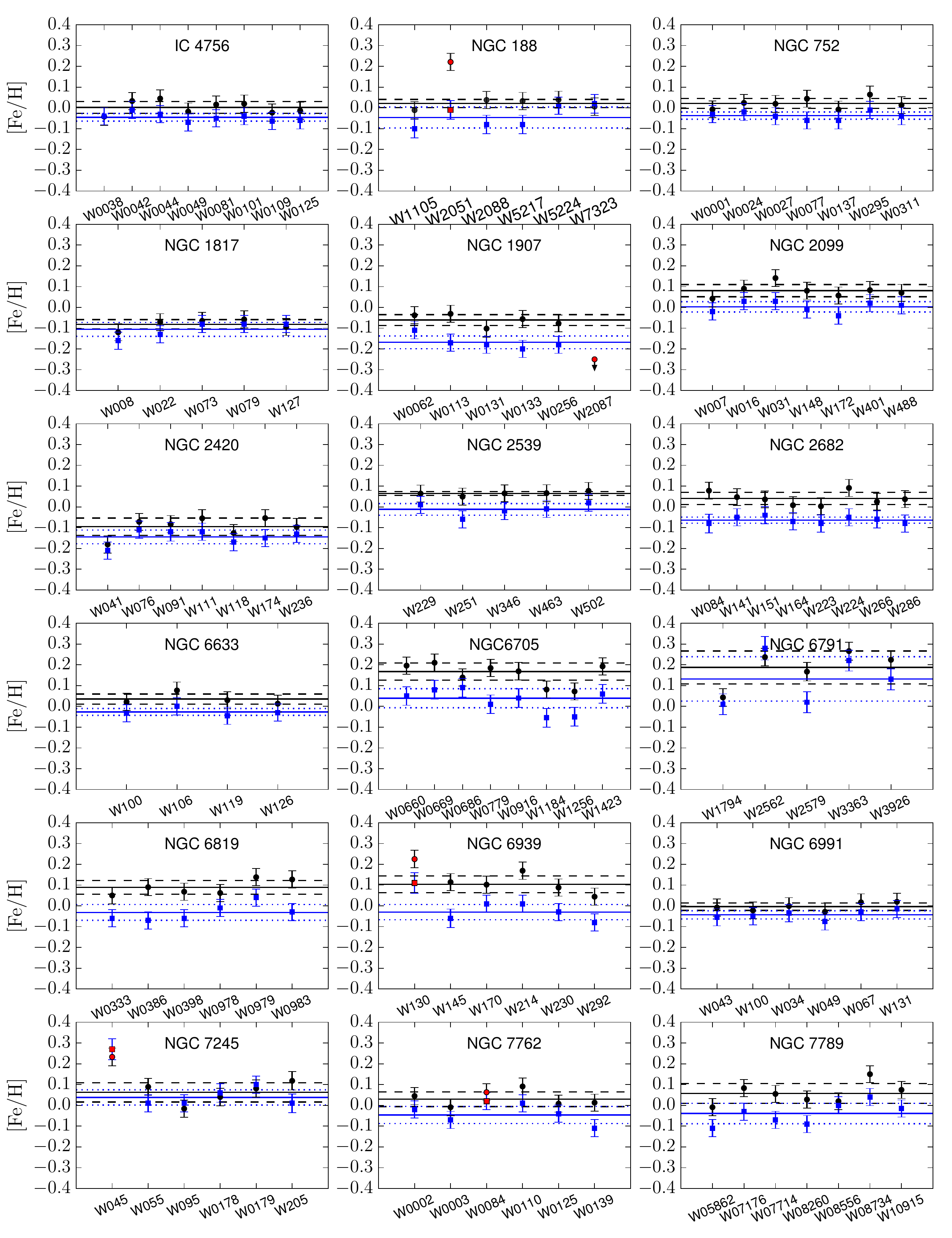}
\caption{Iron abundances obtained for the 18 studied OCs. In black GALA results, in blue iSpec results. Red symbols indicate probable non-members or spectroscopic binaries detected by their radial velocity or [Fe/H]. These stars are not used to compute the mean abundance. NGC~1907 W2087 is indicated with an arrow because it falls out of the plot. The black solid and dashed lines indicate the mean and 1$\sigma$ level of GALA iron abundance, respectively. The blue solid and dotted lines indicate the mean and 1$\sigma$ level of iSpec iron abundance, respectively.}\label{clusters_Fe}
\end{figure*}

We plot in Fig.~\ref{clusters_Fe} the two determinations (GALA and iSpec) of [Fe/H] obtained for the stars in each OC. For the stars that have determinations with the different instruments, we plot the mean value. The cluster averaged [Fe/H] was calculated using only trustful member stars. This means that we exclude those stars with discrepant radial velocities, possible non-members or spectroscopic binaries, or stars that have not converged in the analysis. These stars are marked in red in Fig.~\ref{clusters_Fe}.

We draw special attention to the following stars:

\begin{enumerate}[(i)]
 \item NGC 188 W2051 has a radial velocity above the mean of the cluster, but compatible within 3$\sigma$  (this cluster was not analysed in Paper I). GALA derives a higher [Fe/H] compared with the rest of the stars in the OC. However, iSpec finds it compatible with the rest of the stars. We reject it for safety.
 \item NGC 1907 W2087 was flagged as non-member in Paper I for having a significant difference in radial velocity with respect to the other stars in the cluster. Moreover, both GALA and iSpec obtain a [Fe/H] which differs in more than 0.5 dex from the other stars of the cluster. We confirm that it is a non-member.
 \item NGC 2539 W233 was flagged as spectroscopic binary in Paper I, and previously in the literature. It gave inconsistent results in the analysis by the  two methods: very high gravity and temperature (4.5 dex, 6500 K) in iSpec, and very low microturbulence in GALA compared to the other stars. This is probably because the spectral lines have a distorted shape due to the companion star. Therefore, we do not consider it in the cluster analysis and it is not included in Fig.~\ref{clusters_Fe}.
 \item NGC 2682 W224 has a discrepant radial velocity in Paper I. It was flagged as member spectroscopic binary by \citet{Jacobson+2011b} and \citet{Geller+2015}. The spectral analysis with both GALA and iSpec give results in agreement. Therefore, we consider its results of abundances in the analysis.
 \item NGC 6791 W2604 has a compatible radial velocity with the other stars in this cluster. However, DAOSPEC finds a large line-by-line dispersion when calculating the radial velocity: 3.2 km s$^{-1}$ compared with 1-2.3 km s$^{-1}$ obtained with the other cluster stars. Also, the mean FWHM measured for its lines is significantly higher (13 pixels approximately), compared with the other stars 8.5-10 pixels. A cross-correlation done with iSpec, using a template shows two clear peaks, which indicates that it is probably an spectroscopic binary. Its results of the abundances have large errors and are quite discrepant with the other stars of the cluster. We discard its abundance to calculate the cluster mean.
 \item NGC 6791 W3899 has a compatible radial velocity with the other stars in this cluster but it also shows two peaks in a cross-correlation, which indicates that it is a possible spectroscopic binary. We discard its abundance to calculate the cluster mean and we do not plot it in Fig.~\ref{clusters_Fe}.
 \item NGC 6819 W983 was flagged as spectroscopic binary in Paper I for having variable radial velocity. We could analyse this star by shifting the individual exposures to a common reference frame. It gives satisfactory results with both methods, and compatible Fe abundance. For this reason we consider it in the cluster abundance analysis.
 \item NGC 6939 W130 has a more than 3$\sigma$ discrepant radial velocity with respect to the other cluster members. It gives a around 2$\sigma$ discrepant value of the [Fe/H] so it is probably a non-member. We discard it to calculate the mean abundance.
 \item NGC 7245 W045 has a more than 3$\sigma$ discrepant radial velocity, and has a quite different [Fe/H] from the rest of cluster stars. Its abundance is higher than the rest of the stars by more than 3$\sigma$. So this star is possibly a non-member.
 \item NGC 7762 W0084 had a more than 3$\sigma$ discrepant radial velocity in Paper I, pointing out that it could be a non-member. We do not consider it to compute the cluster abundance.
\end{enumerate} 

The sample of bona fide member stars was used to compute the cluster mean iron abundance. This value and its dispersion is indicated in Table~\ref{tab:meanFe}. The internal dispersions within each cluster are found in the range 0.01-0.08 dex from the EW analysis, and 0.01-0.11 dex from the SS analysis. The largest dispersion for both methods corresponds to NGC~6791 $0.08$ and $0.11$ dex for EW and SS, respectively. This is the faintest OC in our sample with SNR$\sim 50$, while for the others we reach SNR$\sim 70$. This may partly explain the large dispersion.\\
The most metal-rich OCs are NGC~6791 and NGC~6705 according to \texttt{GALA} results, and NGC~6705 is not metal-rich according to \texttt{iSpec}. On the other hand, the most metal-poor clusters are NGC~2420, NGC~1817 and NGC~1907, for both \texttt{GALA} and \texttt{iSpec}. We note that this is the first time chemical abundances are derived from high-resolution spectroscopy for the clusters NGC~6939, NGC~6991 and NGC~7245.

\begin{table}
\centering
  \small
  \caption{\label{tab:meanFe}Iron abundances from GALA and iSpec analysis of the 18 OCs computed as the mean of the bona fide member stars. Dispersions are listed as errors. The number of stars to compute the mean in each cluster is indicated.}
  \setlength\tabcolsep{3.5pt}
\begin{tabular}{ccccccccc}
\hline 
Cluster & [Fe/H]$_{\mathrm{GALA}}$ & [Fe/H]$_{\mathrm{iSpec}}$ &Stars\\
\hline
IC 4756  & $ 0.0   \pm 0.03 $ & $ -0.05 \pm 0.02 $ & 8 \\
NGC 188  & $ 0.02  \pm 0.02 $ & $ -0.05 \pm 0.05 $ & 5 \\
NGC 752  & $ 0.02  \pm 0.02 $ & $ -0.04 \pm 0.02 $ & 7 \\
NGC 1817 & $ -0.08 \pm 0.02 $ & $ -0.11 \pm 0.03 $ & 5 \\
NGC 1907 & $ -0.06 \pm 0.03 $ & $ -0.17 \pm 0.03 $ & 5 \\
NGC 2099 & $ 0.08  \pm 0.03 $ & $ 0.0   \pm 0.02 $ & 7 \\
NGC 2420 & $ -0.1  \pm 0.04 $ & $ -0.14 \pm 0.03 $ & 7 \\
NGC 2539 & $ 0.06  \pm 0.01 $ & $ -0.01 \pm 0.03 $ & 5 \\
NGC 2682 & $ 0.04  \pm 0.03 $ & $ -0.06 \pm 0.01 $ & 8 \\
NGC 6633 & $ 0.04  \pm 0.02 $ & $ -0.03 \pm 0.02 $ & 4 \\
NGC 6705 & $ 0.17  \pm 0.04 $ & $ 0.04  \pm 0.05 $ & 8 \\
NGC 6791 & $ 0.2   \pm 0.08 $ & $ 0.19  \pm 0.11 $ & 6 \\
NGC 6819 & $ 0.09  \pm 0.03 $ & $ -0.03 \pm 0.04 $ & 6 \\
NGC 6939 & $ 0.1   \pm 0.04 $ & $ -0.03 \pm 0.04 $ & 5 \\
NGC 6991 & $ 0.02  \pm 0.02 $ & $ -0.04 \pm 0.02 $ & 6 \\
NGC 7245 & $ 0.06  \pm 0.05 $ & $ 0.04  \pm 0.04 $ & 5 \\
NGC 7762 & $ 0.03  \pm 0.04 $ & $ -0.05 \pm 0.04 $ & 5 \\
NGC 7789 & $ 0.06  \pm 0.05 $ & $ -0.05 \pm 0.04 $ & 7 \\
\hline
\end{tabular}
\end{table}

\section{Comparison with literature}
\begin{figure*}
\centering
\includegraphics[width=\textwidth]{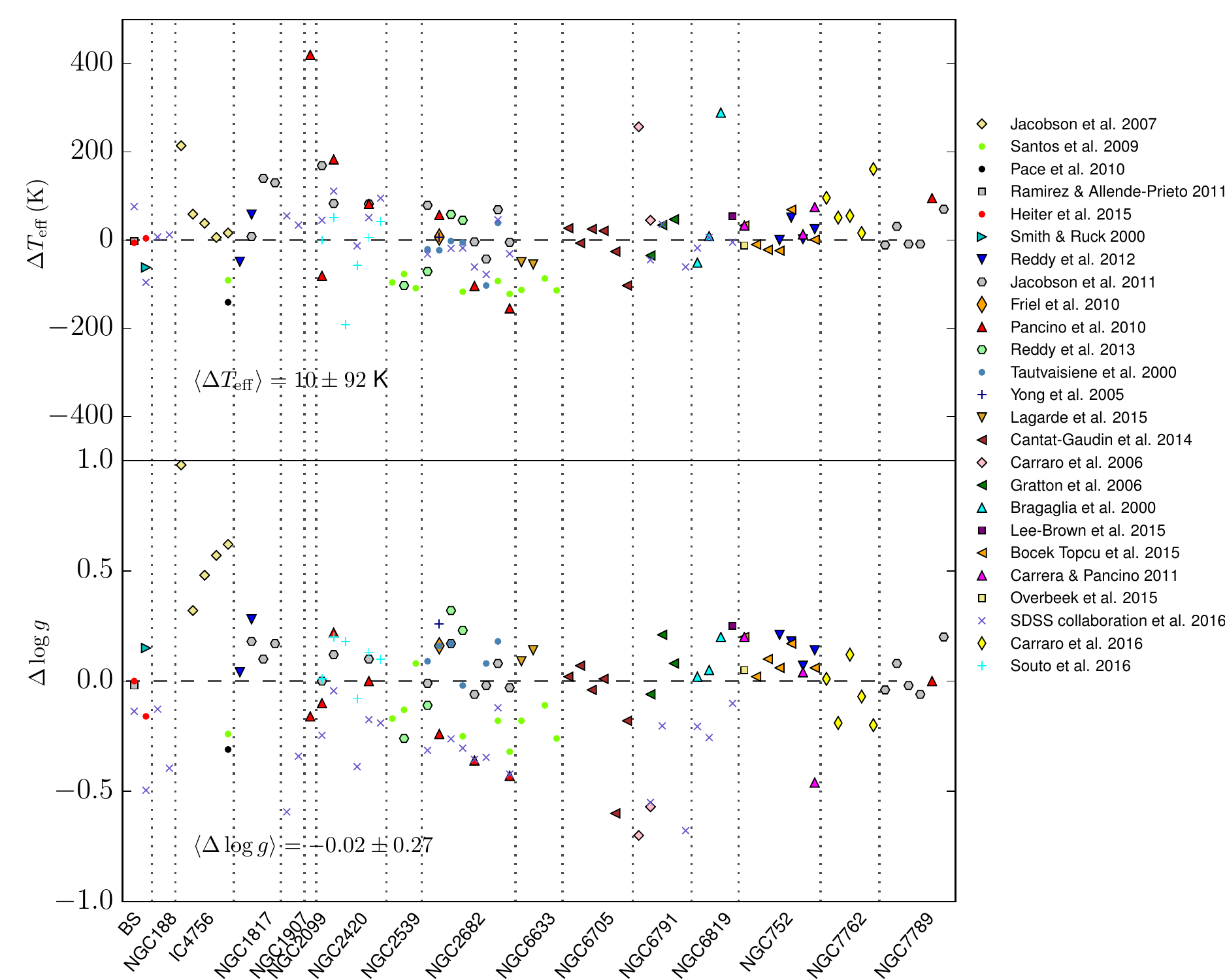}
\caption{Comparison of the derived atmospheric parameters from this study (average values between GALA and iSpec), and previous determinations in the literature. Differences are in the direction this study $-$ literature.}\label{bib_AP}
\end{figure*}

\begin{figure*}
\centering
\includegraphics[width=\textwidth]{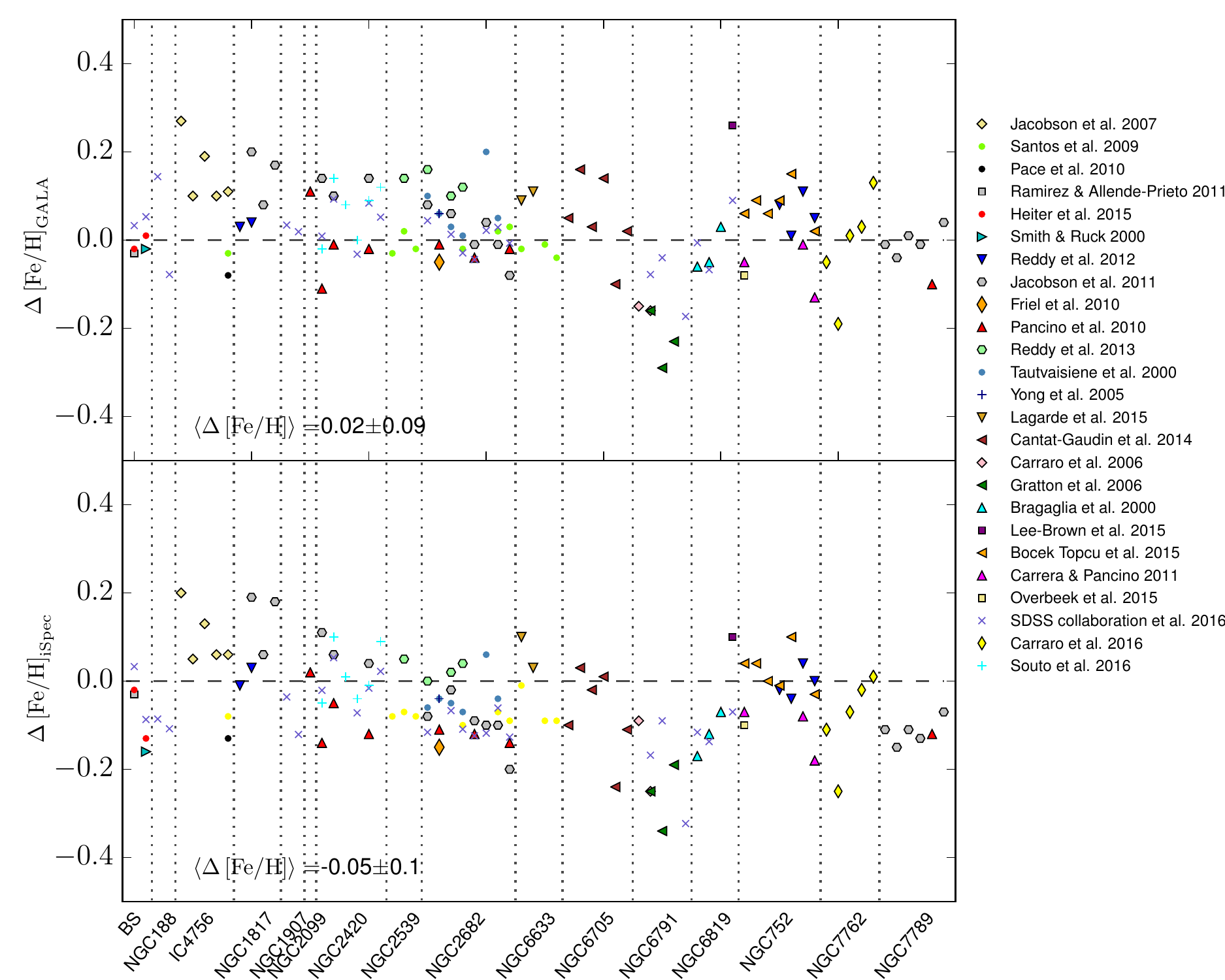}
\caption{Comparison of iron abundances obtained in this study and previous determinations in the literature. In the top pannel we compare values from GALA, in the bottom pannel determinations from iSpec. Differences are in the direction this study $-$ literature.}\label{bib_FeH}
\end{figure*}

Previous works have analysed stars from our sample providing results obtained using different methodologies, resolution, and quality of the spectra. A comparison of our results with those available in the literature provides an independent consistency test for our analysis. We compared the averaged values of $T_{\mathrm{eff}}$, $\log g$, and the two determinations of [Fe/H] with previous measurements in the literature. This is shown in Figs.~\ref{bib_AP} and \ref{bib_FeH}.

In general we find good agreement in effective temperature and surface gravity, with negligible offsets and expected dispersions: $10\pm92$ K, $-0.02\pm0.27$ dex. In metallicity both methods have the same dispersion in comparison with literature with offsets in opposite directions: $0.02\pm0.09$ dex (GALA), $-0.05\pm0.10$ dex (iSpec). These offsets are fully compatible with the quoted dispersions. More importantly, they are consistent with the comparison done in Section~\ref{sec:FeH}, in the sense that we find a systematic difference of $0.07\pm0.05$ dex between the two methods.

There are discrepant cases for particular stars and with some authors, mostly in $\log g$ and [Fe/H], as discussed below. For some of the concerned clusters (IC 4756, NGC 2682, NGC 6791) a detailed metallicity comparison between different authors can also be found in \citet{Heiter+2014}.

\begin{enumerate}[(i)]
 \item \citet{Jacobson+2007} obtained gravities around 0.5 dex lower than ours for IC~4756. However, for the same cluster \citet{Santos+2009} and \citet{Pace+2010} obtain gravities 0.25 dex higher than us. We also have a shift in [Fe/H] of around 0.15/0.1 dex (GALA and iSpec, respectively) with \citet{Jacobson+2007}.
 \item For NGC~6791 \citet{Carraro+2006} finds surface gravities about 0.6 dex higher than ours (2 stars in common), and \citet{Albareti+2016} also finds higher values than us for those stars and high dispersion in the whole cluster. However, \citet{Albareti+2016} finds higher gravities respect to us in the whole sample of common stars. On the contrary, \citet{Gratton+2006} finds very similar results to us for the three stars in common.
 \item \citet{Pancino+2010} finds discrepant gravities, around 0.4 dex higher than us, for the cluster NGC~2682, and more compatible values for the stars in common in: NGC~2099, NGC~2420 and NGC~7789. We also find a quite discrepant value of temperature (400 K higher than them) for the star NGC~2099 W148. Other determinations of gravity of NGC~2682, such as \citet{Jacobson+2011b} and \citet{Tautvaisiene+2000} agree with ours.
 \item We remark the case of NGC~6791, an extensively studied cluster, for which we find lower [Fe/H] than all previous authors with both analysis methods. \citet{Gratton+2006} ($R=29,000$) and \citet{Carretta+2007} ($R=30,000$) find a mean abundance of +0.47 ($\pm0.09$ and $\pm0.12$, respectively), both from the analysis of 4 stars, which is more than 0.25 dex higher than us. \citet{Carraro+2006} found +0.38 using medium resolution spectra ($R=17,000$) of 6 stars, which is still more than 0.15 dex higher. The highest resolution studies are from \citet{Brogaard+2012} ($R=37,000$), \citet{Geisler+2012} ($R=45,000$) and \citet{Boesgaard+2009} ($R=46,000$), which found +0.29, +0.42 and +0.30, respectively. Finally \citet{Albareti+2016} found abundances around 0.15 dex higher than us ($R=22,000$, in the H band). A possible explanation is that all previous studies have lower resolution than us, and this can make a difference for the most metal-rich clusters since they should be more subject to line crowding. Also NGC 6791 stars have the lowest SNR among our sample.
\end{enumerate}

\section{Galactic disk gradients}
In the previous sections we have performed a membership selection based on radial velocities and iron abundances of 18 OCs. We have analysed them in an homogeneous way providing atmospheric parameters and mean iron abundances. The analysed OCs cover a range in Galactocentric radius of 6.8 $<R_{\mathrm{GC}}<$ 10.7 kpc, and span ages between 0.3 and 10.2 Gyr. All the clusters in the sample have $|z|<1$ kpc. Here we discuss the implications of our results on the evolution of the Galactic disk radial metallicity gradient, which is a fundamental constrain for Galactic chemical evolution models. This is a preliminary comparison that will be extended once the full sample of OCs and species are acquired.

\begin{figure*}
\centering
\includegraphics[width=\textwidth]{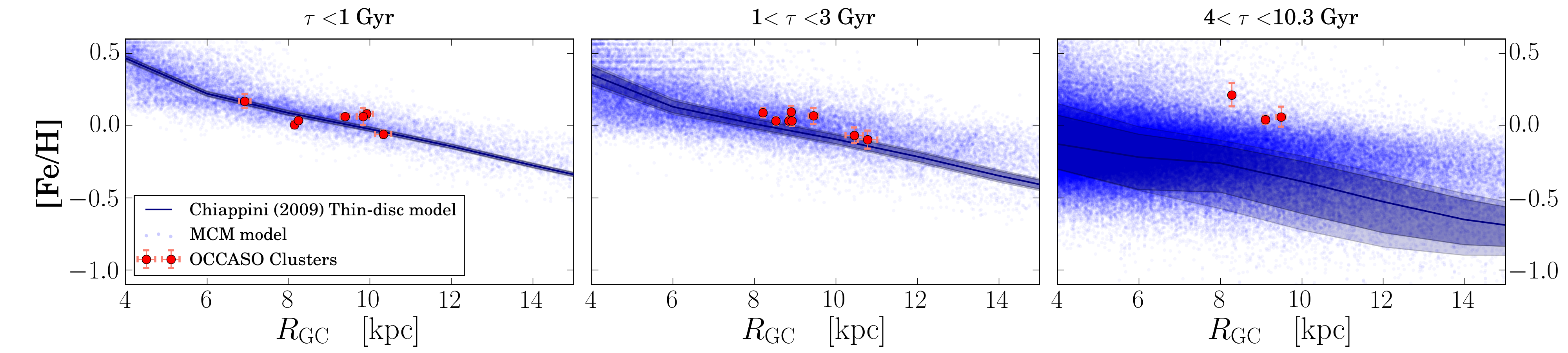}
\caption{[Fe/H] vs $R_{\mathrm{GC}}$ distribution of the 18 OCs in three bins of age. We overplot the pure chemical-evolution model for the thin disc of \citet{Chiappini2009}, and the $N$-body chemo-dynamical model by \citet[MCM]{Minchev+2013,Minchev+2014}.}\label{fig:gradients}
\end{figure*}

In Fig.~\ref{fig:gradients} we show the [Fe/H] vs $R_{\mathrm{GC}}$ distribution of the OCs in 3 bins of age, along with the pure chemical evolution model for the thin disk by \citet{Chiappini2009}, and to the chemo-dynamical thin-disk model by \citet[MCM]{Minchev+2013,Minchev+2014}. The MCM model is a combination of the chemical evolution model of \citet{Chiappini2009} and a high-resolution simulation at a cosmological context, which includes dynamical effects such as radial migration and heating. The abundances of both models are scaled such that the Solar abundance matches the model at the age of the Sun (4.5 Gyr) at the most probable birth position of the Sun (2 kpc closer to the Galactic centre than today; see \citet{Minchev+2013}). This calibration agrees very well with the abundance scale set by local disk Cepheids \citep{Genovali+2013}. The chemical model is shifted by a fix value (+0.1 dex) to fit the Cepheids gradient.

The small uncertainties in iron abundance (Table~\ref{tab:meanFe}) allow us to draw some conclusions. It can be clearly seen that the younger OCs fit perfectly the pure chemical gradient (left panel in Fig.~\ref{fig:gradients}). As OCs get older they start to deviate from the chemical model, and in the oldest bin of age they fall out of it by more than 3$\sigma$. This deviation though, can be explained by the chemo-dynamical model which includes radial mixing, since in fact there are blue points at the position of the two oldest clusters.

\section{Conclusions}
This paper provides the second release of OCCASO, which includes atmospheric parameters ($T_{\text{eff}}$, $\log g$, $\xi$) and [Fe/H] chemical abundances from high-resolution spectra using EW (GALA) and SS (iSpec) methods for 115 stars in 18 OCs.

We made an extensive comparison of the results of both methods to assess our internal consistency and the quoted errors:
\begin{enumerate}[(i)]
\item The comparison between methods of $T_{\text{eff}}$ and $\log g$ per star for the OCs and Arcturus and $\mu$Leo shows that there are no systematic offsets.
\item The comparison of the results obtained by the two methods with the reference values of the GBS also indicates that there are no systematic differences.
\item We calculate atmospheric parameters using Johnson BVI photometry for two OCs: NGC~2420 and NGC~6791. The systematic differences found in the comparison with spectroscopy are inside the errors when varying the assumed $E \left( B-V \right)$, $\left( V_0-M_{\text{V}} \right)$ and [Fe/H], in the photometric analysis.
\end{enumerate}
In all the comparisons we found dispersions of $\sim$60-80 K, 0.15-0.20 dex in $T_{\text{eff}}$ and $\log g$, respectively. The internal dispersion in each cluster from spectroscopy is larger than from photometry.

We calculated [Fe/H] abundances for all OCCASO stars with both methods, using the average values of $T_{\text{eff}}$ and $\log g$. A comparison between the iron abundances from each method showed an offset of $0.07\pm0.05$ dex.

We did several additional tests to investigate the performance of the two methods when calculating iron abundances. We used the GBS sample to derive [Fe/H] in different conditions: (a) fixing $T_{\text{eff}}$ and $\log g$ to their reference value in \citet{Heiter+2015}; (b) fixing also the microturbulence; (c) only using the lines in common to calculate [Fe/H]; (d) only use the common lines and force the synthesis method not to reproduce blends. 

We discussed the [Fe/H] abundances obtained for the stars by OC to perform a more accurate membership selection. With the bona fide member stars we obtained the final values of [Fe/H] per OC. We found cluster dispersions in the range 0.01-0.08 dex from the EW analysis, and 0.01-0.11 dex from the SS analysis. We note that this is the first time chemical abundances are derived from high-resolution spectroscopy for the clusters NGC~6939, NGC~6991 and NGC~7245.

We compared our results with a pure chemical evolution model, and a chemo-dynamical model of the Milky-Way thin disk. We explored the radial gradient in three bins of age obtaining that: the younger OCs fit the gradient drawn by the pure chemical evolution model, and as we go to older ages the metallicity at the traced position can only be explained by the MCM model which adds dynamical effects such as heating and radial migration.

\begin{table*}
\centering
  \scriptsize
  \caption{\label{tab:resultsBS} Results of effective temperature, surface gravity and metallicity for the set of GBS calculated by GALA and iSpec. The reference values are from \citet{Heiter+2015,Jofre+2014}.}
  \setlength\tabcolsep{3.1pt}
\begin{tabular}{ccccccccccc}
Star ID & $T_{\mathrm{eff,ref}}$ & $\log g_{\mathrm{ref}}$ & [Fe/H]$_{\mathrm{ref}}$ & $T_{\mathrm{eff,EW}}$ & $\log g_{\mathrm{EW}}$ & [Fe/H]$_{\mathrm{EW}}$ & $T_{\mathrm{eff,SS}}$ & $\log g_{\mathrm{SS}}$ & [Fe/H]$_{\mathrm{SS}}$ \\
\hline 
HARPS\_HD22879 & $5868$ & $4.27$ & $-0.86$ & $5669\pm23$ & $3.84\pm0.06$ & $-0.960\pm0.045$ & $5861\pm15$ & $4.21\pm0.02$ & $-0.850\pm0.045$ \\ 
NARVAL\_HD22879 & $5868$ & $4.27$ & $-0.86$ & $5698\pm26$ & $3.91\pm0.04$ & $-0.940\pm0.046$ & $5895\pm16$ & $4.28\pm0.03$ & $-0.830\pm0.044$ \\ 
NARVAL\_$\mu$Cas & $5308$ & $4.41$ & $-0.81$ & $5241\pm40$ & $4.19\pm0.05$ & $-0.840\pm0.047$ & $5334\pm15$ & $4.47\pm0.02$ & $-0.810\pm0.044$ \\ 
HARPS\_HD220009 & $4217$ & $1.43$ & $-0.74$ & $4338\pm29$ & $1.91\pm0.05$ & $-0.640\pm0.046$ & $4288\pm5$ & $1.58\pm0.02$ & $-0.710\pm0.048$ \\ 
NARVAL\_HD220009 & $4217$ & $1.43$ & $-0.74$ & $4360\pm38$ & $1.80\pm0.06$ & $-0.610\pm0.047$ & $4274\pm6$ & $1.54\pm0.02$ & $-0.720\pm0.047$ \\ 
HARPS\_$\epsilon$For & $5123$ & $3.52$ & $-0.60$ & $5124\pm26$ & $3.47\pm0.06$ & $-0.540\pm0.045$ & $5001\pm8$ & $3.45\pm0.02$ & $-0.650\pm0.044$ \\ 
ATLAS\_Arcturus & $4286$ & $1.64$ & $-0.52$ & $4354\pm43$ & $1.90\pm0.06$ & $-0.430\pm0.047$ & $4240\pm7$ & $1.50\pm0.02$ & $-0.550\pm0.051$ \\ 
HARPS\_Arcturus & $4286$ & $1.64$ & $-0.52$ & $4345\pm41$ & $1.89\pm0.07$ & $-0.440\pm0.047$ & $4234\pm3$ & $1.42\pm0.02$ & $-0.570\pm0.049$ \\ 
NARVAL\_Arcturus & $4286$ & $1.64$ & $-0.52$ & $4373\pm42$ & $1.79\pm0.09$ & $-0.500\pm0.047$ & $4248\pm5$ & $1.45\pm0.02$ & $-0.590\pm0.050$ \\ 
UVES\_Arcturus-1 & $4286$ & $1.64$ & $-0.52$ & $4387\pm52$ & $1.84\pm0.07$ & $-0.500\pm0.047$ & $4245\pm6$ & $1.52\pm0.02$ & $-0.590\pm0.049$ \\ 
UVES\_Arcturus & $4286$ & $1.64$ & $-0.52$ & $4358\pm52$ & $1.88\pm0.08$ & $-0.490\pm0.047$ & $4240\pm3$ & $1.50\pm0.01$ & $-0.590\pm0.049$ \\ 
ESPADONS\_$\tau$Cet-1 & $5414$ & $4.49$ & $-0.49$ & $5380\pm40$ & $4.43\pm0.04$ & $-0.460\pm0.047$ & $5307\pm5$ & $4.46\pm0.01$ & $-0.490\pm0.044$ \\ 
HARPS\_$\tau$Cet & $5414$ & $4.49$ & $-0.49$ & $5401\pm39$ & $4.48\pm0.05$ & $-0.440\pm0.047$ & $5307\pm10$ & $4.45\pm0.02$ & $-0.500\pm0.044$ \\ 
NARVAL\_$\tau$Cet & $5414$ & $4.49$ & $-0.49$ & $5401\pm47$ & $4.36\pm0.06$ & $-0.450\pm0.047$ & $5314\pm10$ & $4.45\pm0.02$ & $-0.490\pm0.044$ \\ 
ESPADONS\_HD49933-1 & $6635$ & $4.20$ & $-0.41$ & $6551\pm48$ & $3.83\pm0.08$ & $-0.450\pm0.046$ & $6589\pm10$ & $3.97\pm0.02$ & $-0.440\pm0.045$ \\ 
HARPS\_HD49933 & $6635$ & $4.20$ & $-0.41$ & $6495\pm77$ & $3.79\pm0.09$ & $-0.460\pm0.048$ & $6573\pm15$ & $3.93\pm0.04$ & $-0.470\pm0.045$ \\ 
HARPS\_HD107328 & $4496$ & $2.09$ & $-0.33$ & $4417\pm41$ & $1.85\pm0.07$ & $-0.410\pm0.046$ & $4377\pm3$ & $1.69\pm0.02$ & $-0.490\pm0.050$ \\ 
NARVAL\_HD107328 & $4496$ & $2.09$ & $-0.33$ & $4430\pm36$ & $1.90\pm0.08$ & $-0.410\pm0.047$ & $4385\pm4$ & $1.70\pm0.02$ & $-0.490\pm0.050$ \\ 
HARPS\_$\beta$Hyi-w & $5873$ & $3.98$ & $-0.04$ & $5730\pm41$ & $3.67\pm0.06$ & $-0.120\pm0.046$ & $5902\pm10$ & $4.00\pm0.02$ & $-0.050\pm0.043$ \\ 
UVES\_$\beta$Hyi-1 & $5873$ & $3.98$ & $-0.04$ & $5892\pm42$ & $4.06\pm0.05$ & $-0.090\pm0.047$ & $5915\pm11$ & $4.02\pm0.02$ & $-0.050\pm0.044$ \\ 
UVES\_$\beta$Hyi-2 & $5873$ & $3.98$ & $-0.04$ & $5886\pm43$ & $4.06\pm0.04$ & $-0.080\pm0.047$ & $5886\pm18$ & $4.01\pm0.02$ & $-0.070\pm0.043$ \\ 
UVES\_$\beta$Hyi & $5873$ & $3.98$ & $-0.04$ & $5854\pm37$ & $3.94\pm0.04$ & $-0.090\pm0.047$ & $5931\pm8$ & $4.05\pm0.01$ & $-0.040\pm0.044$ \\ 
HARPS\_$\beta$Ara & $4197$ & $1.05$ & $-0.05$ & $4471\pm145$ & $1.63\pm0.26$ & $0.040\pm0.050$ & $4419\pm4$ & $1.13\pm0.02$ & $-0.110\pm0.056$ \\ 
ESPADONS\_Procyon-1 & $6554$ & $4.00$ & $0.01$ & $6626\pm55$ & $3.80\pm0.06$ & $0.000\pm0.046$ & $6439\pm4$ & $3.67\pm0.02$ & $-0.110\pm0.044$ \\ 
HARPS\_Procyon & $6554$ & $4.00$ & $0.01$ & $6632\pm66$ & $3.82\pm0.06$ & $0.030\pm0.046$ & $6404\pm7$ & $3.60\pm0.02$ & $-0.130\pm0.045$ \\ 
NARVAL\_Procyon & $6554$ & $4.00$ & $0.01$ & $6640\pm61$ & $3.74\pm0.10$ & $0.050\pm0.047$ & $6441\pm5$ & $3.68\pm0.02$ & $-0.100\pm0.045$ \\ 
UVES\_Procyon & $6554$ & $4.00$ & $0.01$ & $6572\pm56$ & $3.76\pm0.06$ & $-0.010\pm0.046$ & $6399\pm3$ & $3.61\pm0.01$ & $-0.130\pm0.045$ \\ 
UVES\_Procyon-1 & $6554$ & $4.00$ & $0.01$ & $6608\pm50$ & $3.81\pm0.05$ & $-0.030\pm0.046$ & $6389\pm5$ & $3.57\pm0.02$ & $-0.140\pm0.045$ \\ 
UVES\_Procyon-2 & $6554$ & $4.00$ & $0.01$ & $6513\pm56$ & $3.61\pm0.08$ & $-0.040\pm0.047$ & $6372\pm7$ & $3.55\pm0.03$ & $-0.150\pm0.045$ \\ 
ESPADONS\_18Sco-1 & $5810$ & $4.44$ & $0.03$ & $5858\pm44$ & $4.57\pm0.05$ & $0.080\pm0.047$ & $5814\pm12$ & $4.48\pm0.02$ & $0.080\pm0.043$ \\ 
HARPS\_18Sco & $5810$ & $4.44$ & $0.03$ & $5812\pm37$ & $4.45\pm0.05$ & $0.050\pm0.046$ & $5805\pm16$ & $4.45\pm0.02$ & $0.060\pm0.043$ \\ 
NARVAL\_18Sco & $5810$ & $4.44$ & $0.03$ & $5810\pm43$ & $4.43\pm0.05$ & $0.060\pm0.047$ & $5807\pm12$ & $4.47\pm0.02$ & $0.080\pm0.042$ \\ 
ATLAS\_Sun & $5771$ & $4.44$ & $0.03$ & $5826\pm41$ & $4.51\pm0.04$ & $0.010\pm0.047$ & $5793\pm8$ & $4.48\pm0.01$ & $0.050\pm0.043$ \\ 
HARPS\_Sun-1 & $5771$ & $4.44$ & $0.03$ & $5766\pm45$ & $4.42\pm0.05$ & $0.000\pm0.046$ & $5778\pm11$ & $4.43\pm0.02$ & $0.020\pm0.043$ \\ 
HARPS\_Sun-2 & $5771$ & $4.44$ & $0.03$ & $5740\pm45$ & $4.45\pm0.04$ & $0.000\pm0.046$ & $5786\pm10$ & $4.45\pm0.02$ & $0.020\pm0.043$ \\ 
HARPS\_Sun-3 & $5771$ & $4.44$ & $0.03$ & $5767\pm41$ & $4.40\pm0.06$ & $0.010\pm0.046$ & $5781\pm13$ & $4.43\pm0.02$ & $0.020\pm0.044$ \\ 
HARPS\_Sun-4 & $5771$ & $4.44$ & $0.03$ & $5759\pm44$ & $4.43\pm0.05$ & $0.000\pm0.046$ & $5776\pm8$ & $4.43\pm0.01$ & $0.020\pm0.043$ \\ 
NARVAL\_Sun-1 & $5771$ & $4.44$ & $0.03$ & $5788\pm43$ & $4.50\pm0.05$ & $0.030\pm0.047$ & $5783\pm8$ & $4.46\pm0.01$ & $0.030\pm0.042$ \\ 
NARVAL\_Sun & $5771$ & $4.44$ & $0.03$ & $5757\pm51$ & $4.44\pm0.06$ & $-0.020\pm0.047$ & $5787\pm32$ & $4.45\pm0.05$ & $0.010\pm0.044$ \\ 
UVES\_Sun-1 & $5771$ & $4.44$ & $0.03$ & $5770\pm84$ & $4.47\pm0.05$ & $0.010\pm0.047$ & $5774\pm9$ & $4.45\pm0.01$ & $0.020\pm0.043$ \\ 
UVES\_Sun-2 & $5771$ & $4.44$ & $0.03$ & $5773\pm84$ & $4.39\pm0.05$ & $-0.010\pm0.047$ & $5774\pm20$ & $4.46\pm0.03$ & $0.020\pm0.043$ \\ 
HARPS\_$\delta$Eri-w & $4954$ & $3.76$ & $0.06$ & $4966\pm75$ & $3.73\pm0.04$ & $0.130\pm0.047$ & $5018\pm5$ & $3.70\pm0.01$ & $0.100\pm0.047$ \\ 
NARVAL\_$\delta$Eri & $4954$ & $3.76$ & $0.06$ & $4989\pm46$ & $3.74\pm0.07$ & $0.100\pm0.047$ & $5019\pm7$ & $3.71\pm0.02$ & $0.110\pm0.045$ \\ 
UVES\_$\delta$Eri-1 & $4954$ & $3.76$ & $0.06$ & $4983\pm51$ & $3.76\pm0.05$ & $0.090\pm0.048$ & $5004\pm10$ & $3.70\pm0.02$ & $0.090\pm0.046$ \\ 
UVES\_$\delta$Eri-2 & $4954$ & $3.76$ & $0.06$ & $4959\pm54$ & $3.72\pm0.04$ & $0.110\pm0.049$ & $5005\pm17$ & $3.70\pm0.03$ & $0.090\pm0.045$ \\ 
UVES\_$\delta$Eri & $4954$ & $3.76$ & $0.06$ & $5008\pm48$ & $3.63\pm0.05$ & $0.120\pm0.048$ & $5016\pm6$ & $3.71\pm0.01$ & $0.100\pm0.047$ \\ 
HARPS\_$\beta$Gem & $4858$ & $2.90$ & $0.13$ & $4878\pm37$ & $2.82\pm0.05$ & $0.140\pm0.047$ & $4878\pm5$ & $2.89\pm0.02$ & $0.070\pm0.047$ \\ 
UVES\_$\beta$Gem & $4858$ & $2.90$ & $0.13$ & $4866\pm55$ & $2.93\pm0.07$ & $0.050\pm0.047$ & $4869\pm11$ & $2.98\pm0.02$ & $0.070\pm0.047$ \\ 
ESPADONS\_$\epsilon$Vir & $4983$ & $2.77$ & $0.15$ & $5096\pm51$ & $2.90\pm0.06$ & $0.200\pm0.047$ & $5113\pm7$ & $2.93\pm0.02$ & $0.160\pm0.046$ \\ 
HARPS\_$\epsilon$Vir & $4983$ & $2.77$ & $0.15$ & $5099\pm44$ & $2.91\pm0.05$ & $0.230\pm0.047$ & $5094\pm5$ & $2.85\pm0.02$ & $0.130\pm0.046$ \\ 
NARVAL\_$\epsilon$Vir & $4983$ & $2.77$ & $0.15$ & $5076\pm54$ & $2.91\pm0.07$ & $0.210\pm0.047$ & $5109\pm7$ & $2.93\pm0.02$ & $0.150\pm0.045$ \\ 
ESPADONS\_$\xi$Hya-1 & $5044$ & $2.87$ & $0.16$ & $5005\pm39$ & $2.84\pm0.07$ & $0.090\pm0.047$ & $5088\pm8$ & $3.06\pm0.01$ & $0.120\pm0.046$ \\ 
HARPS\_$\xi$Hya & $5044$ & $2.87$ & $0.16$ & $5055\pm38$ & $2.88\pm0.05$ & $0.140\pm0.047$ & $5081\pm8$ & $3.03\pm0.02$ & $0.110\pm0.046$ \\ 
ESPADONS\_$\beta$Vir-1 & $6083$ & $4.10$ & $0.24$ & $6187\pm85$ & $4.15\pm0.05$ & $0.210\pm0.047$ & $6199\pm9$ & $4.17\pm0.01$ & $0.200\pm0.043$ \\ 
HARPS\_$\beta$Vir & $6083$ & $4.10$ & $0.24$ & $6067\pm109$ & $3.86\pm0.06$ & $0.150\pm0.047$ & $6144\pm12$ & $4.11\pm0.02$ & $0.160\pm0.044$ \\ 
NARVAL\_$\beta$Vir & $6083$ & $4.10$ & $0.24$ & $6183\pm98$ & $4.09\pm0.05$ & $0.230\pm0.047$ & $6186\pm11$ & $4.17\pm0.02$ & $0.200\pm0.043$ \\ 
HARPS\_$\alpha$CenB-w & $5231$ & $4.53$ & $0.22$ & $5211\pm109$ & $4.49\pm0.05$ & $0.210\pm0.047$ & $5172\pm7$ & $4.50\pm0.01$ & $0.240\pm0.045$ \\ 
HARPS\_$\alpha$CenA & $5792$ & $4.31$ & $0.26$ & $5811\pm48$ & $4.44\pm0.05$ & $0.230\pm0.047$ & $5804\pm8$ & $4.32\pm0.01$ & $0.260\pm0.044$ \\ 
HARPS\_$\alpha$CenA-w & $5792$ & $4.31$ & $0.26$ & $5721\pm48$ & $3.86\pm0.06$ & $0.150\pm0.047$ & $5800\pm9$ & $4.31\pm0.01$ & $0.260\pm0.044$ \\ 
UVES\_$\alpha$CenA-1 & $5792$ & $4.31$ & $0.26$ & $5721\pm90$ & $4.08\pm0.08$ & $0.180\pm0.049$ & $5773\pm10$ & $4.30\pm0.02$ & $0.230\pm0.044$ \\ 
ESPADONS\_$\mu$Leo-1 & $4474$ & $2.51$ & $0.25$ & $4426\pm58$ & $2.41\pm0.13$ & $0.300\pm0.050$ & $4488\pm4$ & $2.52\pm0.01$ & $0.200\pm0.053$ \\ 
NARVAL\_$\mu$Leo & $4474$ & $2.51$ & $0.25$ & $4486\pm98$ & $2.35\pm0.16$ & $0.320\pm0.050$ & $4496\pm7$ & $2.54\pm0.01$ & $0.220\pm0.053$ \\ 
HARPS\_$\eta$Boo & $6099$ & $3.79$ & $0.32$ & $5926\pm119$ & $3.23\pm0.09$ & $0.220\pm0.047$ & $6114\pm9$ & $3.89\pm0.02$ & $0.340\pm0.047$ \\ 
NARVAL\_$\eta$Boo & $6099$ & $3.79$ & $0.32$ & $5946\pm87$ & $3.42\pm0.09$ & $0.260\pm0.047$ & $6104\pm14$ & $3.97\pm0.02$ & $0.250\pm0.047$ \\ 
HARPS\_$\mu$Ara & $5902$ & $4.30$ & $0.35$ & $5718\pm44$ & $4.23\pm0.04$ & $0.260\pm0.047$ & $5748\pm12$ & $4.21\pm0.02$ & $0.300\pm0.044$ \\ 
UVES\_$\mu$Ara-1 & $5902$ & $4.30$ & $0.35$ & $5718\pm79$ & $4.14\pm0.06$ & $0.260\pm0.048$ & $5744\pm11$ & $4.25\pm0.02$ & $0.300\pm0.044$ \\ 
UVES\_$\mu$Ara-2 & $5902$ & $4.30$ & $0.35$ & $5804\pm60$ & $4.12\pm0.04$ & $0.300\pm0.048$ & $5737\pm12$ & $4.24\pm0.02$ & $0.300\pm0.044$ \\ 
\hline
\end{tabular}
\end{table*}

\section*{Acknowledgments}
We are greatful to the referee for the suggestions that improved this work.

This work is based on observations made with the Nordic Optical Telescope, operated by the Nordic Optical Telescope Scientific Association, and the Mercator Telescope, operated on the island of La Palma by the Flemish Community, both at the Observatorio del Roque de los Muchachos, La Palma, Spain, of the Instituto de Astrof\'isica de Canarias. This work is also based on observations collected at the Centro Astron\'omico Hispano Alem\'an (CAHA) at Calar Alto, operated jointly by the Max-Planck Institut f\"ur Astronomie and the Instituto de Astrof\'isica de Andaluc\'ia (CSIC).

This research made use of the WEBDA database, operated at the Department of Theoretical Physics and Astrophysics of the Masaryk University, and the SIMBAD database, operated at the CDS, Strasbourg, France.  This work was supported by the MINECO (Spanish Ministry of Economy) - FEDER through grant ESP2016-80079-C2-1-R and ESP2014-55996-C2-1-R and MDM-2014-0369 of ICCUB (Unidad de Excelencia 'Mar\'ia de Maeztu').

RC and CEMV acknowledge support from the IAC (grantP/301204) and from the Spanish Ministry of Economy and Competitiveness (grant AYA2014-56795).

LC acknowledges financial support from the University of Barcelona under the APIF grant, and the financial support by the European Science Foundation (ESF), in the framework of the GREAT Research Networking Programme.

U.H. acknowledges support from the Swedish National Space Board (SNSB/Rymdstyrelsen).

\bibliographystyle{mnras} 
\bibliography{biblio_all}

\begin{thebibliography}{}
\makeatletter
\relax
\def\mn@urlcharsother{\let\do\@makeother \do\$\do\&\do\#\do\^\do\_\do\%\do\~}
\def\mn@doi{\begingroup\mn@urlcharsother \@ifnextchar [ {\mn@doi@}
  {\mn@doi@[]}}
\def\mn@doi@[#1]#2{\def\@tempa{#1}\ifx\@tempa\@empty \href
  {http://dx.doi.org/#2} {doi:#2}\else \href {http://dx.doi.org/#2} {#1}\fi
  \endgroup}
\def\mn@eprint#1#2{\mn@eprint@#1:#2::\@nil}
\def\mn@eprint@arXiv#1{\href {http://arxiv.org/abs/#1} {{\tt arXiv:#1}}}
\def\mn@eprint@dblp#1{\href {http://dblp.uni-trier.de/rec/bibtex/#1.xml}
  {dblp:#1}}
\def\mn@eprint@#1:#2:#3:#4\@nil{\def\@tempa {#1}\def\@tempb {#2}\def\@tempc
  {#3}\ifx \@tempc \@empty \let \@tempc \@tempb \let \@tempb \@tempa \fi \ifx
  \@tempb \@empty \def\@tempb {arXiv}\fi \@ifundefined
  {mn@eprint@\@tempb}{\@tempb:\@tempc}{\expandafter \expandafter \csname
  mn@eprint@\@tempb\endcsname \expandafter{\@tempc}}}

\bibitem[\protect\citeauthoryear{{Alcaino}}{{Alcaino}}{1965}]{Alcaino1965}
{Alcaino} G.,  1965, Lowell Observatory Bulletin, \href
  {http://adsabs.harvard.edu/abs/1965LowOB...6..167A} {6, 167}

\bibitem[\protect\citeauthoryear{{Alonso}, {Arribas}  \&
  {Mart{\'{\i}}nez-Roger}}{{Alonso} et~al.}{1999}]{Alonso+1999}
{Alonso} A.,  {Arribas} S.,   {Mart{\'{\i}}nez-Roger} C.,  1999, \mn@doi [A\&A]
  {10.1051/aas:1999521}, \href
  {http://adsabs.harvard.edu/abs/1999A%26AS..140..261A} {140, 261}

\bibitem[\protect\citeauthoryear{{Andreuzzi}, {Bragaglia}, {Tosi}  \&
  {Marconi}}{{Andreuzzi} et~al.}{2004}]{Andreuzzi+2004}
{Andreuzzi} G.,  {Bragaglia} A.,  {Tosi} M.,   {Marconi} G.,  2004, \mn@doi
  [\mnras] {10.1111/j.1365-2966.2004.07366.x}, \href
  {http://adsabs.harvard.edu/abs/2004MNRAS.348..297A} {348, 297}

\bibitem[\protect\citeauthoryear{{Anstee} \& {O'Mara}}{{Anstee} \&
  {O'Mara}}{1991}]{AnsteeOmara1991}
{Anstee} S.~D.,  {O'Mara} B.~J.,  1991, \mn@doi [\mnras]
  {10.1093/mnras/253.3.549}, \href
  {http://adsabs.harvard.edu/abs/1991MNRAS.253..549A} {253, 549}

\bibitem[\protect\citeauthoryear{{Anthony-Twarog}, {Twarog}, {Kaluzny}  \&
  {Shara}}{{Anthony-Twarog} et~al.}{1990}]{AnthonyTwarog+1990}
{Anthony-Twarog} B.~J.,  {Twarog} B.~A.,  {Kaluzny} J.,   {Shara} M.~M.,  1990,
  \mn@doi [\aj] {10.1086/115436}, \href
  {http://adsabs.harvard.edu/abs/1990AJ.....99.1504A} {99, 1504}

\bibitem[\protect\citeauthoryear{{Anthony-Twarog}, {Twarog}  \&
  {Mayer}}{{Anthony-Twarog} et~al.}{2007}]{Anthonytwarog+2007}
{Anthony-Twarog} B.~J.,  {Twarog} B.~A.,   {Mayer} L.,  2007, \mn@doi [\aj]
  {10.1086/511976}, \href {http://adsabs.harvard.edu/abs/2007AJ....133.1585A}
  {133, 1585}

\bibitem[\protect\citeauthoryear{{Bard} \& {Kock}}{{Bard} \& {Kock}}{1994}]{BK}
{Bard} A.,  {Kock} M.,  1994, Astron. and Astrophys., 282, 1014

\bibitem[\protect\citeauthoryear{{Bard}, {Kock}  \& {Kock}}{{Bard}
  et~al.}{1991}]{BKK}
{Bard} A.,  {Kock} A.,   {Kock} M.,  1991, Astron. and Astrophys., 248, 315

\bibitem[\protect\citeauthoryear{{Barklem} \& {O'Mara}}{{Barklem} \&
  {O'Mara}}{1998}]{BarklemOmara1998}
{Barklem} P.~S.,  {O'Mara} B.~J.,  1998, \mn@doi [\mnras]
  {10.1046/j.1365-8711.1998.01942.x}, \href
  {http://adsabs.harvard.edu/abs/1998MNRAS.300..863B} {300, 863}

\bibitem[\protect\citeauthoryear{{Blackwell}, {Ibbetson}, {Petford}  \&
  {Shallis}}{{Blackwell} et~al.}{1979a}]{BIPS}
{Blackwell} D.~E.,  {Ibbetson} P.~A.,  {Petford} A.~D.,   {Shallis} M.~J.,
  1979a, \mnras, \href {http://cdsads.u-strasbg.fr/abs/1979MNRAS.186..633B}
  {186, 633}

\bibitem[\protect\citeauthoryear{{Blackwell}, {Petford}  \&
  {Shallis}}{{Blackwell} et~al.}{1979b}]{GESB79b}
{Blackwell} D.~E.,  {Petford} A.~D.,   {Shallis} M.~J.,  1979b, \mnras, \href
  {http://adsabs.harvard.edu/abs/1979MNRAS.186..657B} {186, 657}

\bibitem[\protect\citeauthoryear{{Blackwell}, {Petford}, {Shallis}  \&
  {Simmons}}{{Blackwell} et~al.}{1982a}]{GESB82c}
{Blackwell} D.~E.,  {Petford} A.~D.,  {Shallis} M.~J.,   {Simmons} G.~J.,
  1982a, \mnras, \href {http://adsabs.harvard.edu/abs/1982MNRAS.199...43B}
  {199, 43}

\bibitem[\protect\citeauthoryear{{Blackwell}, {Petford}  \&
  {Simmons}}{{Blackwell} et~al.}{1982b}]{GESB82d}
{Blackwell} D.~E.,  {Petford} A.~D.,   {Simmons} G.~J.,  1982b, \mnras, \href
  {http://adsabs.harvard.edu/abs/1982MNRAS.201..595B} {201, 595}

\bibitem[\protect\citeauthoryear{{Blackwell}, {Booth}, {Menon}  \&
  {Petford}}{{Blackwell} et~al.}{1986}]{GESB86}
{Blackwell} D.~E.,  {Booth} A.~J.,  {Menon} S.~L.~R.,   {Petford} A.~D.,  1986,
  \mnras, \href {http://adsabs.harvard.edu/abs/1986MNRAS.220..289B} {220, 289}

\bibitem[\protect\citeauthoryear{{Blanco-Cuaresma}, {Soubiran}, {Jofr{\'e}}  \&
  {Heiter}}{{Blanco-Cuaresma} et~al.}{2014a}]{Blanco+2014}
{Blanco-Cuaresma} S.,  {Soubiran} C.,  {Jofr{\'e}} P.,   {Heiter} U.,  2014a,
  \mn@doi [A\&A] {10.1051/0004-6361/201323153}, \href
  {http://adsabs.harvard.edu/abs/2014A%26A...566A..98B} {566, A98}

\bibitem[\protect\citeauthoryear{{Blanco-Cuaresma}, {Soubiran}, {Heiter}  \&
  {Jofr{\'e}}}{{Blanco-Cuaresma} et~al.}{2014b}]{BlancoCuaresma+2014}
{Blanco-Cuaresma} S.,  {Soubiran} C.,  {Heiter} U.,   {Jofr{\'e}} P.,  2014b,
  \mn@doi [A\&A] {10.1051/0004-6361/201423945}, \href
  {http://adsabs.harvard.edu/abs/2014A%26A...569A.111B} {569, A111}

\bibitem[\protect\citeauthoryear{{Blanco-Cuaresma} et~al.,}{{Blanco-Cuaresma}
  et~al.}{2016a}]{Blanco-Cuaresma2016b}
{Blanco-Cuaresma} S.,  et~al., 2016a, preprint, \href
  {http://adsabs.harvard.edu/abs/2016arXiv160909071B} {} (\mn@eprint {arXiv}
  {1609.09071})

\bibitem[\protect\citeauthoryear{{Blanco-Cuaresma} et~al.,}{{Blanco-Cuaresma}
  et~al.}{2016b}]{Blanco-Cuaresma+2016}
{Blanco-Cuaresma} S.,  et~al., 2016b, in 19th Cambridge Workshop on Cool Stars,
  Stellar Systems, and the Sun (CS19). p.~22 (\mn@eprint {arXiv} {1609.08092}),
  \mn@doi{10.5281/zenodo.155115}

\bibitem[\protect\citeauthoryear{{Boesgaard}, {Jensen}  \&
  {Deliyannis}}{{Boesgaard} et~al.}{2009}]{Boesgaard+2009}
{Boesgaard} A.~M.,  {Jensen} E.~E.~C.,   {Deliyannis} C.~P.,  2009, \mn@doi
  [\aj] {10.1088/0004-6256/137/6/4949}, \href
  {http://adsabs.harvard.edu/abs/2009AJ....137.4949B} {137, 4949}

\bibitem[\protect\citeauthoryear{{Bressan}, {Marigo}, {Girardi}, {Salasnich},
  {Dal Cero}, {Rubele}  \& {Nanni}}{{Bressan} et~al.}{2012}]{Bressan+2012}
{Bressan} A.,  {Marigo} P.,  {Girardi} L.,  {Salasnich} B.,  {Dal Cero} C.,
  {Rubele} S.,   {Nanni} A.,  2012, \mn@doi [\mnras]
  {10.1111/j.1365-2966.2012.21948.x}, \href
  {http://adsabs.harvard.edu/abs/2012MNRAS.427..127B} {427, 127}

\bibitem[\protect\citeauthoryear{{Brogaard} et~al.,}{{Brogaard}
  et~al.}{2012}]{Brogaard+2012}
{Brogaard} K.,  et~al., 2012, \mn@doi [\aap] {10.1051/0004-6361/201219196},
  \href {http://adsabs.harvard.edu/abs/2012A%26A...543A.106B} {543, A106}

\bibitem[\protect\citeauthoryear{{Cantat-Gaudin} \& et. al}{{Cantat-Gaudin} \&
  et. al}{2014}]{cantatgaudin+2014b}
{Cantat-Gaudin} T.,  et. al 2014, \mn@doi [A\&A] {10.1051/0004-6361/201423851},
  \href {http://adsabs.harvard.edu/abs/2014A%26A...569A..17C} {569, A17}

\bibitem[\protect\citeauthoryear{{Cantat-Gaudin} et~al.,}{{Cantat-Gaudin}
  et~al.}{2014}]{cantatgaudin+2014}
{Cantat-Gaudin} T.,  et~al., 2014, \mn@doi [A\&A]
  {10.1051/0004-6361/201322533}, \href
  {http://adsabs.harvard.edu/abs/2014A%26A...562A..10C} {562, A10}

\bibitem[\protect\citeauthoryear{{Cardelli}, {Clayton}  \& {Mathis}}{{Cardelli}
  et~al.}{1989}]{Cardelli+1989}
{Cardelli} J.~A.,  {Clayton} G.~C.,   {Mathis} J.~S.,  1989, \mn@doi [\apj]
  {10.1086/167900}, \href {http://adsabs.harvard.edu/abs/1989ApJ...345..245C}
  {345, 245}

\bibitem[\protect\citeauthoryear{{Carraro}, {Villanova}, {Demarque}, {McSwain},
  {Piotto}  \& {Bedin}}{{Carraro} et~al.}{2006}]{Carraro+2006}
{Carraro} G.,  {Villanova} S.,  {Demarque} P.,  {McSwain} M.~V.,  {Piotto} G.,
   {Bedin} L.~R.,  2006, \mn@doi [ApJ] {10.1086/500801}, \href
  {http://adsabs.harvard.edu/abs/2006ApJ...643.1151C} {643, 1151}

\bibitem[\protect\citeauthoryear{{Carraro}, {Semenko}  \&
  {Villanova}}{{Carraro} et~al.}{2016}]{Carraro+2016}
{Carraro} G.,  {Semenko} E.~A.,   {Villanova} S.,  2016, \mn@doi [\aj]
  {10.3847/1538-3881/152/6/224}, \href
  {http://adsabs.harvard.edu/abs/2016AJ....152..224C} {152, 224}

\bibitem[\protect\citeauthoryear{{Carretta}, {Bragaglia}  \&
  {Gratton}}{{Carretta} et~al.}{2007}]{Carretta+2007}
{Carretta} E.,  {Bragaglia} A.,   {Gratton} R.~G.,  2007, \mn@doi [\aap]
  {10.1051/0004-6361:20065213}, \href
  {http://adsabs.harvard.edu/abs/2007A%26A...473..129C} {473, 129}

\bibitem[\protect\citeauthoryear{{Casamiquela} et~al.,}{{Casamiquela}
  et~al.}{2016}]{Casamiquela+2016}
{Casamiquela} L.,  et~al., 2016, \mn@doi [\mnras] {10.1093/mnras/stw518}, \href
  {http://adsabs.harvard.edu/abs/2016MNRAS.458.3150C} {458, 3150}

\bibitem[\protect\citeauthoryear{{Chiappini}}{{Chiappini}}{2009}]{Chiappini2009}
{Chiappini} C.,  2009, in {Andersen} J.,  {Nordstr{\"o}ara} {m} B.,
  {Bland-Hawthorn} J.,  eds,  IAU Symposium Vol. 254, The Galaxy Disk in
  Cosmological Context. pp 191--196, \mn@doi{10.1017/S1743921308027580}

\bibitem[\protect\citeauthoryear{{Choo} et~al.,}{{Choo}
  et~al.}{2003}]{Choo+2003}
{Choo} K.~J.,  et~al., 2003, \mn@doi [A\&A] {10.1051/0004-6361:20021704}, \href
  {http://adsabs.harvard.edu/abs/2003A%26A...399...99C} {399, 99}

\bibitem[\protect\citeauthoryear{{Den Hartog}, {Ruffoni}, {Lawler},
  {Pickering}, {Lind}  \& {Brewer}}{{Den Hartog} et~al.}{2014}]{GESHRL14}
{Den Hartog} E.~A.,  {Ruffoni} M.~P.,  {Lawler} J.~E.,  {Pickering} J.~C.,
  {Lind} K.,   {Brewer} N.~R.,  2014, preprint, \href
  {http://adsabs.harvard.edu/abs/2014arXiv1409.8142D} {} (\mn@eprint {arXiv}
  {1409.8142})

\bibitem[\protect\citeauthoryear{{Dias}, {Alessi}, {Moitinho}  \&
  {L{\'e}pine}}{{Dias} et~al.}{2002}]{Dias+2002}
{Dias} W.~S.,  {Alessi} B.~S.,  {Moitinho} A.,   {L{\'e}pine} J.~R.~D.,  2002,
  \mn@doi [A\&A] {10.1051/0004-6361:20020668}, \href
  {http://adsabs.harvard.edu/abs/2002A%26A...389..871D} {389, 871}

\bibitem[\protect\citeauthoryear{{Frinchaboy} et~al.,}{{Frinchaboy}
  et~al.}{2013}]{Frinchaboy+2013}
{Frinchaboy} P.~M.,  et~al., 2013, \mn@doi [\apjl]
  {10.1088/2041-8205/777/1/L1}, \href
  {http://adsabs.harvard.edu/abs/2013ApJ...777L...1F} {777, L1}

\bibitem[\protect\citeauthoryear{{Fuhr} \& {Wiese}}{{Fuhr} \&
  {Wiese}}{2006}]{FW06}
{Fuhr} J.~R.,  {Wiese} W.~L.,  2006, \mn@doi [Journal of Physical and Chemical
  Reference Data] {10.1063/1.2218876}, \href
  {http://adsabs.harvard.edu/abs/2006JPCRD..35.1669F} {35, 1669}

\bibitem[\protect\citeauthoryear{{Fuhr}, {Martin}  \& {Wiese}}{{Fuhr}
  et~al.}{1988}]{FMW}
{Fuhr} J.~R.,  {Martin} G.~A.,   {Wiese} W.~L.,  1988, Journal of Physical and
  Chemical Reference Data, Volume 17, Suppl.~4.~New York: American Institute of
  Physics (AIP) and American Chemical Society, 1988, \href
  {http://cdsads.u-strasbg.fr/abs/1988JPCRD..17S....F} {17}

\bibitem[\protect\citeauthoryear{{Geisler}, {Villanova}, {Carraro},
  {Pilachowski}, {Cummings}, {Johnson}  \& {Bresolin}}{{Geisler}
  et~al.}{2012}]{Geisler+2012}
{Geisler} D.,  {Villanova} S.,  {Carraro} G.,  {Pilachowski} C.,  {Cummings}
  J.,  {Johnson} C.~I.,   {Bresolin} F.,  2012, \mn@doi [\apjl]
  {10.1088/2041-8205/756/2/L40}, \href
  {http://adsabs.harvard.edu/abs/2012ApJ...756L..40G} {756, L40}

\bibitem[\protect\citeauthoryear{{Geller}, {Latham}  \& {Mathieu}}{{Geller}
  et~al.}{2015}]{Geller+2015}
{Geller} A.~M.,  {Latham} D.~W.,   {Mathieu} R.~D.,  2015, \mn@doi [\aj]
  {10.1088/0004-6256/150/3/97}, \href
  {http://adsabs.harvard.edu/abs/2015AJ....150...97G} {150, 97}

\bibitem[\protect\citeauthoryear{{Genovali} et~al.,}{{Genovali}
  et~al.}{2013}]{Genovali+2013}
{Genovali} K.,  et~al., 2013, \mn@doi [\aap] {10.1051/0004-6361/201321650},
  \href {http://adsabs.harvard.edu/abs/2013A%26A...554A.132G} {554, A132}

\bibitem[\protect\citeauthoryear{{Gilmore} et~al.,}{{Gilmore}
  et~al.}{2012}]{Gilmore+2012}
{Gilmore} G.,  et~al., 2012, The Messenger, \href
  {http://adsabs.harvard.edu/abs/2012Msngr.147...25G} {147, 25}

\bibitem[\protect\citeauthoryear{{Gratton}, {Bragaglia}, {Carretta}  \&
  {Tosi}}{{Gratton} et~al.}{2006}]{Gratton+2006}
{Gratton} R.,  {Bragaglia} A.,  {Carretta} E.,   {Tosi} M.,  2006, \mn@doi
  [\apj] {10.1086/500729}, \href
  {http://adsabs.harvard.edu/abs/2006ApJ...642..462G} {642, 462}

\bibitem[\protect\citeauthoryear{{Gray} \& {Corbally}}{{Gray} \&
  {Corbally}}{1994}]{Gray+1994}
{Gray} R.~O.,  {Corbally} C.~J.,  1994, \mn@doi [\aj] {10.1086/116893}, \href
  {http://adsabs.harvard.edu/abs/1994AJ....107..742G} {107, 742}

\bibitem[\protect\citeauthoryear{{Grevesse}, {Asplund}  \& {Sauval}}{{Grevesse}
  et~al.}{2007}]{Grevesse+2007}
{Grevesse} N.,  {Asplund} M.,   {Sauval} A.~J.,  2007, \mn@doi [\ssr]
  {10.1007/s11214-007-9173-7}, \href
  {http://adsabs.harvard.edu/abs/2007SSRv..130..105G} {130, 105}

\bibitem[\protect\citeauthoryear{{Gustafsson}, {Edvardsson}, {Eriksson},
  {J{\o}rgensen}, {Nordlund}  \& {Plez}}{{Gustafsson}
  et~al.}{2008}]{Gustafsson2008}
{Gustafsson} B.,  {Edvardsson} B.,  {Eriksson} K.,  {J{\o}rgensen} U.~G.,
  {Nordlund} {\AA}.,   {Plez} B.,  2008, \mn@doi [A\&A]
  {10.1051/0004-6361:200809724}, \href
  {http://adsabs.harvard.edu/abs/2008A%26A...486..951G} {486, 951}

\bibitem[\protect\citeauthoryear{{Harmer}, {Jeffries}, {Totten}  \&
  {Pye}}{{Harmer} et~al.}{2001}]{Harmer+2001}
{Harmer} S.,  {Jeffries} R.~D.,  {Totten} E.~J.,   {Pye} J.~P.,  2001, \mn@doi
  [\mnras] {10.1046/j.1365-8711.2001.04342.x}, \href
  {http://adsabs.harvard.edu/abs/2001MNRAS.324..473H} {324, 473}

\bibitem[\protect\citeauthoryear{{Harris} \& {Harris}}{{Harris} \&
  {Harris}}{1977}]{Harris+1977}
{Harris} G.~L.~H.,  {Harris} W.~E.,  1977, \mn@doi [\aj] {10.1086/112094},
  \href {http://adsabs.harvard.edu/abs/1977AJ.....82..612H} {82, 612}

\bibitem[\protect\citeauthoryear{{Heiter}, {Soubiran}, {Netopil}  \&
  {Paunzen}}{{Heiter} et~al.}{2014}]{Heiter+2014}
{Heiter} U.,  {Soubiran} C.,  {Netopil} M.,   {Paunzen} E.,  2014, \mn@doi
  [\aap] {10.1051/0004-6361/201322559}, \href
  {http://adsabs.harvard.edu/abs/2014A%26A...561A..93H} {561, A93}

\bibitem[\protect\citeauthoryear{{Heiter} et~al.,}{{Heiter}
  et~al.}{2015a}]{Heiter+2015b}
{Heiter} U.,  et~al., 2015a, \mn@doi [\physscr]
  {10.1088/0031-8949/90/5/054010}, \href
  {http://adsabs.harvard.edu/abs/2015PhyS...90e4010H} {90, 054010}

\bibitem[\protect\citeauthoryear{{Heiter}, {Jofr{\'e}}, {Gustafsson}, {Korn},
  {Soubiran}  \& {Th{\'e}venin}}{{Heiter} et~al.}{2015b}]{Heiter+2015}
{Heiter} U.,  {Jofr{\'e}} P.,  {Gustafsson} B.,  {Korn} A.~J.,  {Soubiran} C.,
   {Th{\'e}venin} F.,  2015b, \mn@doi [\aap] {10.1051/0004-6361/201526319},
  \href {http://adsabs.harvard.edu/abs/2015A%26A...582A..49H} {582, A49}

\bibitem[\protect\citeauthoryear{{Hinkel} et~al.,}{{Hinkel}
  et~al.}{2016}]{Hinkel+2016}
{Hinkel} N.~R.,  et~al., 2016, \mn@doi [\apjs] {10.3847/0067-0049/226/1/4},
  \href {http://adsabs.harvard.edu/abs/2016ApJS..226....4H} {226, 4}

\bibitem[\protect\citeauthoryear{{Jacobson}, {Friel}  \&
  {Pilachowski}}{{Jacobson} et~al.}{2007}]{Jacobson+2007}
{Jacobson} H.~R.,  {Friel} E.~D.,   {Pilachowski} C.~A.,  2007, \mn@doi [ApJ]
  {10.1086/520927}, \href {http://adsabs.harvard.edu/abs/2007AJ....134.1216J}
  {134, 1216}

\bibitem[\protect\citeauthoryear{{Jacobson}, {Pilachowski}  \&
  {Friel}}{{Jacobson} et~al.}{2011}]{Jacobson+2011b}
{Jacobson} H.~R.,  {Pilachowski} C.~A.,   {Friel} E.~D.,  2011, \mn@doi [ApJ]
  {10.1088/0004-6256/142/2/59}, \href
  {http://cdsads.u-strasbg.fr/abs/2011AJ....142...59J} {142, 59}

\bibitem[\protect\citeauthoryear{{Jeffries}, {Totten}, {Harmer}  \&
  {Deliyannis}}{{Jeffries} et~al.}{2002}]{Jeffries+2002}
{Jeffries} R.~D.,  {Totten} E.~J.,  {Harmer} S.,   {Deliyannis} C.~P.,  2002,
  \mn@doi [\mnras] {10.1046/j.1365-8711.2002.05788.x}, \href
  {http://adsabs.harvard.edu/abs/2002MNRAS.336.1109J} {336, 1109}

\bibitem[\protect\citeauthoryear{{Jofr{\'e}}, {Heiter}, {Soubiran}  \& et
  al.}{{Jofr{\'e}} et~al.}{2014}]{Jofre+2014}
{Jofr{\'e}} P.,  {Heiter} U.,  {Soubiran} C.,   et al. 2014, \mn@doi [\aap]
  {10.1051/0004-6361/201322440}, \href
  {http://adsabs.harvard.edu/abs/2014A%26A...564A.133J} {564, A133}

\bibitem[\protect\citeauthoryear{{Jofr{\'e}} et~al.,}{{Jofr{\'e}}
  et~al.}{2017}]{Jofre+2016b}
{Jofr{\'e}} P.,  et~al., 2017, \mn@doi [\aap] {10.1051/0004-6361/201629833},
  \href {http://cdsads.u-strasbg.fr/abs/2017A%26A...601A..38J} {601, A38}

\bibitem[\protect\citeauthoryear{{Johnson}}{{Johnson}}{1953}]{Johnson1953}
{Johnson} H.~L.,  1953, \mn@doi [ApJ] {10.1086/145699}, \href
  {http://adsabs.harvard.edu/abs/1953ApJ...117..356J} {117, 356}

\bibitem[\protect\citeauthoryear{{Kharchenko}, {Piskunov}, {R{\"o}ser},
  {Schilbach}  \& {Scholz}}{{Kharchenko} et~al.}{2005a}]{Kharachenko+2005}
{Kharchenko} N.~V.,  {Piskunov} A.~E.,  {R{\"o}ser} S.,  {Schilbach} E.,
  {Scholz} R.-D.,  2005a, \mn@doi [A\&A] {10.1051/0004-6361:20042523}, \href
  {http://adsabs.harvard.edu/abs/2005A%26A...438.1163K} {438, 1163}

\bibitem[\protect\citeauthoryear{{Kharchenko}, {Piskunov}, {R{\"o}ser},
  {Schilbach}  \& {Scholz}}{{Kharchenko} et~al.}{2005b}]{Karchenko+2005}
{Kharchenko} N.~V.,  {Piskunov} A.~E.,  {R{\"o}ser} S.,  {Schilbach} E.,
  {Scholz} R.-D.,  2005b, \mn@doi [\aap] {10.1051/0004-6361:20042523}, \href
  {http://adsabs.harvard.edu/abs/2005A%26A...438.1163K} {438, 1163}

\bibitem[\protect\citeauthoryear{{Kiss}, {Szab{\'o}}, {Szil{\'a}di},
  {Fu{\'e}sz}, {S{\'a}rneczky}  \& {Cs{\'a}k}}{{Kiss} et~al.}{2001}]{Kiss+2001}
{Kiss} L.~L.,  {Szab{\'o}} G.~M.,  {Szil{\'a}di} K.,  {Fu{\'e}sz} G.,
  {S{\'a}rneczky} K.,   {Cs{\'a}k} B.,  2001, \mn@doi [A\&A]
  {10.1051/0004-6361:20010980}, \href
  {http://adsabs.harvard.edu/abs/2001A%26A...376..561K} {376, 561}

\bibitem[\protect\citeauthoryear{{Kock}, {Kroll}  \& {Schnehage}}{{Kock}
  et~al.}{1984}]{KKS84}
{Kock} M.,  {Kroll} S.,   {Schnehage} S.,  1984, \mn@doi [Physica Scripta
  Volume T] {10.1088/0031-8949/1984/T8/013}, \href
  {http://adsabs.harvard.edu/abs/1984PhST....8...84K} {8, 84}

\bibitem[\protect\citeauthoryear{{Kurucz}}{{Kurucz}}{1993}]{Kurucz1993}
{Kurucz} R.,  1993, SYNTHE Spectrum Synthesis Programs and Line Data.~Kurucz
  CD-ROM No.~18.~Cambridge, Mass.: Smithsonian Astrophysical Observatory,
  1993., \href {http://adsabs.harvard.edu/abs/1993KurCD..18.....K} {18}

\bibitem[\protect\citeauthoryear{{Kurucz}}{{Kurucz}}{2005}]{Kurucz2005}
{Kurucz} R.~L.,  2005, Memorie della Societa Astronomica Italiana Supplementi,
  \href {http://adsabs.harvard.edu/abs/2005MSAIS...8...14K} {8, 14}

\bibitem[\protect\citeauthoryear{{Kurucz}}{{Kurucz}}{2007}]{K07}
{Kurucz} R.~L.,  2007, Robert L. Kurucz on-line database of observed and
  predicted atomic transitions

\bibitem[\protect\citeauthoryear{{Maciejewski} \& {Niedzielski}}{{Maciejewski}
  \& {Niedzielski}}{2007}]{Maciejewski+2007}
{Maciejewski} G.,  {Niedzielski} A.,  2007, \mn@doi [A\&A]
  {10.1051/0004-6361:20066588}, \href
  {http://adsabs.harvard.edu/abs/2007A%26A...467.1065M} {467, 1065}

\bibitem[\protect\citeauthoryear{{May}, {Richter}  \& {Wichelmann}}{{May}
  et~al.}{1974}]{MRW}
{May} M.,  {Richter} J.,   {Wichelmann} J.,  1974, \aaps, \href
  {http://cdsads.u-strasbg.fr/abs/1974A%26AS...18..405M} {18, 405}

\bibitem[\protect\citeauthoryear{{McNamara} \& {Solomon}}{{McNamara} \&
  {Solomon}}{1981}]{McNamara+1981}
{McNamara} B.~J.,  {Solomon} S.,  1981, A\&A,Suplement, \href
  {http://adsabs.harvard.edu/abs/1981A%26AS...43..337M} {43, 337}

\bibitem[\protect\citeauthoryear{{Mel{\'e}ndez} \& {Barbuy}}{{Mel{\'e}ndez} \&
  {Barbuy}}{2009}]{MB09}
{Mel{\'e}ndez} J.,  {Barbuy} B.,  2009, \mn@doi [\aap]
  {10.1051/0004-6361/200811508}, \href
  {http://adsabs.harvard.edu/abs/2009A%26A...497..611M} {497, 611}

\bibitem[\protect\citeauthoryear{{Minchev}, {Chiappini}  \& {Martig}}{{Minchev}
  et~al.}{2013}]{Minchev+2013}
{Minchev} I.,  {Chiappini} C.,   {Martig} M.,  2013, \mn@doi [\aap]
  {10.1051/0004-6361/201220189}, \href
  {http://adsabs.harvard.edu/abs/2013A%26A...558A...9M} {558, A9}

\bibitem[\protect\citeauthoryear{{Minchev}, {Chiappini}  \& {Martig}}{{Minchev}
  et~al.}{2014}]{Minchev+2014}
{Minchev} I.,  {Chiappini} C.,   {Martig} M.,  2014, \mn@doi [\aap]
  {10.1051/0004-6361/201423487}, \href
  {http://adsabs.harvard.edu/abs/2014A%26A...572A..92M} {572, A92}

\bibitem[\protect\citeauthoryear{{Mochejska} \& {Kaluzny}}{{Mochejska} \&
  {Kaluzny}}{1999}]{Mochejska+1999}
{Mochejska} B.~J.,  {Kaluzny} J.,  1999, Acta Astronomica, \href
  {http://adsabs.harvard.edu/abs/1999AcA....49..351M} {49, 351}

\bibitem[\protect\citeauthoryear{{Montgomery}, {Marschall}  \&
  {Janes}}{{Montgomery} et~al.}{1993}]{Montgomery+1993}
{Montgomery} K.~A.,  {Marschall} L.~A.,   {Janes} K.~A.,  1993, \mn@doi [ApJ]
  {10.1086/116628}, \href {http://adsabs.harvard.edu/abs/1993AJ....106..181M}
  {106, 181}

\bibitem[\protect\citeauthoryear{{Mucciarelli}, {Pancino}, {Lovisi}, {Ferraro}
  \& {Lapenna}}{{Mucciarelli} et~al.}{2013}]{Mucciarelli+2013}
{Mucciarelli} A.,  {Pancino} E.,  {Lovisi} L.,  {Ferraro} F.~R.,   {Lapenna}
  E.,  2013, \mn@doi [ApJ] {10.1088/0004-637X/766/2/78}, \href
  {http://adsabs.harvard.edu/abs/2013ApJ...766...78M} {766, 78}

\bibitem[\protect\citeauthoryear{{Nilakshi} \& {Sagar}}{{Nilakshi} \&
  {Sagar}}{2002}]{Nilakshi+2002}
{Nilakshi} N.,  {Sagar} R.,  2002, \mn@doi [\aap] {10.1051/0004-6361:20011492},
  \href {http://adsabs.harvard.edu/abs/2002A%26A...381...65N} {381, 65}

\bibitem[\protect\citeauthoryear{{O'Brian}, {Wickliffe}, {Lawler}, {Whaling}
  \& {Brault}}{{O'Brian} et~al.}{1991}]{BWL}
{O'Brian} T.~R.,  {Wickliffe} M.~E.,  {Lawler} J.~E.,  {Whaling} W.,   {Brault}
  J.~W.,  1991, Journal of the Optical Society of America B Optical Physics, 8,
  1185

\bibitem[\protect\citeauthoryear{{Pace}, {Danziger}, {Carraro}, {Melendez},
  {Fran{\c c}ois}, {Matteucci}  \& {Santos}}{{Pace} et~al.}{2010}]{Pace+2010}
{Pace} G.,  {Danziger} J.,  {Carraro} G.,  {Melendez} J.,  {Fran{\c c}ois} P.,
  {Matteucci} F.,   {Santos} N.~C.,  2010, \mn@doi [A\&A]
  {10.1051/0004-6361/200913029}, \href
  {http://adsabs.harvard.edu/abs/2010A%26A...515A..28P} {515, A28}

\bibitem[\protect\citeauthoryear{{Pancino}, {Carrera}, {Rossetti}  \&
  {Gallart}}{{Pancino} et~al.}{2010}]{Pancino+2010}
{Pancino} E.,  {Carrera} R.,  {Rossetti} E.,   {Gallart} C.,  2010, \mn@doi
  [A\&A] {10.1051/0004-6361/200912965}, \href
  {http://adsabs.harvard.edu/abs/2010A%26A...511A..56P} {511, A56}

\bibitem[\protect\citeauthoryear{{Pandey}, {Sharma}, {Upadhyay}, {Ogura},
  {Sandhu}, {Mito}  \& {Sagar}}{{Pandey} et~al.}{2007}]{Pandey+2007}
{Pandey} A.~K.,  {Sharma} S.,  {Upadhyay} K.,  {Ogura} K.,  {Sandhu} T.~S.,
  {Mito} H.,   {Sagar} R.,  2007, \mn@doi [PASJ] {10.1093/pasj/59.3.547}, \href
  {http://adsabs.harvard.edu/abs/2007PASJ...59..547P} {59, 547}

\bibitem[\protect\citeauthoryear{{Platais}, {Kozhurina-Platais}, {Mathieu},
  {Girard}  \& {van Altena}}{{Platais} et~al.}{2003}]{Platais+2003}
{Platais} I.,  {Kozhurina-Platais} V.,  {Mathieu} R.~D.,  {Girard} T.~M.,
  {van Altena} W.~F.,  2003, \mn@doi [\aj] {10.1086/379677}, \href
  {http://adsabs.harvard.edu/abs/2003AJ....126.2922P} {126, 2922}

\bibitem[\protect\citeauthoryear{{Pr{\v s}a} et~al.,}{{Pr{\v s}a}
  et~al.}{2016}]{Prvsa+2016}
{Pr{\v s}a} A.,  et~al., 2016, \mn@doi [\aj] {10.3847/0004-6256/152/2/41},
  \href {http://adsabs.harvard.edu/abs/2016AJ....152...41P} {152, 41}

\bibitem[\protect\citeauthoryear{{Raassen} \& {Uylings}}{{Raassen} \&
  {Uylings}}{1998}]{RU}
{Raassen} A.~J.~J.,  {Uylings} P.~H.~M.,  1998, \aap, \href
  {http://adsabs.harvard.edu/abs/1998A%26A...340..300R} {340, 300}

\bibitem[\protect\citeauthoryear{{Richter} \& {Wulff}}{{Richter} \&
  {Wulff}}{1970}]{RW70}
{Richter} J.,  {Wulff} P.,  1970, \aap, \href
  {http://adsabs.harvard.edu/abs/1970A%26A.....9...37R} {9, 37}

\bibitem[\protect\citeauthoryear{{Rosvick} \& {Vandenberg}}{{Rosvick} \&
  {Vandenberg}}{1998}]{Rosvick+1998}
{Rosvick} J.~M.,  {Vandenberg} D.~A.,  1998, \mn@doi [AJ] {10.1086/300304},
  \href {http://adsabs.harvard.edu/abs/1998AJ....115.1516R} {115, 1516}

\bibitem[\protect\citeauthoryear{{Ruffoni}, {Den Hartog}, {Lawler}, {Brewer},
  {Lind}, {Nave}  \& {Pickering}}{{Ruffoni} et~al.}{2014}]{R14}
{Ruffoni} M.~P.,  {Den Hartog} E.~A.,  {Lawler} J.~E.,  {Brewer} N.~R.,  {Lind}
  K.,  {Nave} G.,   {Pickering} J.~C.,  2014, \mn@doi [\mnras]
  {10.1093/mnras/stu780}, \href
  {http://adsabs.harvard.edu/abs/2014MNRAS.441.3127R} {441, 3127}

\bibitem[\protect\citeauthoryear{{SDSS Collaboration} et~al.,}{{SDSS
  Collaboration} et~al.}{2016}]{Albareti+2016}
{SDSS Collaboration} et~al., 2016, preprint, \href
  {http://adsabs.harvard.edu/abs/2016arXiv160802013S} {} (\mn@eprint {arXiv}
  {1608.02013})

\bibitem[\protect\citeauthoryear{{Salaris}, {Weiss}  \& {Percival}}{{Salaris}
  et~al.}{2004}]{Salaris+2004}
{Salaris} M.,  {Weiss} A.,   {Percival} S.~M.,  2004, \mn@doi [\aap]
  {10.1051/0004-6361:20031578}, \href
  {http://adsabs.harvard.edu/abs/2004A%26A...414..163S} {414, 163}

\bibitem[\protect\citeauthoryear{{Sandage}, {Lubin}  \& {VandenBerg}}{{Sandage}
  et~al.}{2003}]{Sandage+2003}
{Sandage} A.,  {Lubin} L.~M.,   {VandenBerg} D.~A.,  2003, \mn@doi [\pasp]
  {10.1086/378243}, \href {http://adsabs.harvard.edu/abs/2003PASP..115.1187S}
  {115, 1187}

\bibitem[\protect\citeauthoryear{{Santos}, {Lovis}, {Pace}, {Melendez}  \&
  {Naef}}{{Santos} et~al.}{2009}]{Santos+2009}
{Santos} N.~C.,  {Lovis} C.,  {Pace} G.,  {Melendez} J.,   {Naef} D.,  2009,
  \mn@doi [A\&A] {10.1051/0004-6361:200811093}, \href
  {http://adsabs.harvard.edu/abs/2009A%26A...493..309S} {493, 309}

\bibitem[\protect\citeauthoryear{{Sbordone}, {Bonifacio}, {Castelli}  \&
  {Kurucz}}{{Sbordone} et~al.}{2004}]{Sbordone+2004}
{Sbordone} L.,  {Bonifacio} P.,  {Castelli} F.,   {Kurucz} R.~L.,  2004,
  Memorie della Societa Astronomica Italiana Supplementi, \href
  {http://adsabs.harvard.edu/abs/2004MSAIS...5...93S} {5, 93}

\bibitem[\protect\citeauthoryear{{Smiljanic} et~al.,}{{Smiljanic}
  et~al.}{2014}]{Smiljanic+2014}
{Smiljanic} R.,  et~al., 2014, \mn@doi [\aap] {10.1051/0004-6361/201423937},
  \href {http://adsabs.harvard.edu/abs/2014A%26A...570A.122S} {570, A122}

\bibitem[\protect\citeauthoryear{{Stetson}}{{Stetson}}{2000}]{Stetson2000}
{Stetson} P.~B.,  2000, \mn@doi [\pasp] {10.1086/316595}, \href
  {http://adsabs.harvard.edu/abs/2000PASP..112..925S} {112, 925}

\bibitem[\protect\citeauthoryear{{Stetson} \& {Pancino}}{{Stetson} \&
  {Pancino}}{2008}]{Stetson+2008}
{Stetson} P.~B.,  {Pancino} E.,  2008, \mn@doi [PASP] {10.1086/596126}, \href
  {http://adsabs.harvard.edu/abs/2008PASP..120.1332S} {120, 1332}

\bibitem[\protect\citeauthoryear{{Stetson}, {Bruntt}  \& {Grundahl}}{{Stetson}
  et~al.}{2003}]{Stetson+2003}
{Stetson} P.~B.,  {Bruntt} H.,   {Grundahl} F.,  2003, \mn@doi [PASP]
  {10.1086/368337}, \href {http://adsabs.harvard.edu/abs/2003PASP..115..413S}
  {115, 413}

\bibitem[\protect\citeauthoryear{{Subramaniam} \& {Bhatt}}{{Subramaniam} \&
  {Bhatt}}{2007}]{Subramaniam+2007}
{Subramaniam} A.,  {Bhatt} B.~C.,  2007, \mn@doi [\mnras]
  {10.1111/j.1365-2966.2007.11648.x}, \href
  {http://adsabs.harvard.edu/abs/2007MNRAS.377..829S} {377, 829}

\bibitem[\protect\citeauthoryear{{Subramaniam} \& {Sagar}}{{Subramaniam} \&
  {Sagar}}{1999}]{Subramaniam+1999}
{Subramaniam} A.,  {Sagar} R.,  1999, \mn@doi [\aj] {10.1086/300716}, \href
  {http://adsabs.harvard.edu/abs/1999AJ....117..937S} {117, 937}

\bibitem[\protect\citeauthoryear{{Sung}, {Bessell}, {Lee}, {Kang}  \&
  {Lee}}{{Sung} et~al.}{1999}]{Sung+1999}
{Sung} H.,  {Bessell} M.~S.,  {Lee} H.-W.,  {Kang} Y.~H.,   {Lee} S.-W.,  1999,
  \mn@doi [MNRAS] {10.1046/j.1365-8711.1999.02961.x}, \href
  {http://adsabs.harvard.edu/abs/1999MNRAS.310..982S} {310, 982}

\bibitem[\protect\citeauthoryear{{Tautvai{\v s}iene}, {Edvardsson}, {Tuominen}
  \& {Ilyin}}{{Tautvai{\v s}iene} et~al.}{2000}]{Tautvaisiene+2000}
{Tautvai{\v s}iene} G.,  {Edvardsson} B.,  {Tuominen} I.,   {Ilyin} I.,  2000,
  A\&A, \href {http://adsabs.harvard.edu/abs/2000A%26A...360..499T} {360, 499}

\bibitem[\protect\citeauthoryear{{Vogel}, {Braden}, {Deliyannis}, {Steinhauer}
  \& {Jacobson}}{{Vogel} et~al.}{2003}]{Vogel+2003}
{Vogel} K.~W.,  {Braden} E.~K.,  {Deliyannis} C.~P.,  {Steinhauer} A.,
  {Jacobson} H.,  2003, in American Astronomical Society Meeting Abstracts.
  p.~1228

\bibitem[\protect\citeauthoryear{{Wolnik}, {Berthel}  \& {Wares}}{{Wolnik}
  et~al.}{1970}]{WBW70}
{Wolnik} S.~J.,  {Berthel} R.~O.,   {Wares} G.~W.,  1970, \mn@doi [\apj]
  {10.1086/150735}, \href {http://adsabs.harvard.edu/abs/1970ApJ...162.1037W}
  {162, 1037}

\bibitem[\protect\citeauthoryear{{Wolnik}, {Berthel}  \& {Wares}}{{Wolnik}
  et~al.}{1971}]{WBW}
{Wolnik} S.~J.,  {Berthel} R.~O.,   {Wares} G.~W.,  1971, \mn@doi [\apjl]
  {10.1086/180733}, \href {http://cdsads.u-strasbg.fr/abs/1971ApJ...166L..31W}
  {166, L31+}

\makeatother
\end{thebibliography}



\appendix

\bsp	

\label{lastpage}
\end{document}